\documentclass[a4paper,twosided]{article}

\usepackage{amssymb}
\usepackage{subfigure}

\usepackage[utf8]{inputenc}
\usepackage{amsmath,amssymb}
\usepackage{graphicx}
\usepackage{subfigure}
\usepackage{color}
\usepackage{array}

\usepackage{algorithm}
\usepackage{algorithmic}

\begin{document}

\title{On the Degree Distribution of Faulty Peer-to-Peer Overlays}
\author{Stefano Ferretti\\
Department of Computer Science, University of Bologna\\ 
Mura Anteo Zamboni 4, 40127, Bologna, Italy\\
sferrett@cs.unibo.it}
\date{}
\maketitle

% \begin{frontmatter}


% % \title{Title\tnoteref{label1}}
% % \tnotetext[label1]{}
% \fnref{Tel: +39 051 2094845, Fax: +39 051 2094510}}


\begin{abstract}
This paper presents an analytical framework to model fault-tolerance in unstructured peer-to-peer overlays, represented as complex networks. We define a distributed protocol peers execute for managing the overlay and reacting to node faults. Based on the protocol, evolution equations are defined and manipulated by resorting to generating functions. Obtained outcomes provide insights on the nodes' degree probability distribution. 
From the study of the degree distribution, it is possible to estimate other important metrics of the peer-to-peer overlay, such as the diameter of the network.
We study different networks, characterized by three specific desired degree distributions, i.e.~nets with nodes having a fixed desired degree, random graphs and scale-free networks. All these networks are assessed via the analytical tool and simulation as well. Results show that the approach can be factually employed to dynamically tune the average attachment rate at peers so that they maintain their own desired degree and, in general, the desired network topology.
\end{abstract}

% \begin{keyword}
% Peer-to-Peer \sep Complex Networks \sep Degree Distribution
% %% keywords here, in the form: keyword \sep keyword
% 
% %% PACS codes here, in the form: \PACS code \sep code
% 
% %% MSC codes here, in the form: \MSC code \sep code
% %% or \MSC[2008] code \sep code (2000 is the default)
% 
% \end{keyword}

% \linenumbers

\section{Introduction}
\label{sec:intro}

The mechanics of complex networks represent an insightful research domain for those that try to understand the behavior and the characteristics of a network by looking at its general (statistical) properties. Basically, the focus concerns the organization and the interaction among multiple nodes in a dynamical system  \cite{Barabasi2000,Verlag03structuralproperties,Newman03thestructure}. 
% The graphs resulting from the construction of the map of interaction among nodes, together with their properties, allow  to model different 
The theory and methods of analysis can be applied in the same fashion to 
existing real and abstract networks belonging to several domains, e.g.~biology, sociology, physics, computer science \cite{Nature2001_Amaral,broder,JeoMasBarOlt01,Faloutsos:1999,Price:1965}.
Examples of statistical properties of common interest are the probability that nodes have a certain degree (i.e.~the number of neighbours connected to a given node), the probability that a node has links with the friends of its friends (which allows to understand how much the network is organized in clusters), the average number of second (third, etc.) neighbours (which provides insights on the size of the network component of a given node), the network diameter, etc.
All these metrics reveal some features of a given network, such as its ability to disseminate information and/or propagate viruses, its resilience to nodes' departure, its connectivity \cite{Newman03thestructure,2000Nature_Albert,newmanHandbook,cohen2000res,leonard,wu}. 

As for computer networks, modeling peer-to-peer overlays as complex nets allows to understand the level of reliability, scalability and tolerance to faults of these overlays. This is basically the purpose of this paper. Specifically, we provide a framework to model self-organizing, unstructured peer-to-peer architectures with periodical faults.
% . In essence, 
Nodes of the network simply correspond to peers, while edges represent a communication connection between two peers \cite{androu,narada,holzer,mobiopp,ioannidis,pompili,ShahabiK05,wang,donet}. (Since nodes of the modeled network represent peers in the distributed system, hereinafter the terms \textit{node} and \textit{peer} will be used as synonyms.) 
In general, a peer-to-peer network is characterized by specifying: i) the system model, i.e.~the environment of execution of the peers, together with the types of faults they are subject to; and ii) the distributed communication protocol, i.e.~how peers connect and interact with other nodes in the net.

The peer-to-peer network is unstructured, in the sense that the overlay is constructed based on some general desired topology that does not depend on the contents being disseminated through the net \cite{EberspacherS05a}. 
Rather, local choices are made by each peer to manage its connections. This may lead to a non-optimal organization of the overlay, from the view-point of the content distribution. However, the costs for managing such overlays are very limited.
Thus, unstructured systems may have better performances in highly dynamic environments \cite{SchmidW07}.

The system is composed of a set of peers that may fail during the evolution of the network.
Node failures are modeled as random variables characterized by an average failure rate, as usual. 
A node failure does not cause the complete removal of the peer from the network. Rather, the peer loses all its links.
Based on the protocol we define, peers react to these disconnections by actively creating novel links with their non-neigh\-bo\-urs, trying to maintain a specific desired degree. As mentioned, the overlay is unstructured; thus, it is assumed that a self-organizing mechanism is employed to govern the network dynamics. Hence, local decisions are taken by peers to manage disconnections, without the intervention of a central entity \cite{holzer}. 
The procedure related to the discovery of a non-neighbour and the creation of a novel edge is periodically executed, based on another rate.

Once having defined the system model and the distributed protocol peers execute, we provide a mathematical analysis on the evolution of the nodes' degree. This is accomplished by introducing an infinite set of differential equations. Then, these equations are turned into a single differential equation by exploiting generating functions. Its solution allows to calculate the nodes' degree probability.

The novelty of this proposal is essentially due to the dynamic behavior of peers. Classic works on complex networks usually concentrate on node removals without the possibility to resort to some counter-mechanism to be executed, corresponding to a dynamic and continuous reconfiguration of the network \cite{Newman03thestructure,broder,2000Nature_Albert,newmanHandbook}. Indeed, a ``passive'' behavior may perfectly model a viral propagation of diseases in human contact nets, denial of services in computer nets, and general sudden attacks in a network, which do not evolve during the period of the attack (or rather, the system evolution proceeds at a pace significantly slower than the attack). Conversely, this kind of approaches cannot model the typical interactions of self-organizing peer-to-peer architectures, 
commonly exploited in unstructured overlay management techniques,
with peers being programmed to dynamically react to (or prevent) possible node faults.
The framework provided in this paper allows to determine the degree distribution at peers in presence of node faults (and link creation) which occur during the whole system evolution. Concurrently, all the reasoning related to complex networks theory can be applied.

We compare the mathematical model with results obtained from a simulative assessment that mimics the corresponding distributed protocol. We vary the nodes' desired degree distribution. Specifically, we study three classic (desired) network topologies: i) uniform networks where all nodes have the same desired degree, ii) random graphs, iii) scale-free networks. Results show that the two different (theoretical and simulative) approaches provide similar outcomes, hence confirming the correctness of the proposal. Not only, they provide insights on the degree that peers succeed to maintain in presence of node faults. In fact, being the network continuously affected by node faults, and being nodes able to create novel links based on local (self-regulated) choices, it turns out that peers can maintain their own desired degree only when a high attachment rate is utilized (w.r.t.~the failure rate). 
Once the degree distribution has been calculated, given the system settings, it is possible to estimate
% of a given network may evolve, due to the experienced average failure and the average attachment rate, allows to (statistically) calculate 
the average number of second (third, etc.) neighbours, as well as the average size of the component a peer is connected to. In particular, we estimate the diameter of the considered networks.

Of course, ensuring that peers have an actual degree equal (or similar) to their desired degree is mandatory to guarantee that the structure of the peer-to-peer network corresponds to its desired topology.
Hence, the provided analytical tool can be factually exploited at peers to dynamically identifying a proper attachment rate they might maintain during the distributed interactions, based on the experienced node failure rate.
Simple algorithms may be thus implemented, that allow to adapt the attachment rate.

The remainder of the paper is organized as follows. Section \ref{sec:protocol} presents the distributed protocol we consider. Section \ref{sec:model} describes the analytical modeling of such a protocol. In Section \ref{sec:res}, results coming from a simulation study are outlined. These outcomes are compared with the numerical results obtained through the presented model. Finally, Section \ref{sec:conc} provides some concluding remarks.

\section{The Distributed Protocol}
\label{sec:protocol}

Consider a distributed system composed of a set of peers $\Pi$. Communication amo\-ng peers occurs through an overlay network. The system is faulty, in the sense that nodes may fail during their interaction with other ones. When a node failure happens, the peer loses all its links with its neighbours. After the failure, the peer is instantaneously able to create novel link connections, i.e.~the time needed by the peer to restart its local system and re-join the network is assumed to be negligible.

In the model, we consider node faults, rather that link faults,
since in an unstructured peer-to-peer system it is more likely that a peer fails, rather than a single edge of the graph permanently fails. A node may fail because of a voluntarily action taken by the user that decides to leave the network, or when the peer remains isolated from the rest of the network, due for instance to some technical problems which prevent that node to communicate with its Internet Service Provider, or when it loses its network coverage (hence losing all its connections with the rest of the world). 
Conversely, while still possible the removal of a single link in a peer-to-peer overlay network (with both peers remaining active) should be less frequent. Of course, TCP/UDP connections among two hosts, representing the transport-layer implementation of a link among two peers, may be interrupted due to several reasons. However, from a networking point of view, several techniques can be exploited such as, for instance,
% Such claim is further reinforced by the actual P2P communications may resort to 
session-layer protocols, which augment the reliability of an end-to-end communication 
% with respect to standard transport-level communications 
\cite{aict,disio_ghini}.

Due to the dynamic and evolving nature of the network, we enable peers to create novel links with non-neighbours; this is accomplished through a local, random choice taken by the peer. 
% The rate probability of creation of a novel link is controlled by the parameter $\alpha$. 
Peers have a specific chosen degree and try to maintain it during the system evolution, in spite of nodes' faults. In substance, nodes select a \emph{desired degree} ($dd$), whose value
% Such a desired degree is basically a maximum possible degree value peer have. 
% The selected $dd$ value 
might depend on the specific characteristics of the node, e.g.~computational and network capacities, role of the node in the network. 
When modeling the network, to characterize its desired topology, $dd$ values will be assigned to nodes by utilizing some statistical distribution. As an example, for the sake of load balancing, 
peers' $dd$s could be forced to assume values within a limited range (or a single value). Instead, the use of other desired degree distributions, such as power laws (typical of scale-free nets), would mimic hybrid multi-level peer-to-peer networks with the presence of hubs/super-peers.

Based on their $dd$, during the system evolution peers that have an actual degree lower than such a value periodically start a discovery process to find a novel neighbour. We assume that when a peer asks another one to establish a novel link in the overlay, the latter refuses it only if its actual degree is equal to its $dd$. Otherwise, it accepts the link creation.

%%%%%%%%%%%%%%%%%%

\begin{algorithm}[t]
\caption{Distributed Protocol: Attachment Process} 
\label{alg:attachment}
\begin{algorithmic}
\STATE \textbf{vars}: \textit{actualDegree}: current degree of the node executing the protocol
\STATE \qquad \mbox{   } $dd$: desired degree of the node executing the protocol
\STATE %
\STATE \textbf{precondition}: \textit{actualDegree} $< dd$%
\STATE %
\STATE found = \textit{false};
\WHILE{($\neg$ found)}%
  \STATE p = \textit{NonNeighbourDiscovery}(); %
  \STATE \textit{sendLinkCreationRequest}(p);
  \STATE ans = \textit{receiveAnswer}();
  \IF{(ans == ``ok'')}
    \STATE found = \textit{true};
    \STATE \textit{createNovelLink}(p);
    \STATE \textit{actualDegree} ++;
  \ENDIF
\ENDWHILE
\STATE \textit{waitRandomTime}();
\end{algorithmic}
\end{algorithm}

The distributed protocol discussed above is summarized in Algorithms \ref{alg:attachment}-\ref{alg:answer}. Basically, when the actual degree of a node is lower than $dd$ (the precondition in Algorithm \ref{alg:attachment}), a discovery process is activated to find novel neighbours. Algorithm \ref{alg:attachment} does not report a specific implementation of the discovery of a non-neighbour, since several alternatives are possible, not strictly dependent on the protocol under consideration. 
We just basically assume that the selection of the new neighbour is accomplished by randomly picking up a peer, as made in most unstructured peer-to-peer overlay networks \cite{guclu,HaridasanvanRenesse08,keidar,linlin,Qi}. 
To find the novel node, a distributed oracle (or some approximation of it, obtained through local interactions) is employed which provides the complete list of active peers. 
% Possible solutions that factually implement such a discovery service range from the use of a single (centralized or distributed) lobby service, to more sophisticated peer-to-peer approaches, like those based on Distributed Hash Tables (DHTs) \cite{kelaskar,pastry}.
Once a novel peer has been found, a request is sent to that peer. If a positive answer is received, a novel link is created. Otherwise, the node looks for another peer. Note that in the pseudo-code a random sleep has been inserted, to state that such procedure should be periodically executed while the node seeks to reach an actual degree equal to its $dd$.

\begin{algorithm}[t]
\caption{Distributed Protocol: Upon Request for a Novel Link} 
\label{alg:answer}
\begin{algorithmic}
\STATE \textbf{vars}: \textit{actualDegree}: current degree of the node executing the protocol
\STATE \qquad \mbox{   } $dd$: desired degree of the node executing the protocol
\STATE %
\STATE \textbf{precondition}: message received for link creation%
\STATE %
\STATE p = \textit{sendingPeer}(); %
\IF{(actualDegree $<dd$)}
  \STATE \textit{sendPositiveAnswer}(p);
  \STATE \textit{createNovelLink}(p);
  \STATE actualDegree ++;
\ELSE
  \STATE \textit{sendNegativeAnswer}(p);
\ENDIF
\end{algorithmic}
\end{algorithm}

Algorithm \ref{alg:answer} is executed upon request for a novel link from a non-neighbour. The behavior is quite simple, if the receiving node has an actual degree lower than its $dd$, it accepts the request and a novel link is created. Otherwise, it refuses the request.

\section{Modeling the System as a Complex Network}
\label{sec:model}

In this section, we show that the presented system can be modeled as a complex network, through the use of differential equations and generating functions. 
% We discuss the methodology to analyze these kinds of models and obtain a degree distribution that refers to the distributed protocol being employed.
% Basically, in the model each node has assigned a degree probability distribution $D_k=P(deg=k)$, i.e.~the probability that a node $n$ has a specific number $k$ of nodes connected to it (the \emph{neighbours} of $n$). 
Nodes' failures are modeled as random variables characterized by an average rate $\phi$. Moreover, we assume that
the rate of creation of a novel link is controlled by the parameter $\alpha$. It is the difference between $\alpha$ and $\phi$ that determines how peers react to failures. The attachment and failure rates $\alpha, \phi$ do not depend on any specific characteristics of the peers (e.g.~node degree). 
% Rather, random choices are made, being $\alpha, \phi$ not dependent on the node degree. 
This means that the model does not consider any form of preferential attachment, which would privilege nodes with higher (lower) degrees \cite{Newman03thestructure}, neither that nodes with higher (lower) degrees are likely to fail, i.e.~those nodes that have much (less) workload in the communication network. 

% In particular, the choice $\alpha=\phi$ models, from a statistical point of view, a system where peers react to neighbours' faults by actively searching for novel neighbours, with the same rate of faults. Conversely, $\alpha<\phi$ may correspond to a lazy reaction by peers, while $\alpha>\phi$ models a system where peers actively search novel neighbours even not in presence of faults, but to prevent them.

\subsection{Preliminaries and Methodology}

Here, a general overview is provided on the methodology employed to model the distributed protocol. 
% The approach is in general not that different to works which model dynamical systems as complex networks. 
The idea is to define the evolution equations describing how the system evolves in time. In practice, for each possible degree, a differential equation is defined which characterizes the probability that a peer, having such a degree, may change its state.
The model will be composed of an infinite set of simultaneous linear differential equations (one for each possible degree). These equations will be turned into a single differential equation by exploiting generating functions.

A probability generating function is of the form $F(x,t) = \sum_{i\geq 0} D_i(t) x^i$, where $D_i(t)$ is the set of coefficients composing the power series (in our case, these coefficients are the probabilities of having a certain degree $i$, at time $t$), while $x$ is a dummy variable, employed for pure algebraic purposes. 
$F(x,t)$ captures all the information present in the original sequence $D_i(t)$, as each of these probabilities can be recovered by simple differentiation:
\begin{eqnarray}
D_i(t) = [x_i]F =  \frac{1}{i!}\frac{\partial^i F}{\partial x^i}\Big\rvert_{x=0}.\nonumber
% \label{}
\end{eqnarray}
The notation $[x_i]F$ represents the coefficient associated to the term $x^i$ in the power series.

In general, many properties can be obtained by evaluating some manipulation of the generating function, at $x=1$. For instance, having probabilities as coefficients of the power series, a check to perform is to assess whether the sum of all coefficients in $F$ equals $1$, i.e~$F(1,t) = 1$. Moreover, the average of the coefficients composing the generating function can be measured by evaluating the partial derivative with respect to $x$, $F_x = \frac{\partial F}{\partial x}$ at $x=1$, i.e.~$F_x(1,t) = \sum_i i D_i(t)$.

Other useful algebraic properties, which will be used in the rest of the paper, and easy to verify, are the following ones
\begin{equation}
\sum_{i\geq 0}(i+1) D_{i+1}(t) x^i= F_x \qquad \sum_{i\geq 0} i D_i(t) x^i = x F_x, \qquad
\sum_{i \geq 0} D_{i-1}(t) x^i = x F.
\label{eq:genfun_prop}
\end{equation}
Then, rules of power series state that if $\big[x_i]A = a_i$, $\big[x_i]B = b_i$
\begin{equation}\big[x_i] \frac{A}{1-x} = \sum_{j=0}^i a_j,\qquad \big[x_i] A \cdotp B = \sum_{j=0}^i a_j b_{i-j}.
\label{eq:rules_p_series}
\end{equation}

The use of generating functions will hence allow to consider a single differential equation which comprises all the evolution equations of the model. From its solution it will be possible to extract the elements of the power series, i.e.~the degree distribution.

In the following, we will also consider the system in its steady state, i.e.~in the limit $t\rightarrow \infty$. This in fact enables to calculate the probability that a node has a given degree in the stationary state. Moreover, it avoids the presence of the partial derivative of the generating function with respect to the time variable $t$, hence simplifying the mathematical analysis and the related discussion.

\subsection{The Protocol in Differential Equations}

Let $D_{i,j}(t)=P(deg=i | dd=j, \textrm{ at time } t)$ denote the probability that a given node at time $t$ has degree equal to $i$, knowing that its desired degree is $j$. Note that, following the protocol, peers with an actual degree equal to their desired degree do not accept novel links; hence, a probability higher than $0$ is possible only when $j\geq i$. In general, the evolution of the degree of a given peer can be modeled, using $D_{i,j}(t)$, as
\begin{eqnarray}
\frac{\partial D_{i,j}(t)}{\partial t} = \left\{ \begin{array}{ll}
	    \textstyle \phi (i+1) D_{i+1,j}(t) + \phi\delta_{i,0} + 2\alpha D_{i-1,j}(t) +
		\\ \quad \quad -[\phi (i+1) + 2\alpha ] D_{i,j}(t) & i<j\\
& \\
		\phi\delta_{i,0} + 2\alpha D_{i-1,i}(t) + -\phi (i+1) D_{i,i}(t) & i=j\\ 
& \\
                                                0 &  i>j\\ 
\end{array} \right.
\label{eq:m2_P_ev}
\end{eqnarray}
In (\ref{eq:m2_P_ev}), a distinction is made between three cases, depending on the values of $i$ and $j$. 
The case $i<j$ corresponds to the case when the node has a degree lower than its desired degree. Hence,
the first term on the right of the equation corresponds to the probability that the considered peer has degree equal to $i+1$ and one of the $i+1$ neighbours fails. As a consequence, the node passes from a degree equal to $i+1$ to $i$. The second term considers the probability that the peer fails, thus increasing the number of nodes in the network with degree equal to $0$. The third term accounts for the probability that the peer has degree $i-1$, and it either decides to create a novel connection with a non-neighbour, thus increasing its degree of one novel edge, or also that another peer asks the considered one to become neighbours. Note that in this case we do not insert any limit on the number of non-neighbours, assuming that the total number of nodes is high (or tends to $\infty$); such an  assumption is quite common in complex networks theory \cite{Newman03thestructure}. The remaining terms have the same meaning of the preceding ones, but account for those cases when the node has degree $i$, and itself or one of its $i$ neighbours fail (hence, its degree downgrades to $0$ or $i-1$, respectively), or when a new edge is created between the considered peer and another one, and the peer already has $i$ neighbours (hence, its degree upgrades to $i+1$).  
The case $i=j$ considers only those transitions discussed above that correspond to degrees equal to $i$ or $i-1$, avoiding the probability of having a transition from (to) a degree equal to $i+1>j$ (again, not possible). 
As previously stated, the case $i>j$ (i.e.~an actual degree higher than the desired degree) is not possible due to the protocol executed by peers; hence, the probability is $0$. As a final remark, in (\ref{eq:m2_P_ev}) it is assumed that the probability that two transitions occur simultaneously is negligible, as usual.

As mentioned, it might be interesting to consider the system in its steady state, assuming the existence of the limit $D_{i,j} = \lim_{t \rightarrow \infty} D_{i,j}(t)$, which implies that the variation on the probability to have a certain degree goes to $0$, i.e.~$\frac{\partial D_{i,j}(t)}{\partial t}=0$. 
% This shall eliminate the derivative with respect to $t$ in the final differential equation. 
Equation (\ref{eq:m2_P_ev}) thus becomes
\begin{eqnarray}
    \phi (i+1) D_{i,i} = \phi\delta_{i,0} + 2\alpha D_{i-1,i} & i=j
\label{eq:m2_boundary}
\end{eqnarray}
\begin{eqnarray}
    [\phi (i+1) + 2\alpha ] D_{i,j} = \phi (i+1) D_{i+1,j} 
% + \nonumber \\
    + \phi\delta_{i,0} + 2\alpha D_{i-1,j} & i<j
\label{eq:m2_P_ev_steady}
\end{eqnarray}
To solve these equations using generating functions, consider for the moment the auxiliary system of equations obtained by ignoring the limit imposed by the desired degree. Let hence use different coefficients $\hat{D}_{i,j}$ (it will be possible to derive $D_{i,j}$, once having determined $\hat{D}_{i,j}$). The equations to manage are
\begin{eqnarray}
    [\phi (i+1) + 2\alpha ] \hat{D}_{i,j} = \phi (i+1) \hat{D}_{i+1,j} 
% + \nonumber \\
    + \phi\delta_{i,0} + 2\alpha \hat{D}_{i-1,j}.
\label{eq:m2_D}
\end{eqnarray}

There are two indexes associated to coefficients $\hat{D}_{i,j}$, i.e.~the actual and the desired degree of a given node. Therefore, we employ a $2$-variable generating function 
$$F(x,y) = \sum_{i,j\geq 0} \hat{D}_{i,j} x^i y^j,$$ 
where $x$ controls the actual degree of the peer, while $y$ controls the desired degree of the node. 

Now, multiply (\ref{eq:m2_D}) by $x^i$ and $y^j$ and sum over all $i,j\geq0$. The result is that the infinite set of simultaneous differential equations is turned into a single, novel differential equation for the generating function $F$,
\begin{eqnarray}
\phi (x-1) F_x + [\phi - 2\alpha(x-1) ] F = \frac{\phi}{1-y}.
\label{eq:m2_pde}
\end{eqnarray}
Such an equation is obtained by exploiting properties of generating functions (\ref{eq:genfun_prop}) and observing that $\sum_{i, j\geq 0} \delta_{i,0}x^i y^j = \frac{1}{1-y}.$
As mentioned, $F_t$ is not present since we are considering the system directly in the steady state. It is possible to verify that a solution of this differential equation is
\begin{eqnarray}
F(x,y) & = & {\textstyle\frac{\textstyle \phi }{\textstyle 2\alpha (1-x)(1-y)}} - \frac{F_0 e^{\frac{2\alpha x}{\phi}}}{1-x},
\label{eq:F}
\end{eqnarray}
where $F_0$ is an initial function to be determined, based on the boundary conditions.

\subsection{Degree Probability}

The obtained function $F$ is an unfortunate one, since it is not defined for $x=1$, and we already mentioned that many properties might have been obtained by evaluating some manipulation of $F$ measured at $x=1$. However, given (\ref{eq:F}), the elements composing the generating function can be extracted by employing classic results of power series.
In particular, we may first assume that $F_0$ can be expanded in power series, i.e.~$F_0(y) = \sum_{j \geq 0} c_j y^j.$
Then, observe that
$$ \frac{\textstyle \phi}{\textstyle 2\alpha (1-x)(1-y)} = \frac{\phi}{2\alpha} \sum_{i,j} x^i y^j,$$
and, due to the mentioned rules (\ref{eq:rules_p_series}) of power series, we have
$$ \frac{F_0 e^{\frac{2\alpha x}{\phi}}}{1-x} = \sum_{j \geq 0} c_j y^j \sum_{i \geq 0} e_i\Big(\frac{2 \alpha}{\phi}\Big) 
% \sum_{k=0}^i \frac{1}{k!} \Big(\frac{2 \alpha}{\phi}\Big)^k 
x^i,$$
where $e_n(r)$ is the exponential sum function $e_n(r) = \sum_{k=0}^n \frac{r^k}{k!}$.
By combining these results, a general formula is obtained for the elements of the auxiliary system, which is
\begin{eqnarray}
\hat{D}_{i,j} = [x_i y_j]F = \frac{\phi}{2 \alpha} - c_j e_i\Big(\frac{2 \alpha}{\phi}\Big).
\label{eq:hat_d}
\end{eqnarray}

It is now possible to calculate $D_{i.j}$ from $\hat{D}_{i,j}$, by determining coefficients $c_j$ in (\ref{eq:hat_d}), such that $D_{i,j}=\hat{D}_{i,j}$ when $i\leq j$, and also in order to satisfy the boundary equation (\ref{eq:m2_boundary}), considering the case $i=j$. In particular, when $i=j$, comparison of equations (\ref{eq:m2_boundary}) and (\ref{eq:m2_D}), shows that if $D_{i,i}=\hat{D}_{i,i}$ is true, then it must be 
$$2 \alpha \hat{D}_{i,i} = \phi (i+1) \hat{D}_{i+1,i}.$$
From this last equation, coefficients $c_i$ are determined,
$$c_i = \frac{\phi}{2 \alpha} \frac{\phi(i+1) - 2 \alpha}{[\phi(i+1) - 2 \alpha] e_i\Big(\frac{2 \alpha}{\phi}\Big) + \frac{\phi}{i!} \Big(\frac{2 \alpha}{\phi}\Big)^{i+1}}.$$
Thus,
\begin{eqnarray}
D_{i,j} = \frac{\phi}{2 \alpha} - c_j e_i\Big(\frac{2 \alpha}{\phi}\Big).
% \sum_{k=0}^i \frac{1}{k!} \Big(\frac{2 \alpha}{\phi}\Big)^k.
\label{eq:m2_d}
\end{eqnarray}

Now, $D_{i,j}$ represents the probability that a node has an actual degree equal to $i$, knowing that its desired degree is $j$. To find the probability $D_i$ that a node has degree $i$, it is thus sufficient to employ the formula
$$ D_i = \sum_j P(deg=i | dd=j) P(dd=j) = \sum_j D_{i,j} P(dd=j),$$
once having specified a desired degree distribution $P(dd=j)$, $j>0$, during the design of the peer-to-peer system.

\subsection{Nodes at Distance $m$, Network Diameter}
\label{sec:mean_neigh}

Once having obtained a degree probability distribution for the considered network, interesting measures to calculate are the mean number of first, second neighbours, and generally the number of neighbours at distance $m$ from a given chosen peer. These metrics have in fact a great importance to understand how, and how fast, the network is able to disseminate information in a peer-to-peer network.

Of course, having the degree probability distribution, the average number of first neighbours $z_1$ of a given peer, i.e.~the mean degree, can be calculated as $z_1 = \langle k \rangle = \sum_k k D_k$. 
Then, an important result is that if the network exhibits a small clustering, the probability that one of the second neighbours of a peer is also a first neighbour of it, is negligible in (very) large networks \cite{newmanHandbook}. This allows to easily calculate the mean number of second neighbours as $z_2 = \sum_k (k-1)k D_k = \langle k^2 \rangle - \langle k \rangle$. In general, the number of neighbours at distance $m$, can be estimated as $z_m = (z_2 / z_1)^{m-1}z_1$. Moreover, when $z_2 > z_1$ the net exhibits a giant component which, roughly speaking, connects the majority of nodes in the network (the reader may refer to \cite{newmanHandbook} for a complete discussion).

A method to construct a network with small clustering, regardless of the desired degree distribution, is as follows. For each node $i$ in the network, assign its desired degree $dd_i$, following a desired degree distribution. Then add to it $dd_i$ stubs, representing the end of the links it would like to maintain. Finally, create links by randomly connecting stubs of different nodes. This is the approach we adopt to create and simulate networks with different desired topologies (as discussed in the next section). Using these networks, it is hence easy to calculate $z_m$ values. The reader might argue that these nets do not represent ``real'' existing peer-to-peer systems. Indeed, one might think at several examples of peer-to-peer architectures which do have clusters. In such a case, the obtained results represent upper bounds of the real estimations of $z_m$. 

In any case, 
% some considerations are in order here \cite{newmanHandbook}. First, when $z_2 > z_1$, the majority of the network is connected (there is a giant component). Second, 
when $z_2 /z_1 \gg 1$, there is an average distance $l$ representing the number of hops needed to reach a node, starting from another one \cite{newmanHandbook}. Put in other words, the number of nodes reachable within $l$ hops is almost the total number of nodes in the network $|\Pi|$; hence we have 
\begin{equation}
|\Pi| \simeq z_l = \Big(\frac{z_2}{z_1}\Big)^{l-1}z_1 \Rightarrow l \simeq \frac{\log(|\Pi|/z_1)}{\log(z_2/z_1)} +1. 
\label{eq:diam}
\end{equation}
In \cite{newmanHandbook} it is argued that based on empirical results, estimations obtained using this last formula are close to correct measurements for several real networks. Hence, we will use (\ref{eq:diam}) in Section \ref{sec:res}, to have an estimation of the diameter of our considered peer-to-peer overlays.

%%%%%%%%%%%%%%%%%%%%
%% RESULTS
%%%%%%%%%%%%%%%%%%%%

\section{Experimental Assessment}
\label{sec:res}

This section presents an assessment performed to validate the model discussed in the previous section and evaluate the ability of the outlined peer-to-peer system to cope with node faults. A comparison is performed between the analytical model and results obtained through a simulation of the distributed protocol. As shown in the reminder of the section, the two approaches provide very similar outcomes. 
% These results are not compared with real measurements, obtained through analysis of an existing peer-to-peer system. However, 
The employed approaches are very different, being the former purely analytic while the latter a simulator that mimics the distributed protocol executed by a number of peers. Hence, the similarity on the obtained results confirms that the final equation of the mathematical model can be easily employed to characterize the fault-tolerance and thus the reliability of  a system having a defined desired topology.

As to the desired degree distribution, we consider three different distributions and vary their related parameters. Namely, the three considered scenarios are: i) a fixed desired degree distribution, which would  produce a uniform graph with all nodes having the same number of links; ii) a classic random graph where nodes are connected with others with a certain probability \cite{newmanHandbook}; iii) a power law distribution, which would create a scale-free network \cite{Barabasi2000,Newman03thestructure,Aiello00arandom,simutools}.
% These choices are motivated as follows.

\subsection{On the Simulator}

A discrete-event simulator has been built to model the defined distributed protocol. It has been implemented in C code, by exploiting the GNU Scientific Library (GSL), a library that provides implementation of several mathematical routines for numerical and statistical analysis, such as pseudo-random generators \cite{gsl}. The simulator provides the possibility of generating a varying number of nodes. During the initialization phase, a random network is created based on the chosen desired degree distribution. Different techniques can be employed to create such a random network \cite{newmanHandbook,simutools,BenderC78}. As already discussed in Section \ref{sec:mean_neigh}, in this case once having assigned a specific desired degree to each node, based on the specific desired distribution, a random mapping is made so that links are created until each node has reached its own desired degree.
% This is a common technique to generate random graphs with a given degree sequence. 
Hence, at the beginning of the evolution nodes already have the number of links they would like to maintain (this generally affects only the transient part of the simulation).

The simulator creates a network with a fixed number of nodes. This eases the measurement of the degree nodes have in time, without the need to consider novel nodes that join the network during the execution of the protocol. Hence, once a peer fails, it is not removed from the network; rather, all its links are removed. From that moment, the node will try to create novel links with novel peers, searching to reach its desired degree.

After the network initialization phase, the evolution of the network starts. Nodes' failures and the discovery of other nodes for the creation of novel links have been implemented as Poisson processes, whose rates are regulated by the parameters $\alpha$ and $\phi$, respectively. 
The shown results represent the status of the system after a specified simulation time. The length of the simulation was $10^4$ simulation steps. When not differently stated, the number of nodes was set equal to $1000$. For each specific configuration, we ran $30$ different experiments. Shown outcomes correspond to average results.

\subsection{Degree Distribution of Fixed Desired Degree Networks}

The first type of generated networks was based on a fixed $dd$, i.e.~peers have the same value of desired degree $dd = n$. Forcing peers to have the same desired degree $dd$ allows to model those classic scenarios in peer-to-peer environments where the software running on peers is configured to have a given number of links in the overlay, i.e.~$dd$. This is quite common in real peer-to-peer systems and it is usually accomplished for load balancing purposes \cite{wang}.

The model restricts the event space to the case when all nodes' desired degree is constant, $dd = n$; an obvious consequence is that $D_{i,j} = 0, j\neq n$. Moreover, due to the distributed protocol, $D_{i,j} = 0, i \geq j$. Hence, the sum of all the values of $D_{i,n}$ when $i$ is varied, restricts to $\sum_{i\leq n} D_{i,n} = P(deg =i | dd = n) = 1$. In this case, we can hence simply consider in the model the values of $D_{i,n} = P(deg =i | dd = n)$, for a fixed $n$.

% \subsection{Results}

\begin{figure}[t]
   \centering
   \includegraphics[angle=270,width=.45\linewidth]{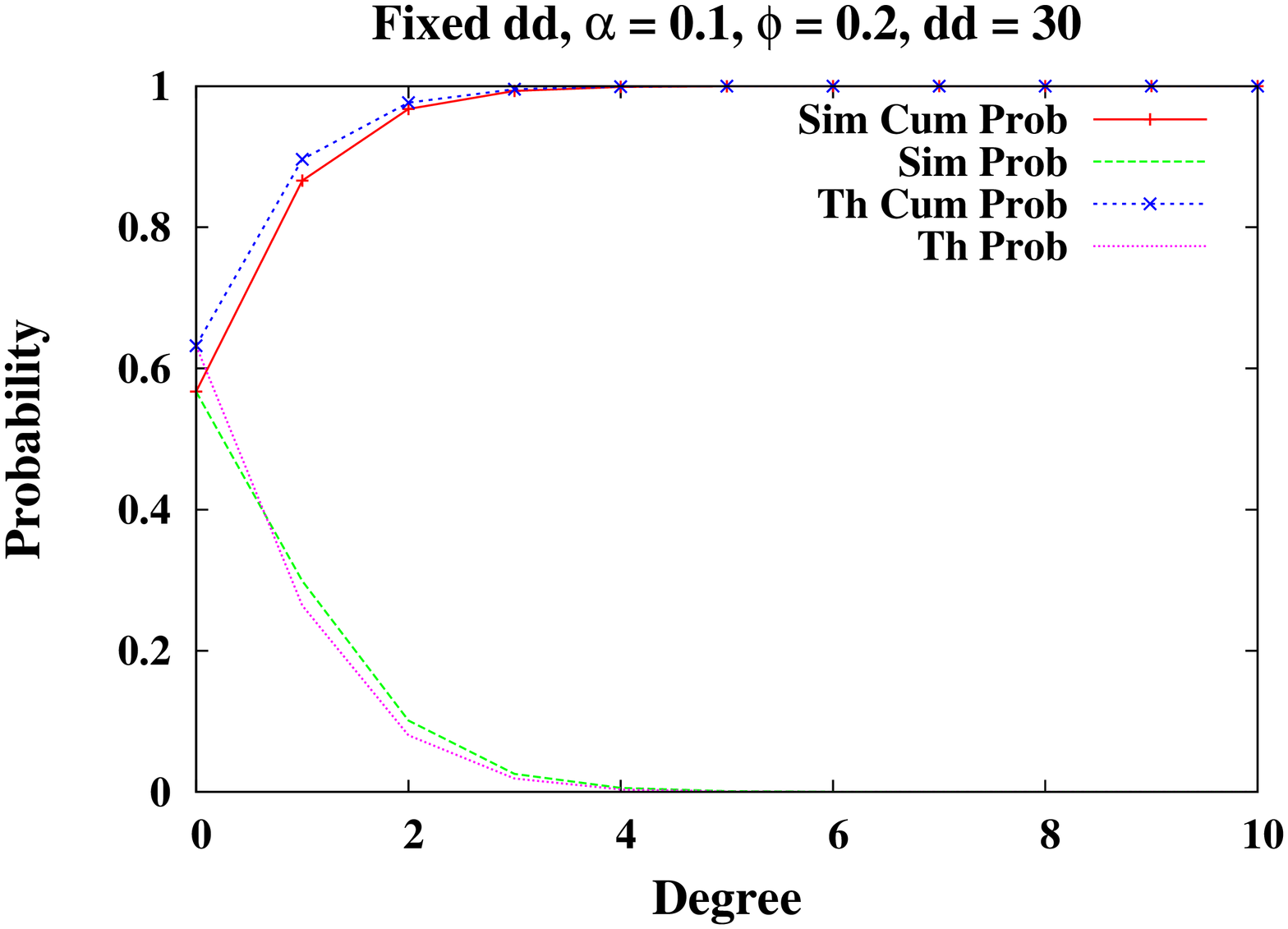}
   \includegraphics[angle=270,width=.45\linewidth]{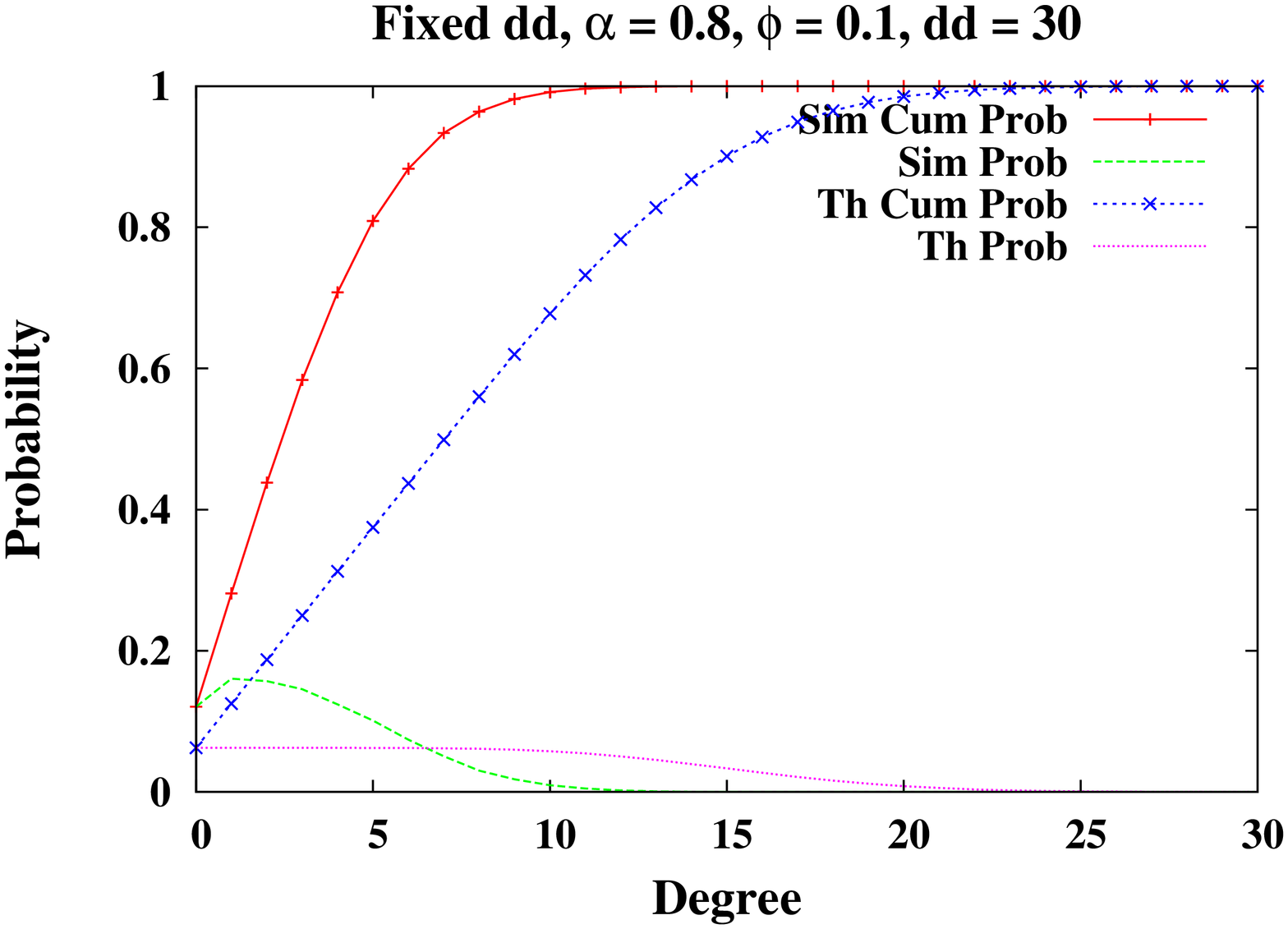}
   \includegraphics[angle=270,width=.45\linewidth]{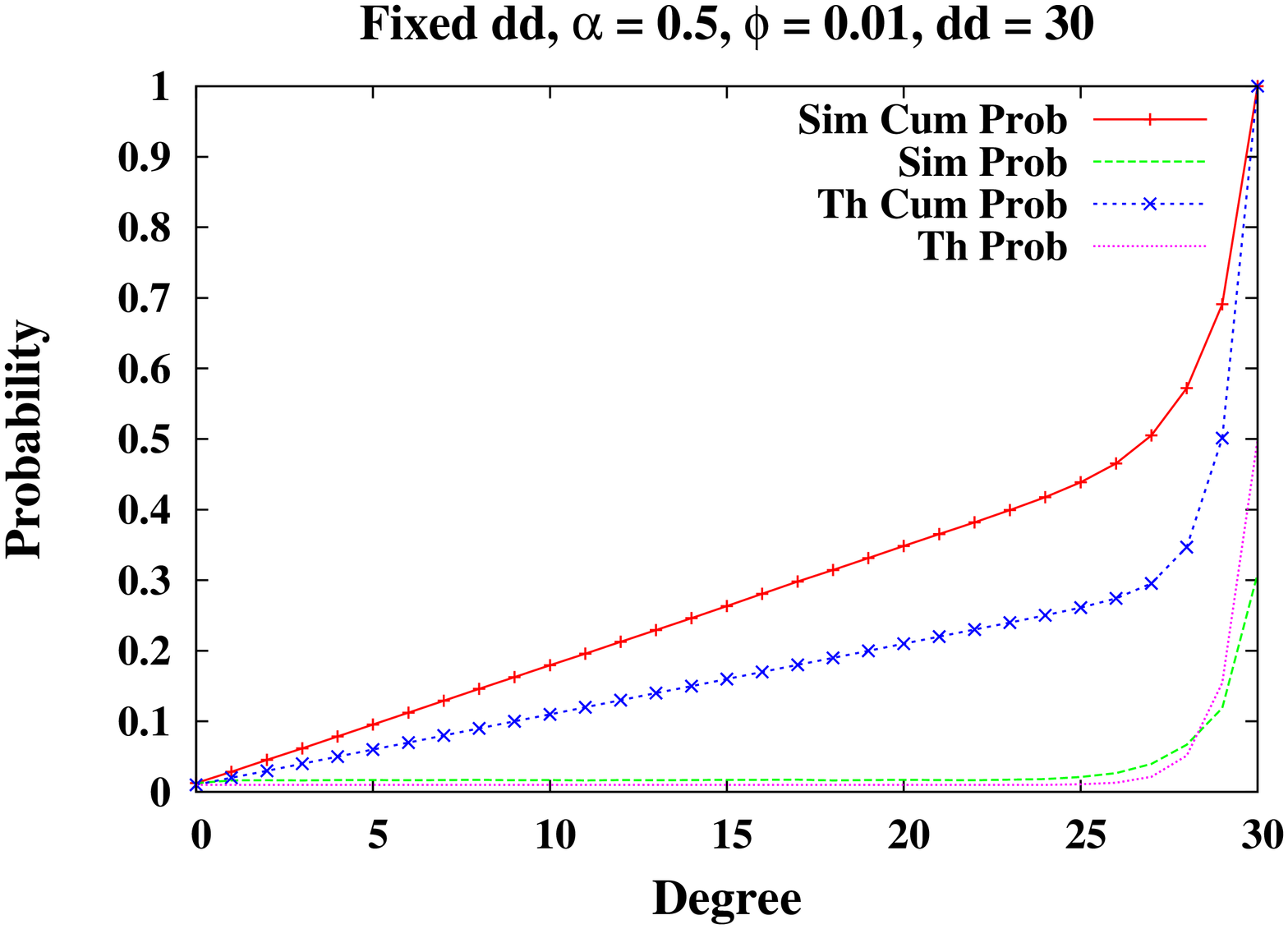}
   \includegraphics[angle=270,width=.45\linewidth]{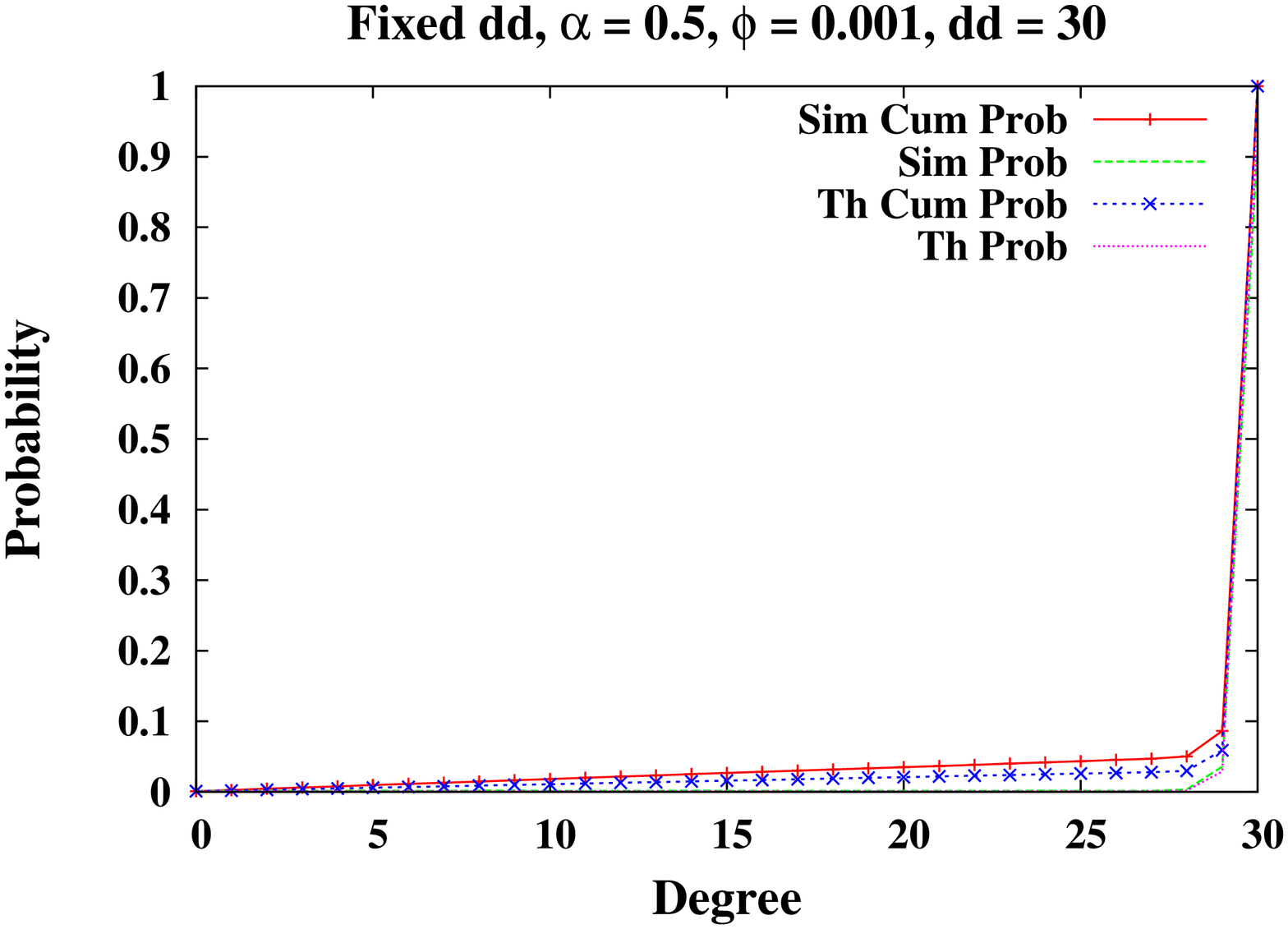}
   \caption{Degree probability and cumulative degree probability; results obtained through simulation (Sim) and the mathematical modeling (Th); $\alpha = 0.1, \phi = 0.2, dd=30$}
   \label{fig:fig1}
\end{figure}

\begin{figure}[t]
   \centering
   \includegraphics[angle=270,width=.45\linewidth]{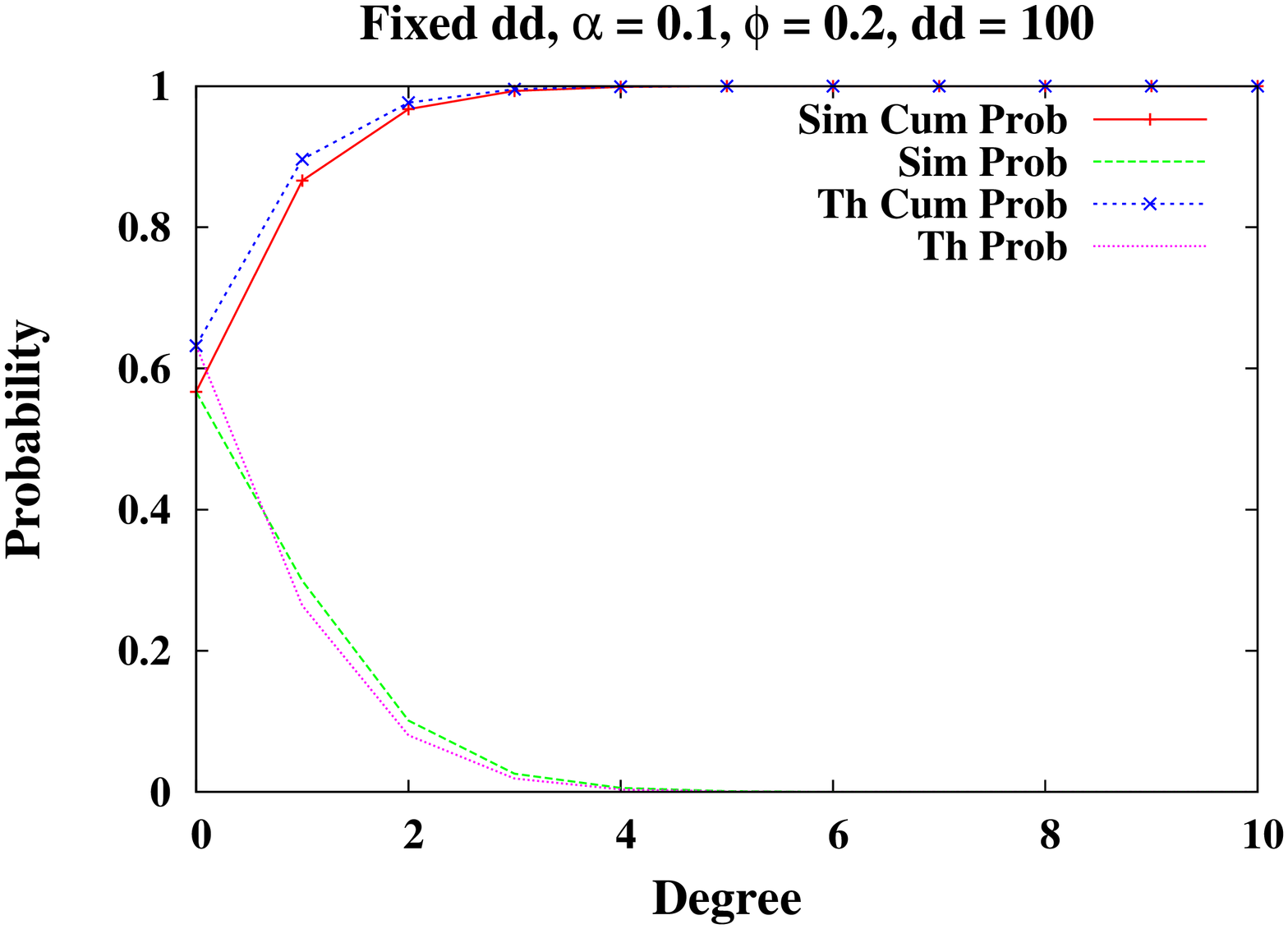}
   \includegraphics[angle=270,width=.45\linewidth]{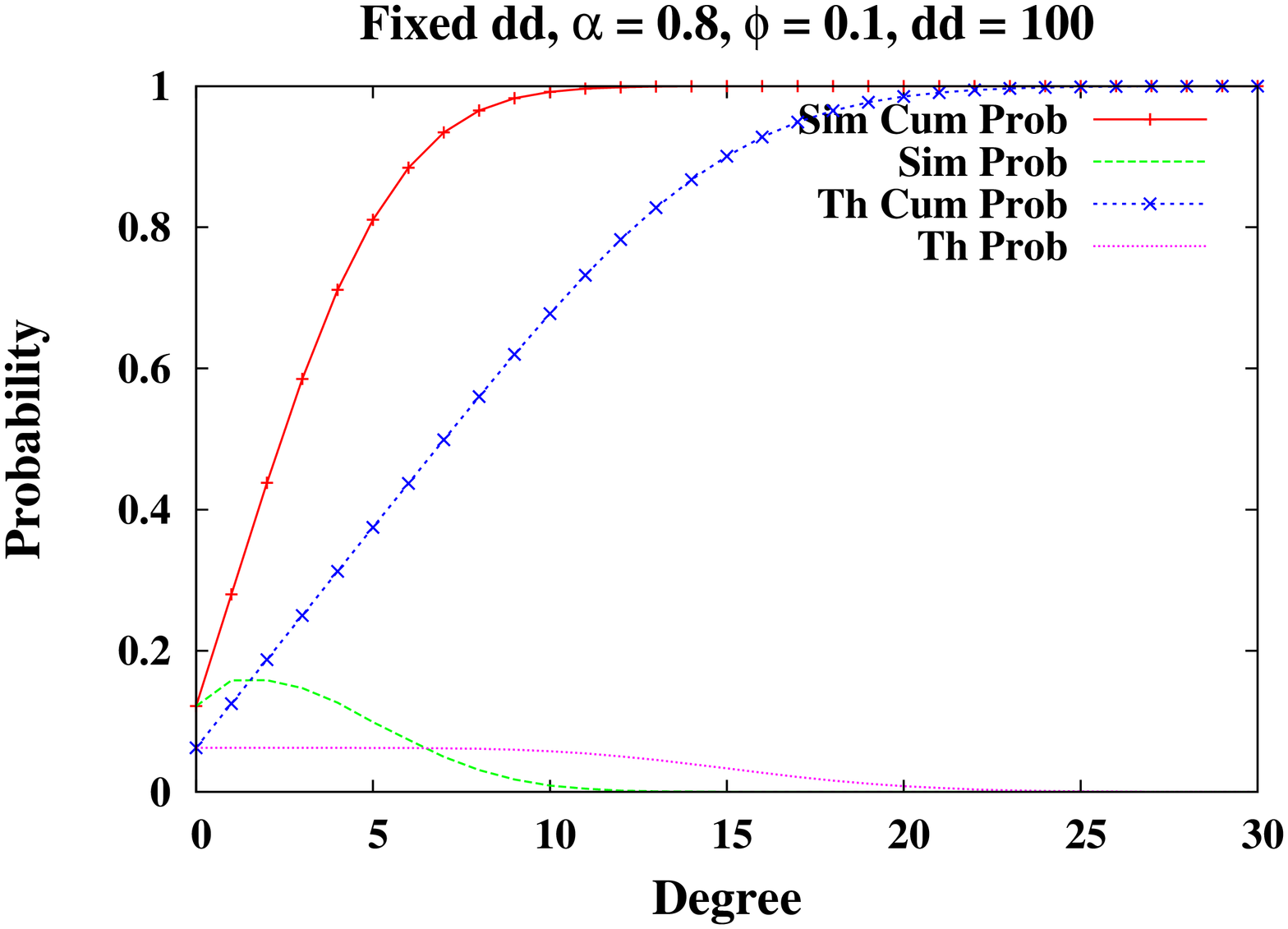}
   \includegraphics[angle=270,width=.45\linewidth]{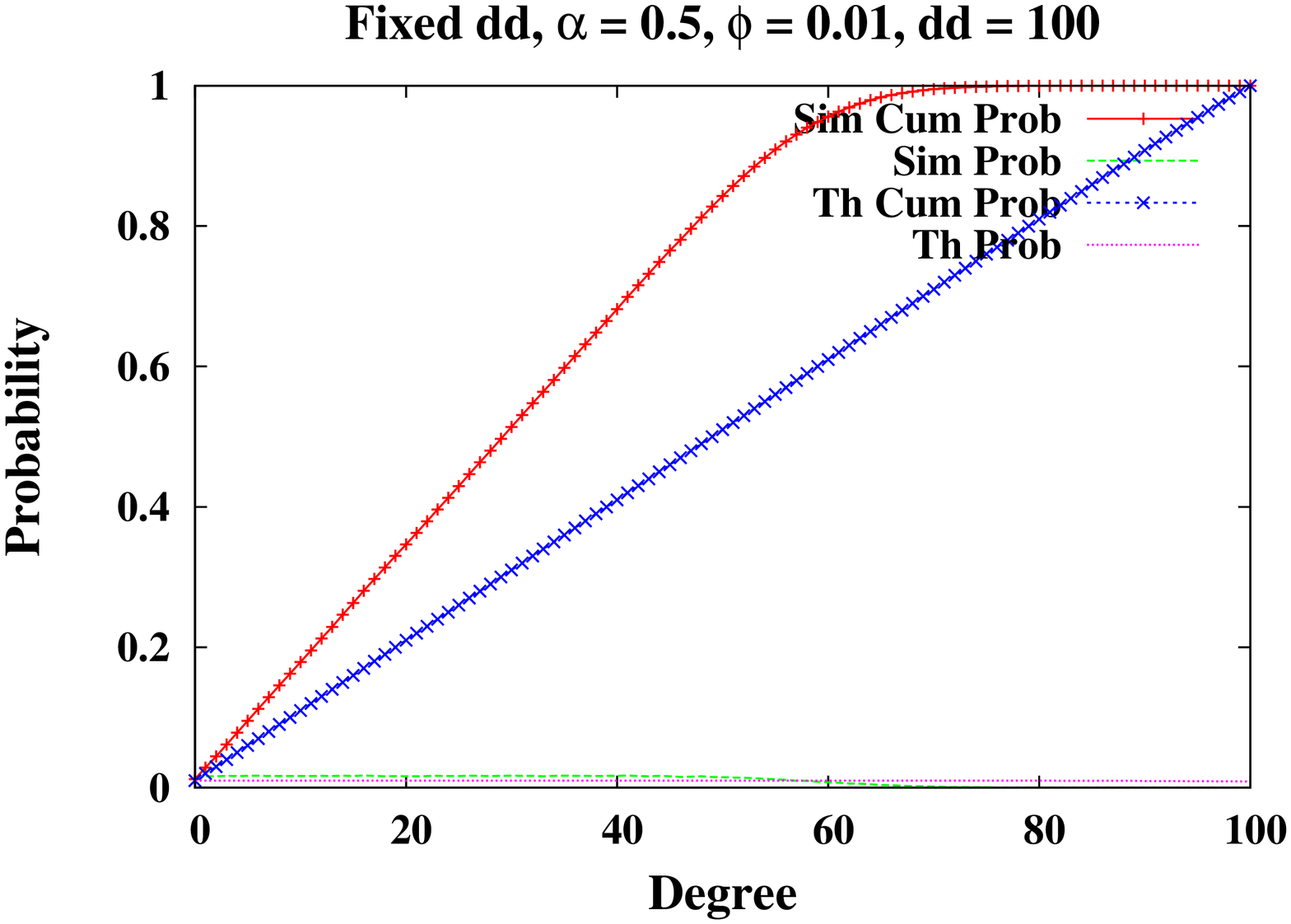}
   \includegraphics[angle=270,width=.45\linewidth]{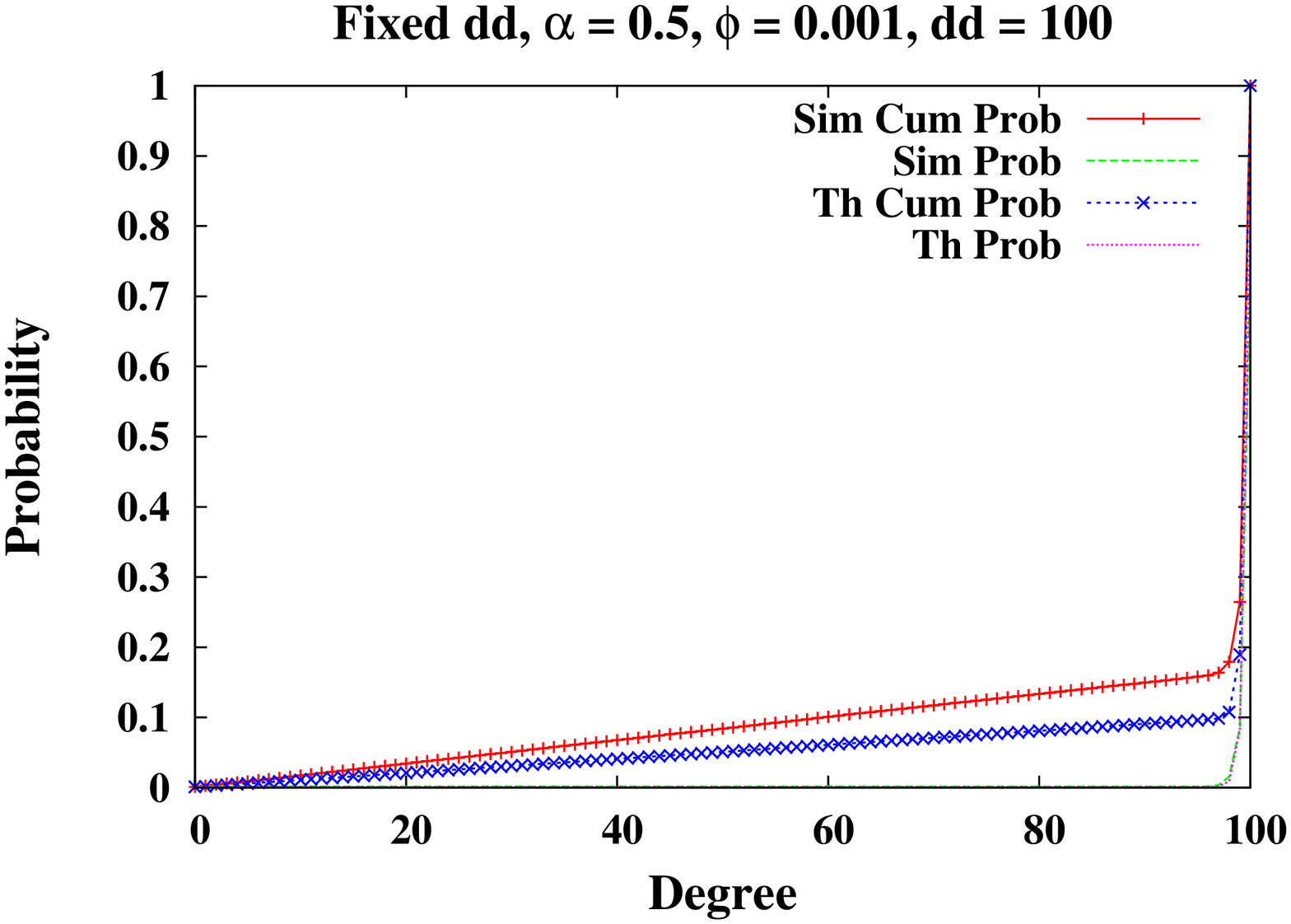}
   \caption{Degree probability and cumulative degree probability; results obtained through simulation (Sim) and the mathematical modeling (Th); $\alpha = 0.1, \phi = 0.2, dd=100$}
   \label{fig:fig2}
\end{figure}

Figures \ref{fig:fig1}-\ref{fig:fig2} show the probability that a given node has a certain degree, based on the parameters $\alpha, \phi$.
% , which controls the rate at which a peer decides to create a novel edge with a non-neighbour node, and on the parameter $\phi$, related to nodes' failures. 
All figures report both the node degree probability itself, as well as the cumulative probability, i.e.~the probability that a node has a degree less or equal to the considered value. For these two metrics, two measurements are reported, obtained by using Equation (\ref{eq:m2_d}) and through simulation.
We concentrate on two different types of networks, corresponding to two desired degree values, i.e.~$dd=30$ (Figure \ref{fig:fig1}) and $dd=100$ (Figure \ref{fig:fig2}). 
As shown below, the two networks have similar behaviors for the selected values of the rates $\alpha, \phi$; the same holds for other similar $dd$s.
% Values for the parameters $\alpha$ and $\phi$ were varied. Of course, due to space constraints only charts relating to some selected couples of values for $\alpha,\phi$ are reported. However, similar results were obtained for other similar choices of couples $\alpha, \phi$.

By looking at figures, a first consideration is that similar results are obtained using simulation and the mathematical model. Then, very different outcomes are measured, depending on $\alpha,\phi$ values. In particular, when the value of the failure rate $\phi$ is higher than attachment rate $\alpha$, in the steady state only low degree values have a probability significantly higher than $0$. This can be appreciated by looking at the first chart of Figures \ref{fig:fig1}-\ref{fig:fig2}, where $\alpha=0.1, \phi=0.2$. 
In both cases, degree values that take some non-negligible probabilities are those that range in the interval $0-6$. The cumulative probabilities, in the considered scenarios, reach values near to $1$ at very low values. This basically means that in the steady state almost all peers tend to have experienced some failures and they do not succeed in maintaining the desired network topology. As mentioned before, our assumption is that peers instantaneously come back in the system and try to create some novel links, yet without being able to gain some noticeable degree. This is due to the low value of $\alpha$. Moreover, since non negligible values are very well below the considered desired degrees, the obtained charts reported in Figures \ref{fig:fig1} and \ref{fig:fig2}
% , which refer to different desired network topologies, 
are mostly equal (but they are indeed slightly different), since the $dd$ value does not act as a bound for the link creation. 
These first discussed results demonstrate that peers must be able to react to changing conditions of the system and self-orga\-ni\-ze. In fact, $\alpha$ can be interpreted as a basic parameter that regulates how a peer is active in the network.

Things start to change when $\alpha$ takes values higher than $\phi$. These settings mimic those situations according to which peers actively create links, more rapidly than failure rates. The second charts in Figures \ref{fig:fig1}-\ref{fig:fig2} show results when $\alpha=0.8$, while keeping $\phi$ equal to $0.1$, lower than $\alpha$. In this case, non-negligible degree probabilities may be observed for degree values higher than those obtained before, yet still without reaching the desired degree (this is more evident when $dd=100$). It may be observed that, in this particular scenario, results from the simulation and the mathematical modeling are not perfectly identical, but slight differences can be appreciated. In substance, simulations show that nodes tend to have a lower degree than that predicted by the mathematical modeling. Nevertheless, obtained results are well below the nodes' desired degree.

Results completely change when $\phi$ is selected quite below the value of $\alpha$. In these scenarios, in the steady state the probability that a node has a certain degree is mostly uniform for all degrees in the range between $0$ and the nodes' desired degree. This can be appreciated by looking at the two final charts of the considered figures.
In particular, with the following setting $\alpha = 0.5, \phi = 0.01, dd=30$, it is quite probable that in the steady state nodes have their desired degree, while with $dd=100$ probabilities of degree values lower than $dd$ are almost uniformly distributed. When $\phi=0.001$, instead, the probability of having a degree equal to $dd$ in the steady-state reaches a high value also if $dd=100$. In substance, under this setting, the desired network topology is maintained in the steady state.

\begin{figure}
   \centering
   \includegraphics[angle=270,width=.7\linewidth]{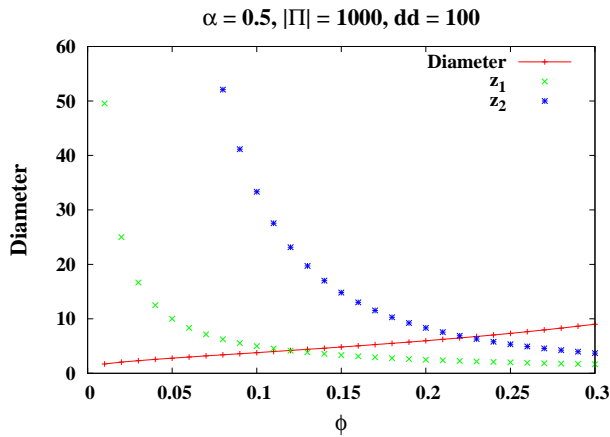}
   \caption{Diameter, average number of first and second neighbours of fixed desired degree networks, when varying $\phi$, calculated using Equation (\ref{eq:diam})}
   \label{fig:diam_dd}
\end{figure}

Figure \ref{fig:diam_dd} shows the estimated diameter of the networks obtained when running the distributed protocol with an average attachment rate $\alpha=0.5$, while varying the value of $\phi$, assuming a network composed of $1000$ nodes. The chart also reports the average number of first neighbours $z_1 = \langle k \rangle$ and of the second neighbours $z_2$ (measured through the analytical model). It can be observed that the number of second neighbours is higher than the number of first neighbours, when $\alpha > \phi$. Hence, when employed on large networks, the protocol allows the creation of a giant component. Note that when $\phi$ has low values, the diameter is very limited and nodes succeed in maintaining a very high degree value, since the network is composed of only $1000$ nodes, while the desired degree of each peer is equal to $100$. This confirms that a proper attachment rate may guarantee that contents can be rapidly disseminated through the overlay, whatever the communication strategy employed on top of it. Then, as the failure rate grows, there is a growth also on the network diameter. It is however worth noticing that as $\phi$ grows, the ratio $z_2 /z_1$ decreases. Thus, the estimation of the network diameter might be less reliable.

%%%%%%%%%%%%%%%%%%%%

\subsection{Degree Distribution of Random Graphs}

Here, we consider random graphs to model the desired degree distribution of networks. This is a generalization of the approach described above, with peers all having the same probability to attach to other links. 
In substance, when a random graph is generated, a link between each pair of peers is created with a certain probability $p$. The average degree is thus $\langle k \rangle = p |\Pi|$. It is well known that when the number of peers $|\Pi|$ is large, nodes' degrees of random graphs may be well characterized using a Poisson distribution $\frac{\langle k \rangle^i e^{-\langle k \rangle}}{i!}$.
Several works employ this construction tool for generating random graphs \cite{newmanHandbook}.

\begin{figure}
   \centering
   \includegraphics[angle=270,width=.45\linewidth]{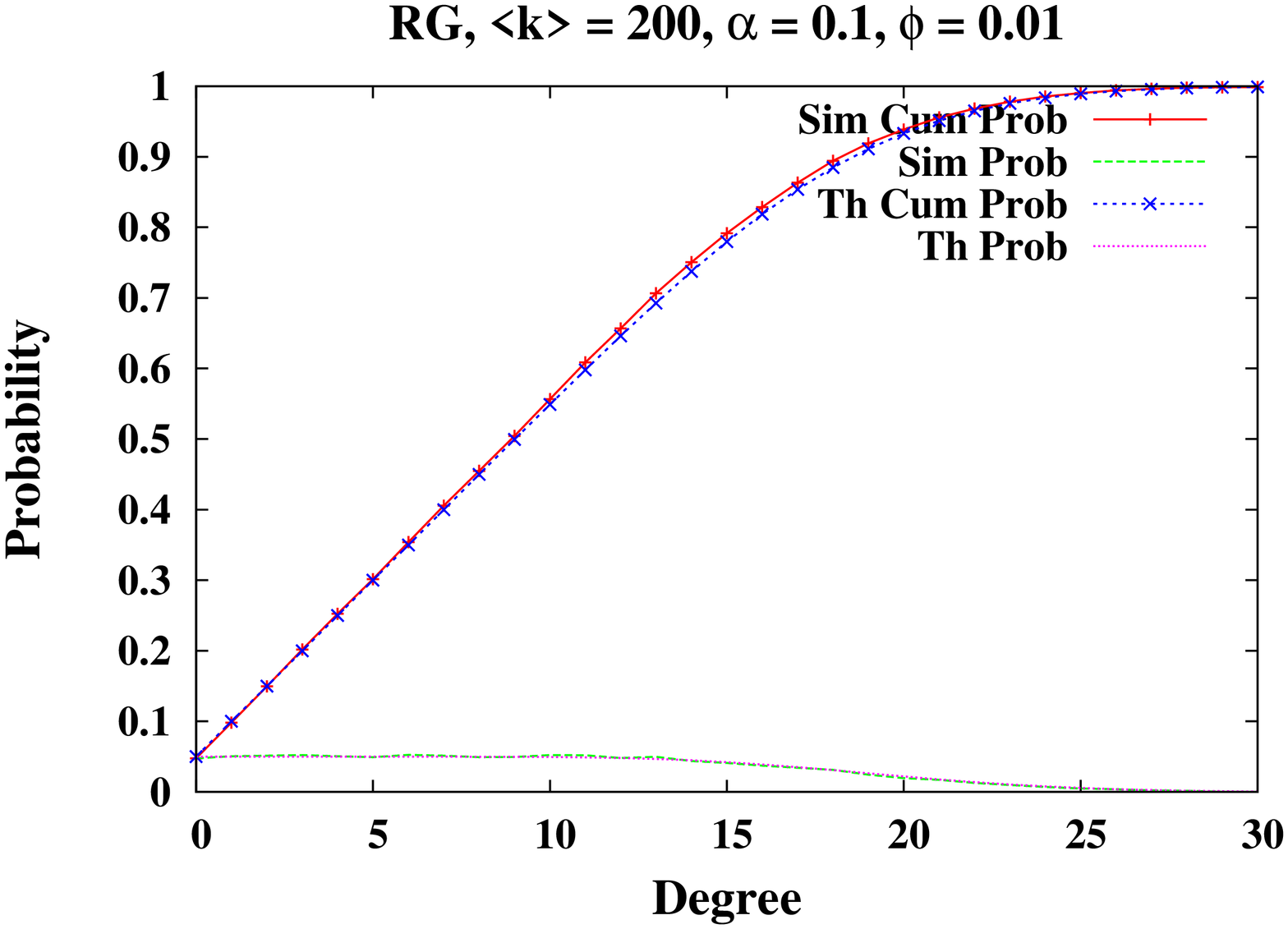}
   \includegraphics[angle=270,width=.45\linewidth]{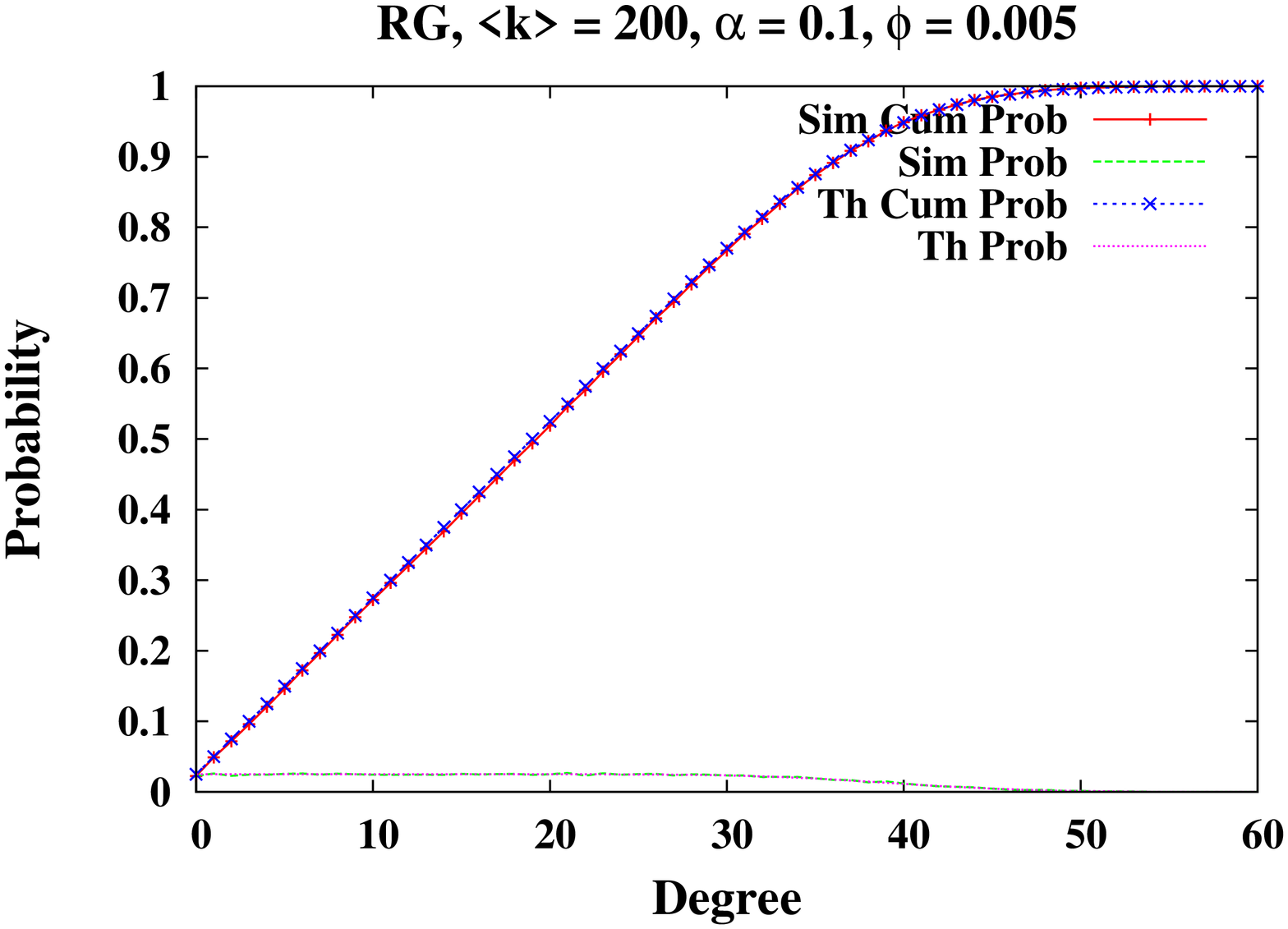}
   \includegraphics[angle=270,width=.45\linewidth]{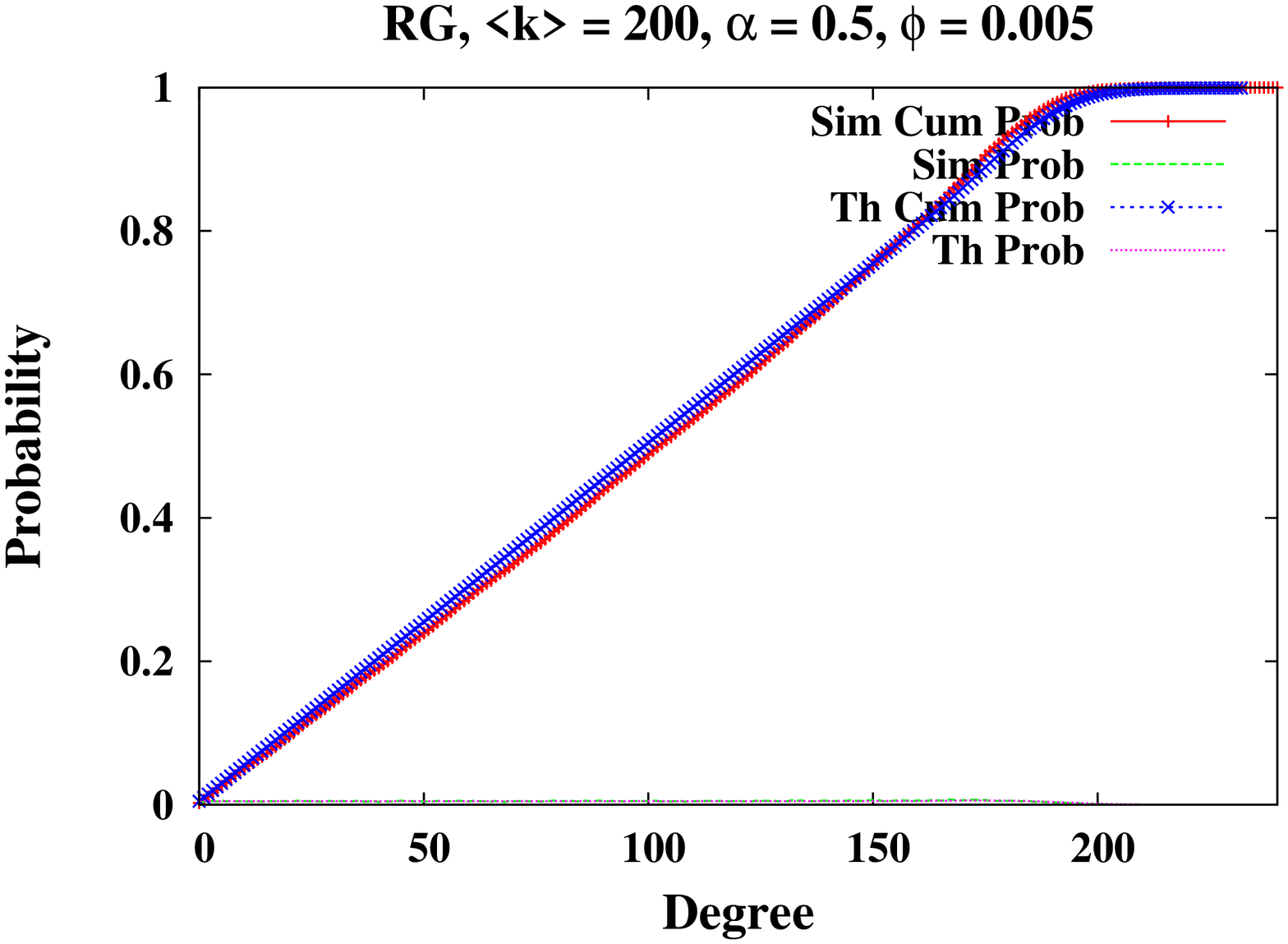}
   \includegraphics[angle=270,width=.45\linewidth]{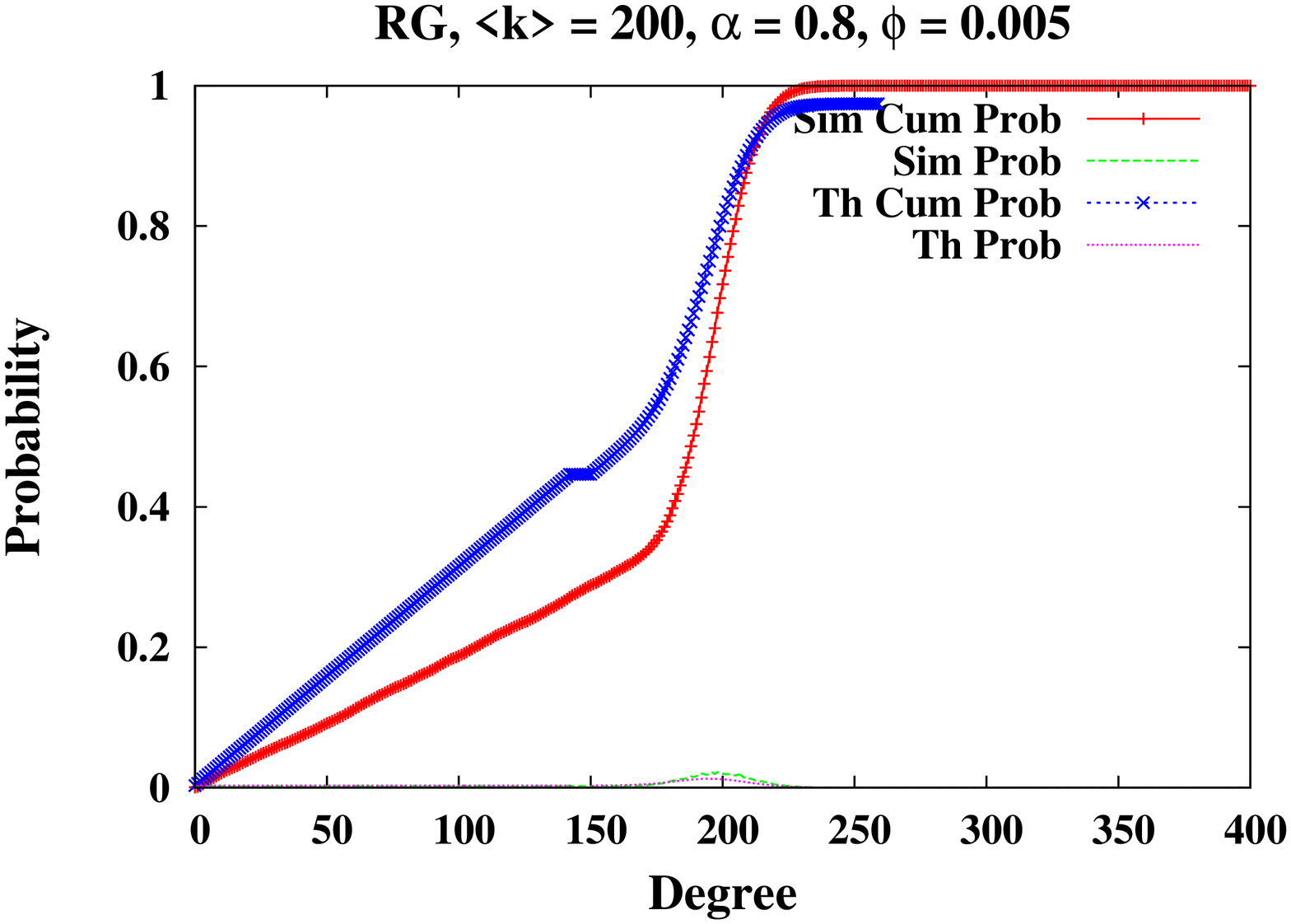}
   \caption{Degree probability varying $\alpha, \phi$; results obtained through simulation (Sim) and the mathematical modeling (Th); Random Graph model $p=0.2, |\Pi|=1000, \langle k \rangle = 200$}   
   \label{fig:fig_rg3}
\end{figure}

\begin{figure}
   \centering
   \includegraphics[angle=270,width=.45\linewidth]{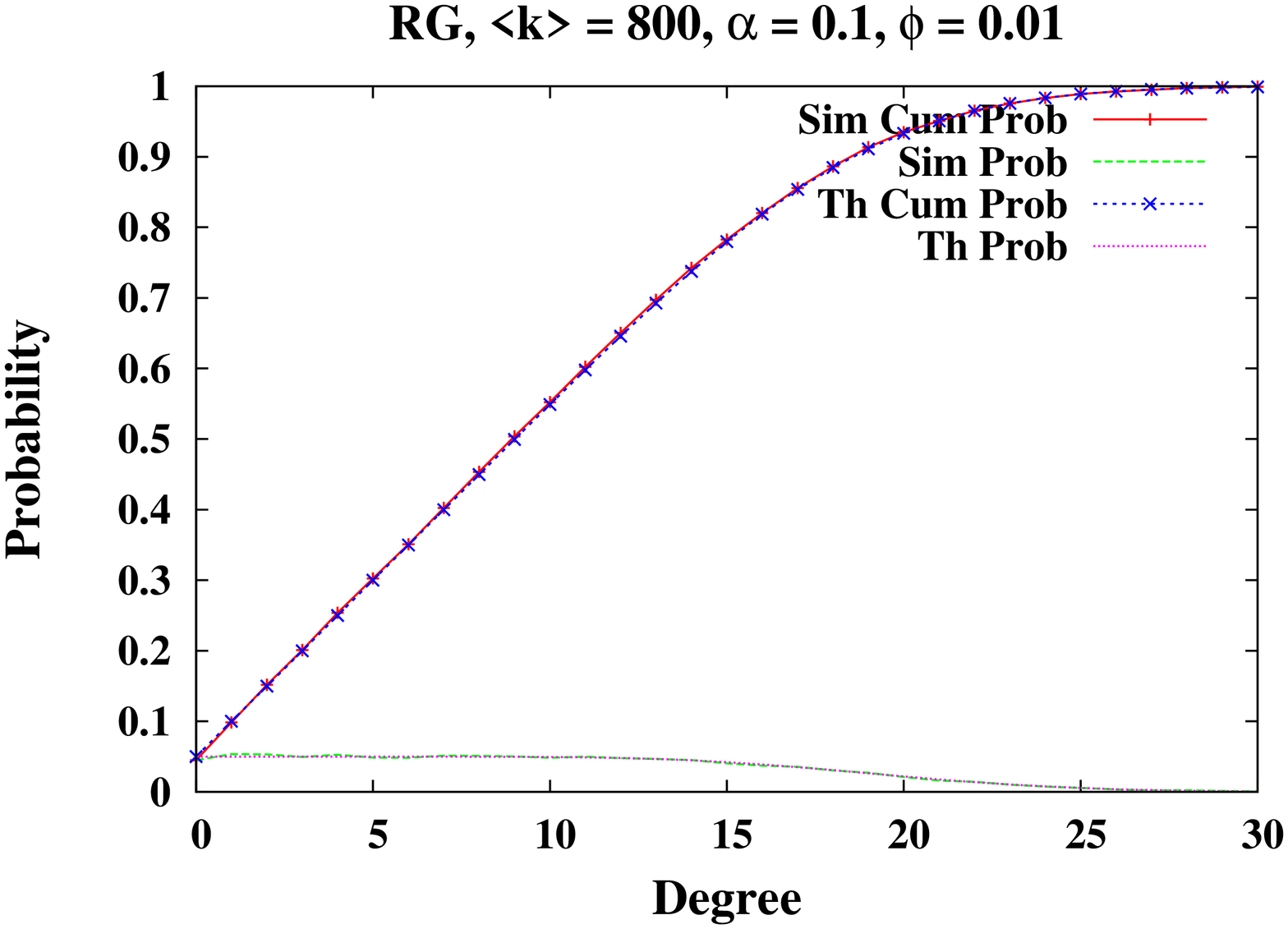}
   \includegraphics[angle=270,width=.45\linewidth]{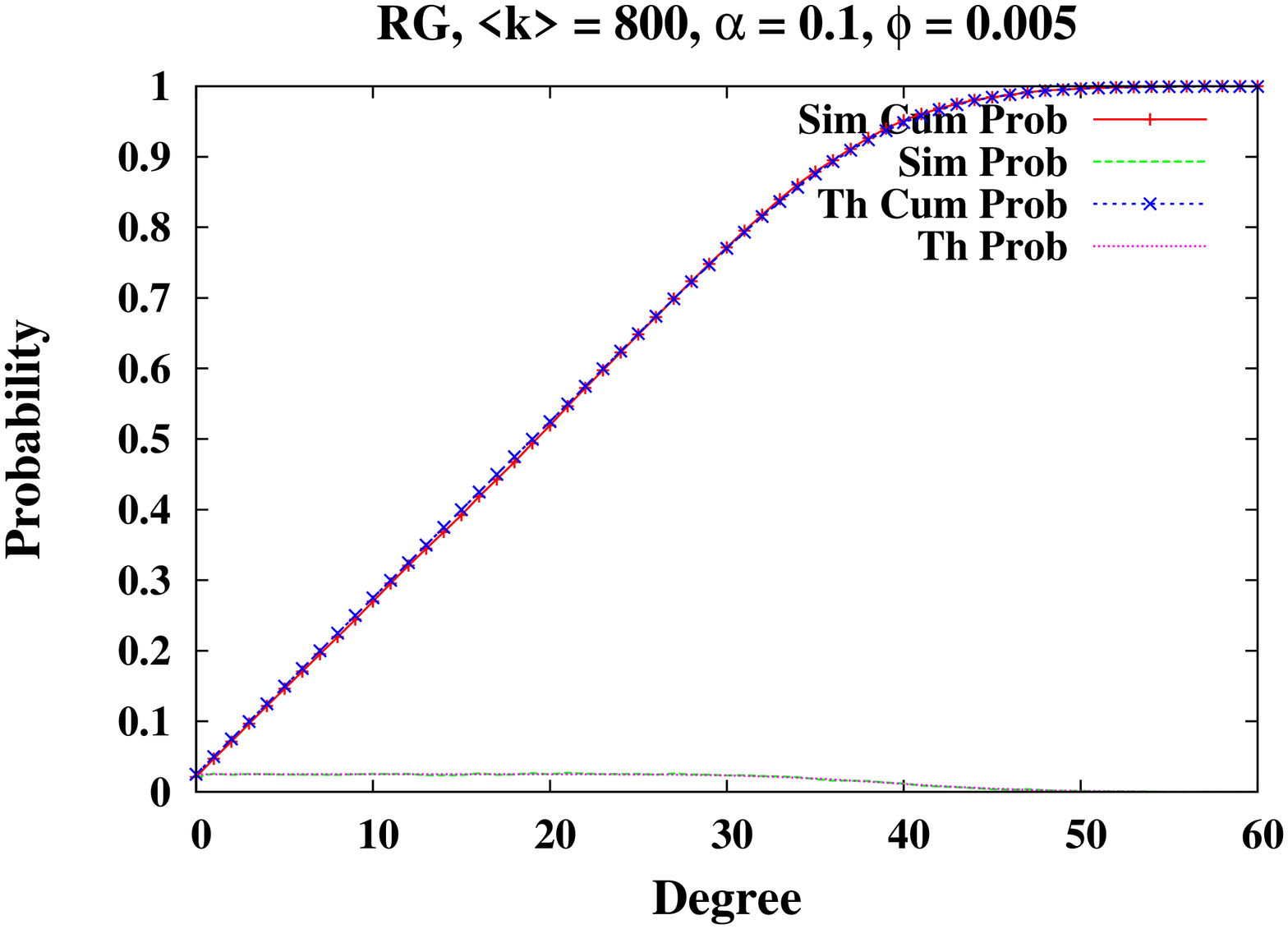}
   \includegraphics[angle=270,width=.45\linewidth]{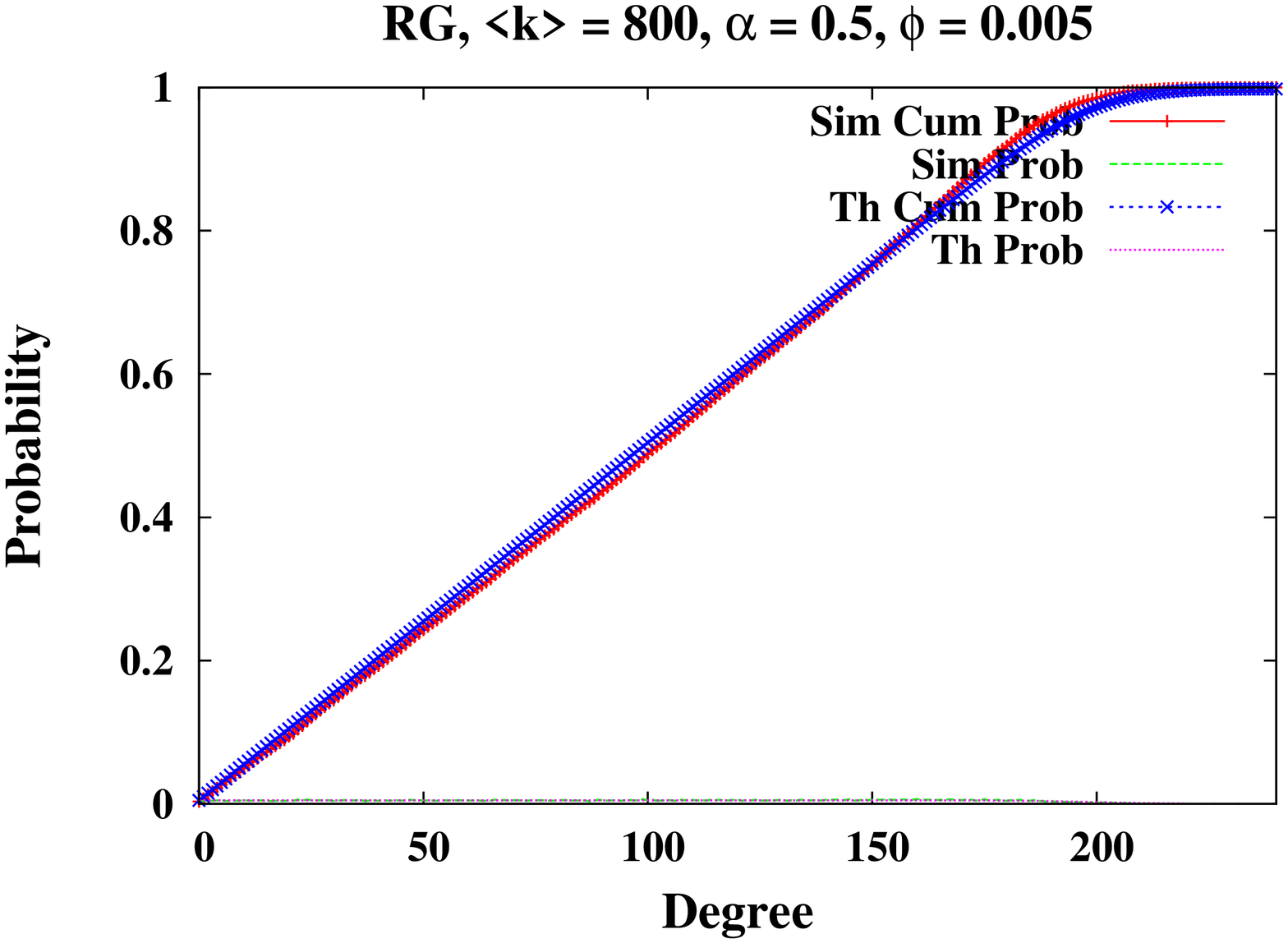}
   \includegraphics[angle=270,width=.45\linewidth]{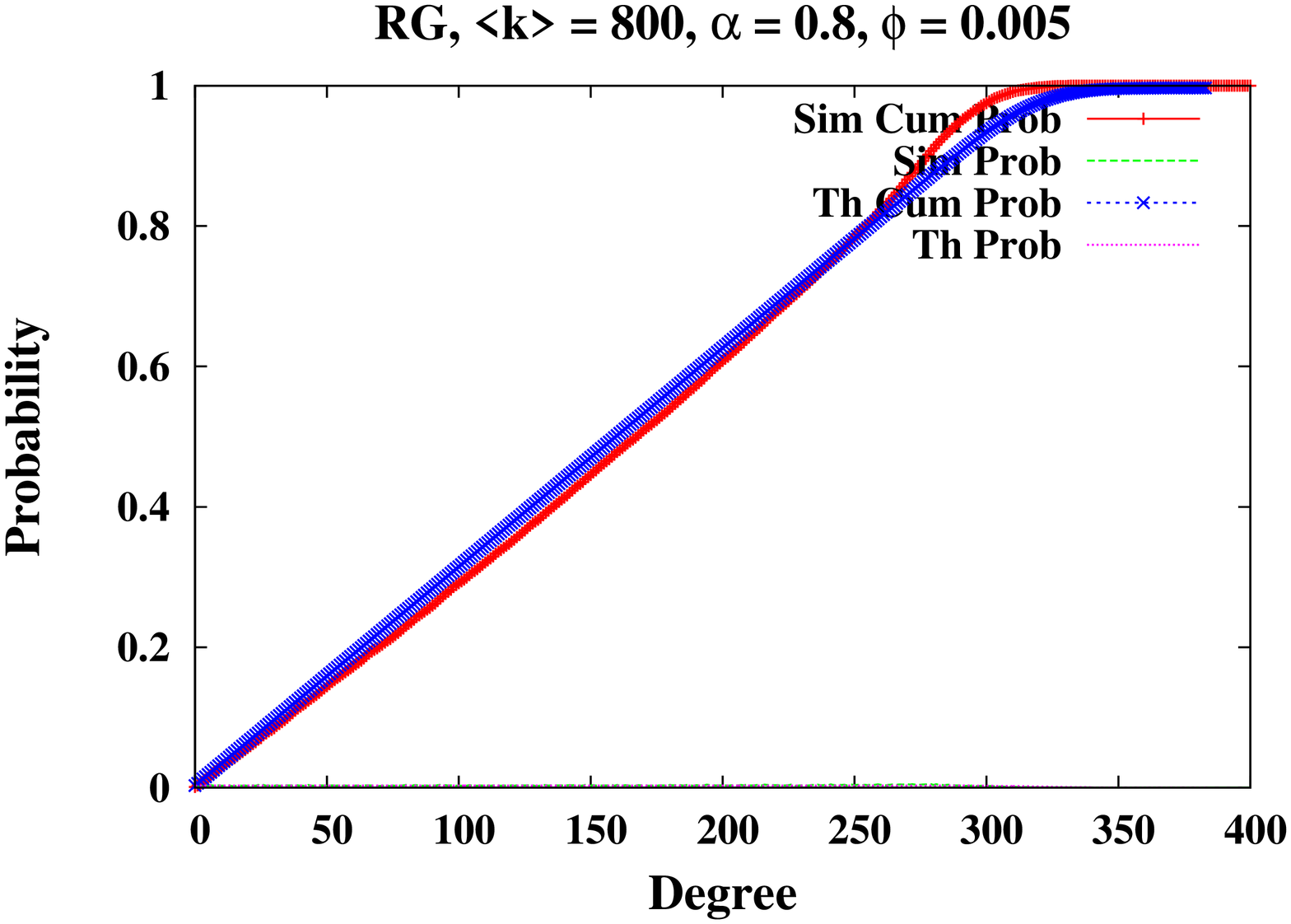}
   \caption{Degree probability varying $\alpha, \phi$; results obtained through simulation (Sim) and the mathematical modeling (Th); Random Graph model $p=0.8, |\Pi|=1000, \langle k \rangle = 800$}   
   \label{fig:fig_rg4}
\end{figure}

Figure \ref{fig:fig_rg3} shows the degree distribution through the analytical model (and simulation) obtained in the steady state (after the mentioned number of simulation steps), when the desired degree distribution 
% (that corresponds to the initial distribution peers have when starting the simulations) 
models a random graph with a probability $p=0.2$ and with a number of nodes $|\Pi| = 1000$. Figure \ref{fig:fig_rg4}, instead, reports results when $p=0.8$. 
As shown in both figures, when parameters are set as $\alpha=0.1, \phi=0.01$, a non-negligible probability is obtained only for values lower than $30$, being nodes not able to reach the average desired degrees. Similar outcomes are measured when $\phi$ is decreased down to $0.005$; in this case, non-negligible values are obtained for degrees up to $50$. Hence, in this case the desired topology is lost in the steady state.

The two considered types of random graphs behave differently when the setting is $\alpha =0.5, \phi=0.005$ (third chart of Figures \ref{fig:fig_rg3}-\ref{fig:fig_rg4}). In fact, as shown in Figure \ref{fig:fig_rg3}, with $p=0.2$, in the steady state peers have a non-negligible probability to reach degrees near the average degree $\langle k \rangle=200$. Conversely, in the latter setting ($\langle k \rangle=800$, Figure \ref{fig:fig_rg4}) the chosen value of $\alpha$ does not permit to maintain the nodes' desired degree. Similar considerations can be made for the last considered setting $\alpha =0.8, \phi=0.005$. In this case, when $p=0.2$ a peak is obtained on the degree probability for the average value $200$. Hence, the network topology is maintained for $p=0.2$, but not for $p=0.8$.
% (or networks with fixed desired degrees, which are basically a simplification of random graphs) 
These results once again confirm that the value of $\alpha$ must be properly tuned based on the average nodes' desired degree and the failure rate.

%%%%%%%%%%%%%%%%%%%%%%%%%%%%%%%%%%%%%%

\begin{figure}
   \centering
   \includegraphics[angle=270,width=.7\linewidth]{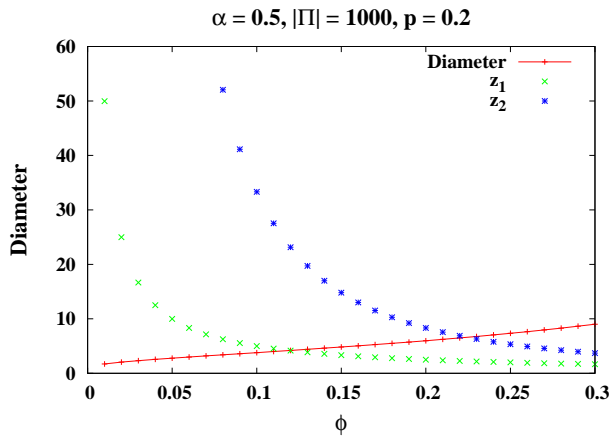}
   \caption{Diameter and average number of first neighbours of random graphs, when varying $\phi$, calculated using Equation (\ref{eq:diam})}
   \label{fig:diam_rg}
\end{figure}

Figure \ref{fig:diam_rg} shows the estimated diameter (and average number of first  and second neighbours $z_1, z_2$) of the considered random graphs, obtained when $\alpha=0.5$, while varying $\phi$, again assuming a network composed of $1000$ nodes. 
Similar considerations can be made with respect to those made for uniform graphs. 
That is, the diameter grows with $\phi$, hence confirming that a proper attachment rate must be employed to face with failures and guarantee that contents can be rapidly disseminated through the overlay, whatever the communication strategy employed on top of it.
% Also in this case, being the average desired degree high, with respect to the total number of nodes, when $\phi$ has low values, the network diameter has very small values; this values grows with $\phi$, as expected.

%%%%%%%%%%%%%%%%%%%%

\subsection{Degree Distribution of Scale Free Networks}

Scale free networks gained a lot of interest in recent years, since it has been empirically noticed that power law degree distributions $\mathit{D_k \sim k^{-\alpha}}$ 
% % (where $\mathit{P_k}$ is the distribution probability that a node has a degree equal to $\mathit{k}$) 
are quite good to model several types of real networks \cite{Barabasi2000,Verlag03structuralproperties,Faloutsos:1999,Price:1965,Adamic03localsearch,dobrescu,conf/nca/GarbinatoRT07}.
These networks are often referred as scale-free networks \cite{Newman03thestructure,simutools}.
They are characterized by the presence of hubs, i.e.~nodes with degrees higher than the average, that have an important impact on the connectivity of the net. Several works assert that scale-free networks are quite resilient to random node faults, due to the presence of hubs \cite{2000Nature_Albert,newmanHandbook}. Indeed, the majority of nodes are those with small degree; thus, it is more likely that these ones will fail, while the probability that all hubs are eliminated is almost negligible. 
% This is generally not true for other kinds of networks.
% On the other hand, studies have shown that if one selects only hubs as the faulty nodes in a scale-free network, the network rapidly becomes not connected, with several isolated graphs.
% 
% Another important consideration is that, due to the mentioned similarity to several existing networks, past works specifically concentrate their attention to scale-free nets. When turning to peer-to-peer systems, a question is if the use of scale-free networks is the proper way to model them. Of course, the answer is that it depends on the peer-to-peer network itself. For example, i

The interest on scale-free networks in this work relates to the fact that several peer-to-peer systems are indeed scale-free networks. Gnutella is a main example \cite{Adamic03localsearch}. Moreover, other peer-to-peer architectures exploit super-peers, which strongly resemble those hubs of scale-free networks \cite{cooper,garbacki,lin,pyun}.
% \footnote{This is in contrast with many approaches that voluntarily avoid the presence of hubs, which usually assume a dominant role in an overlay net. In fact, approaches based on Distributed Hash Tables commonly build a network where the degree of the nodes is kept uniform \cite{pastry,chord}. Same holds for other overlay networks that try to balance the load among peers \cite{ioannidis,wang}. These nets are hence more similar to the approaches previously considered.}

To build scale-free networks, our simulator implements a construction method which has been proposed in \cite{Aiello00arandom}. 
The interesting aspect of this algorithm is that it differs from other proposals, which build networks with a power law distribution by continuously adding novel nodes and edges, hence having networks that grow in time
\cite{Barabasi2000,barab99}. Conversely, the method in \cite{Aiello00arandom} employs a network of fixed size, characterized by two parameters $a, b$. Given $a, b$, a network is built whose number of nodes depends on these two parameters. More specifically, the number of nodes $y$ which have a degree $x$ is $\lfloor\frac{e^a}{x^b}\rfloor$. Thus, the total number of nodes of the generated network is 
$$|\Pi| = \sum_{x=1}^{\lfloor e^{\frac{a}{b}}\rfloor} \frac{e^a}{x^b},$$
being $\lfloor e^{\frac{a}{b}}\rfloor$ the maximum possible degree of the network, since it must be that $0 \leq \log{y} = a - b \log{x}$.
Once the number of nodes and their degrees have been determined, edges are randomly created among nodes until reaching their desired degrees.
We remind that, for each node in the network, such an initial degree is set as the desired degree $dd$ of the node.

\begin{figure}
   \centering
   \includegraphics[angle=270,width=0.8\linewidth]{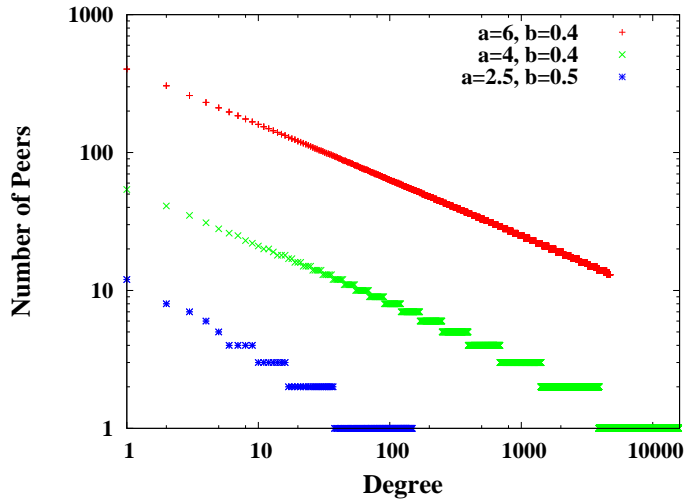}
   \caption{Degree Distribution of some scale-free networks using the construction method proposed in \cite{Aiello00arandom}}
   \label{fig:fig_rete_Aiello}
\end{figure}

Figure \ref{fig:fig_rete_Aiello} shows some examples of networks built with our simulator, implementing the construction method proposed in \cite{Aiello00arandom}. In particular, the chart reports, for three different settings of $a, b$, the number of nodes which have a given degree, in a log-log scale. It is possible to appreciate how such distributions are almost linear in a log-log scale, hence confirming they all follow some power law function.

Next Figures \ref{fig:fig_sf9}-\ref{fig:fig_sf14} show the resulting degree distribution obtained through the analytical model and through simulation, when employed over scale-free networks. For each setting, we report the degree distribution both in a linear scale (with the cumulative probability) and in a log-log scale. The latter type of charts allows to easily understand whether in the steady state the network maintains scale-free properties (i.e.~networks have a power law degree distribution) when running the distributed protocol.
% , based on the parameters $\alpha, \phi$. 
In this case, five different types of networks are considered, obtained by employing the following pairs of parameters, i.e.~$a = 3, b= 0.5$ (forming scale-free networks with a number of nodes $|\Pi| = 777$, Figure \ref{fig:fig_sf9}), $a = 4.5, b= 0.8$ ($|\Pi| = 876$, Figure \ref{fig:fig_sf11}), $a = 5, b= 0.9$ ($|\Pi| = 1079$, Figure \ref{fig:fig_sf12}), $a = 3.2, b= 0.5$ ($|\Pi| = 1167$, Figure \ref{fig:fig_sf13}), $a = 3.2, b= 0.45$ ($|\Pi| = 2196$, Figure \ref{fig:fig_sf14}).
For these networks, values of $\alpha, \phi$ were varied.

Results show that indeed scale-free properties are maintained, in the steady state, when high attachment rates are selected
% , hence confirming to have a reliable topology 
(see the two last scenarios in the various figures, with $\phi=0.005$, while $\alpha=0.5, 0.8$, respectively). Conversely, values of $\alpha$ reported in the first two scenarios of each figure ($\alpha=0.1, \phi=0.01, 0.005$) demonstrate that when the attachment rate is not sufficiently rapid to repair failures, the typical topology of a scale-free network is lost. In fact, the degree distribution in the log-log scale is not linear. These results are common to all the considered networks.

The reliability of scale-free nets was already demonstrated in other works \cite{Verlag03structuralproperties,Newman03thestructure,2000Nature_Albert,dumitriu}. However, they usually considered attacks while keeping the network almost static, without the possibility to react to these nodes/links removals. (The main reason is that these models are often employed for studying, for instance, the spread of viruses or general percolation properties in a net.) Our assessment demonstrates that the simple proposed distributed protocol enables the maintenance of scale-free topologies also when nodes are subjected to periodical failures. Once the desired topology of the network has been specified and each node has its own assigned degree, it suffices to employ an adequate attachment rate to randomly select novel neighbours.
% \footnote{To avoid any confusion between the random peer selection of the distributed protocol and the preferential attachment that may be employed to build a scale-free network, note that we are not stating that a simple random attachment allows to create a scale-free network \cite{Barabasi2000,2000Nature_Albert}. Rather, the desired degree distribution of peers follows a power law distribution. Then, the attachment rate of the distributed protocol allows a given peer to randomly select another one when it needs some additional link it previously lost (the distributed protocol does not depend on the desired network topology).}
As already mentioned, when nodes are randomly selected to fail, there is a low probability that a major portion of hubs of the network is removed from the net (since there are few hubs in the network, with respect to other nodes) \cite{Newman03thestructure,newmanHandbook}. Rather, it is more likely that peers which fail are non-hubs with low degrees. Under these circumstances, hubs that lose some neighbours have time to react to these failures by finding novel nodes to link with. This allows to maintain a scale-free topology.

%%%%%%%%%%%%%%%%%%%%%%%%%%%%%%%%%%%%%%%%%%%%%%%%%%%%%%%%%%%%%%%%%%
\begin{figure}
   \centering
   \includegraphics[angle=270,width=.45\linewidth]{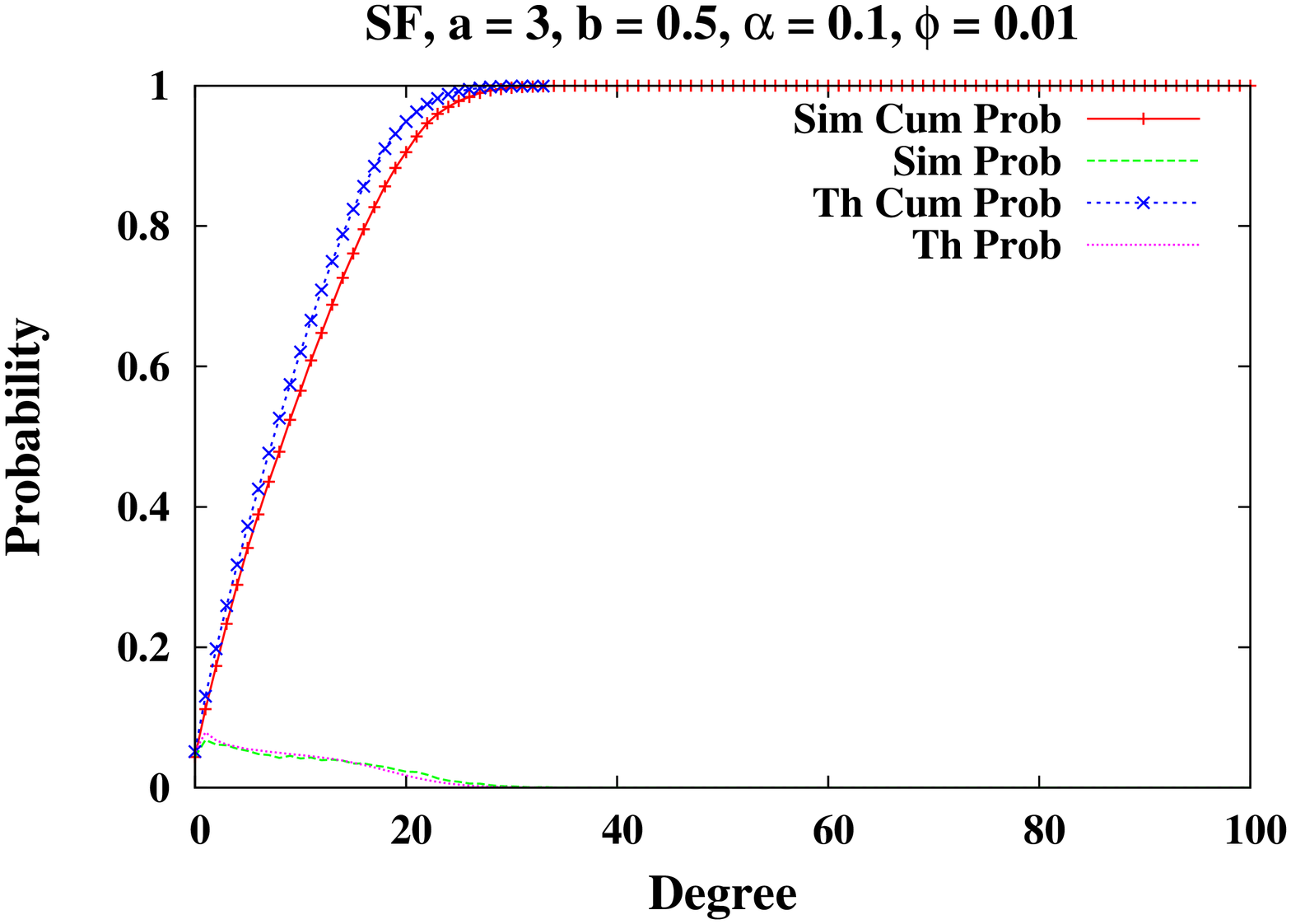}
   \includegraphics[angle=270,width=.45\linewidth]{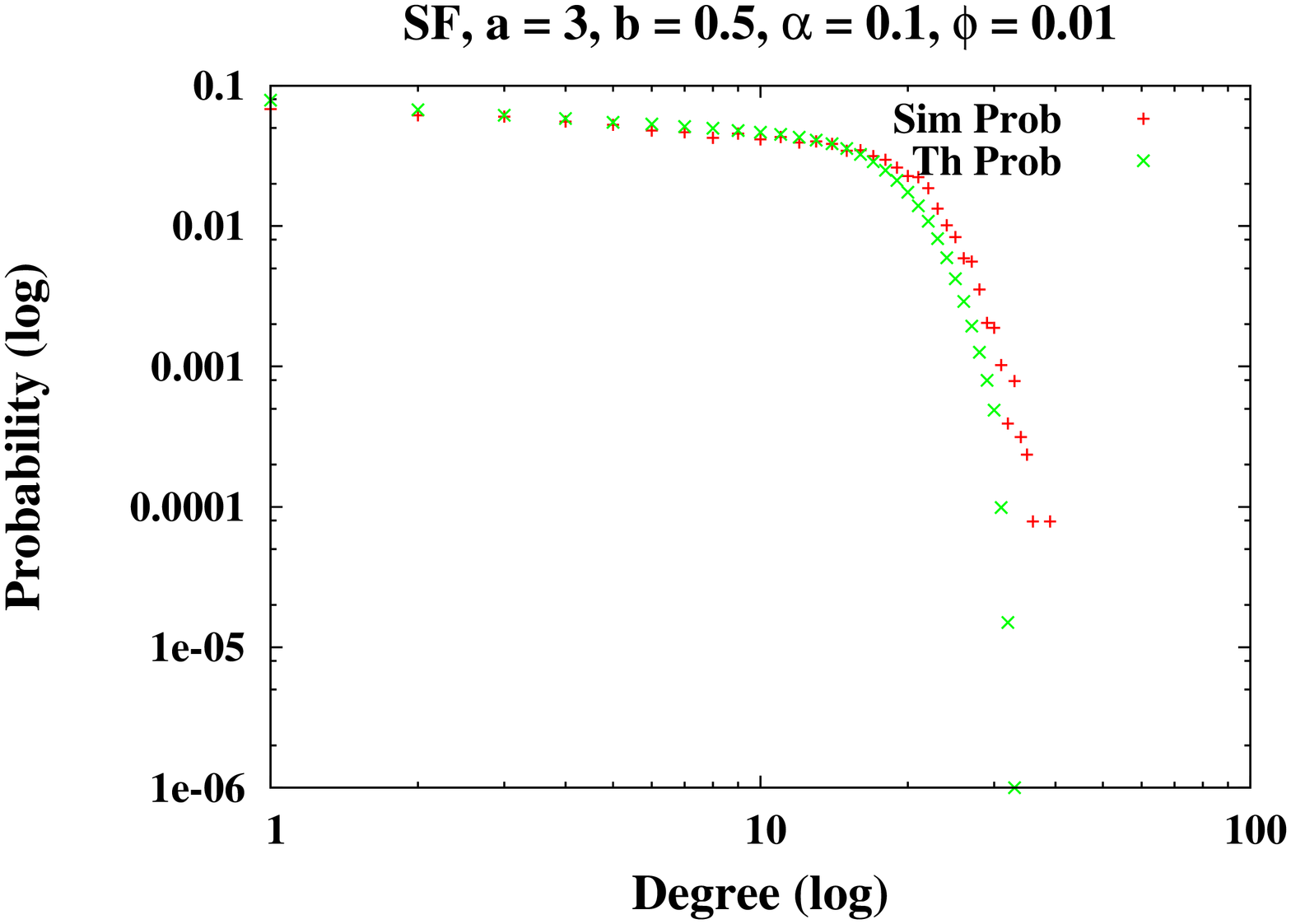}
   \includegraphics[angle=270,width=.45\linewidth]{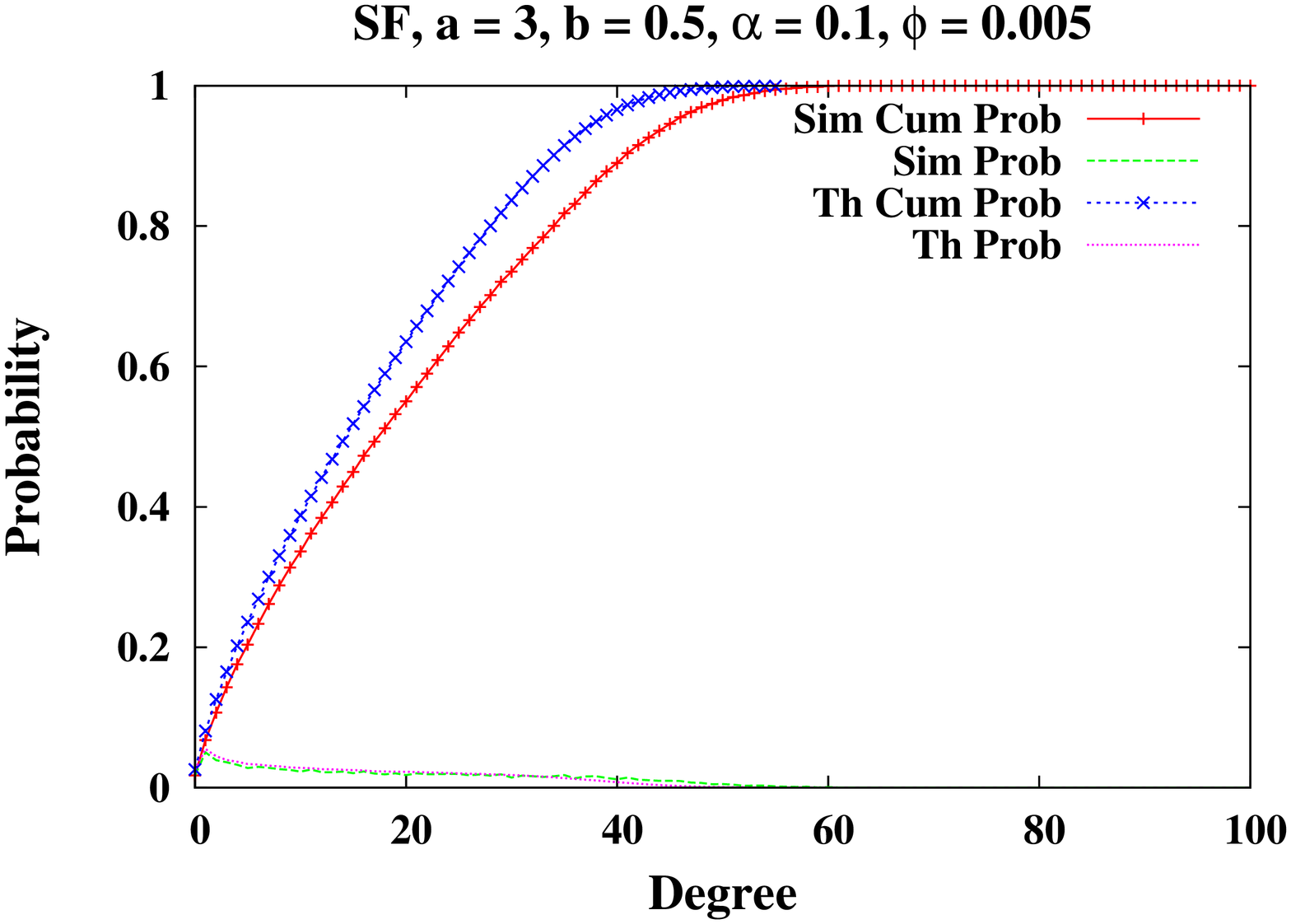}
   \includegraphics[angle=270,width=.45\linewidth]{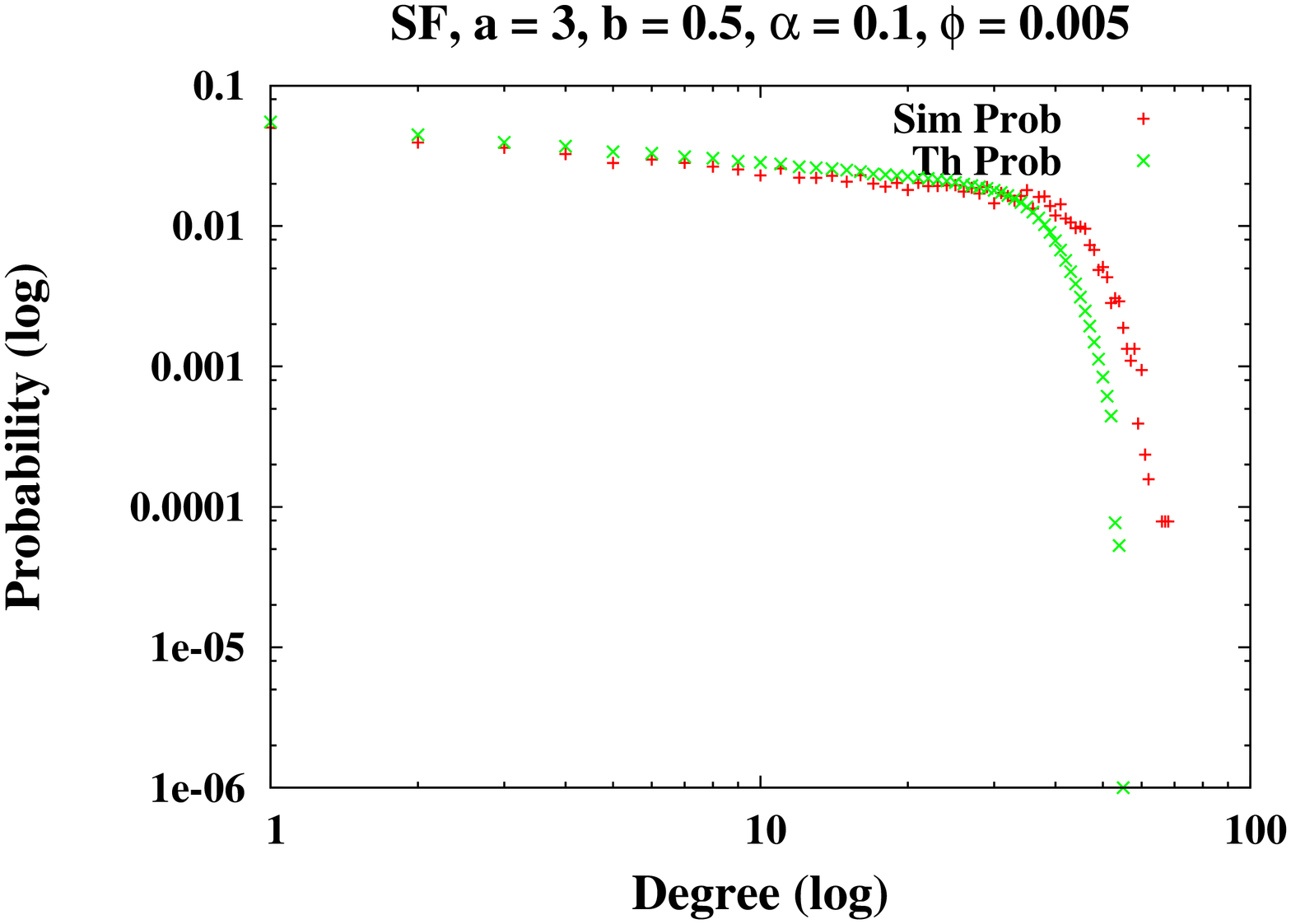}
   \includegraphics[angle=270,width=.45\linewidth]{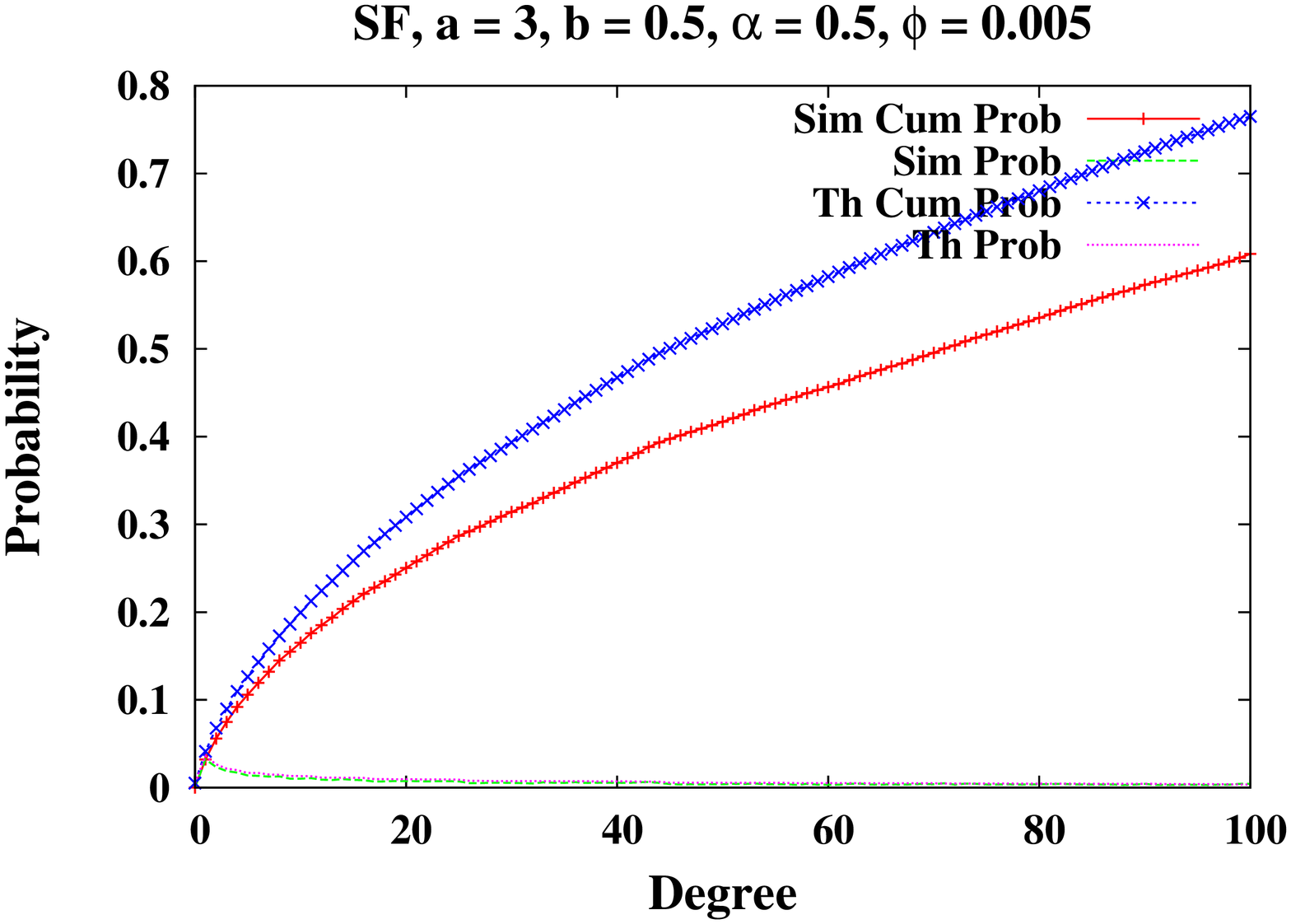}
   \includegraphics[angle=270,width=.45\linewidth]{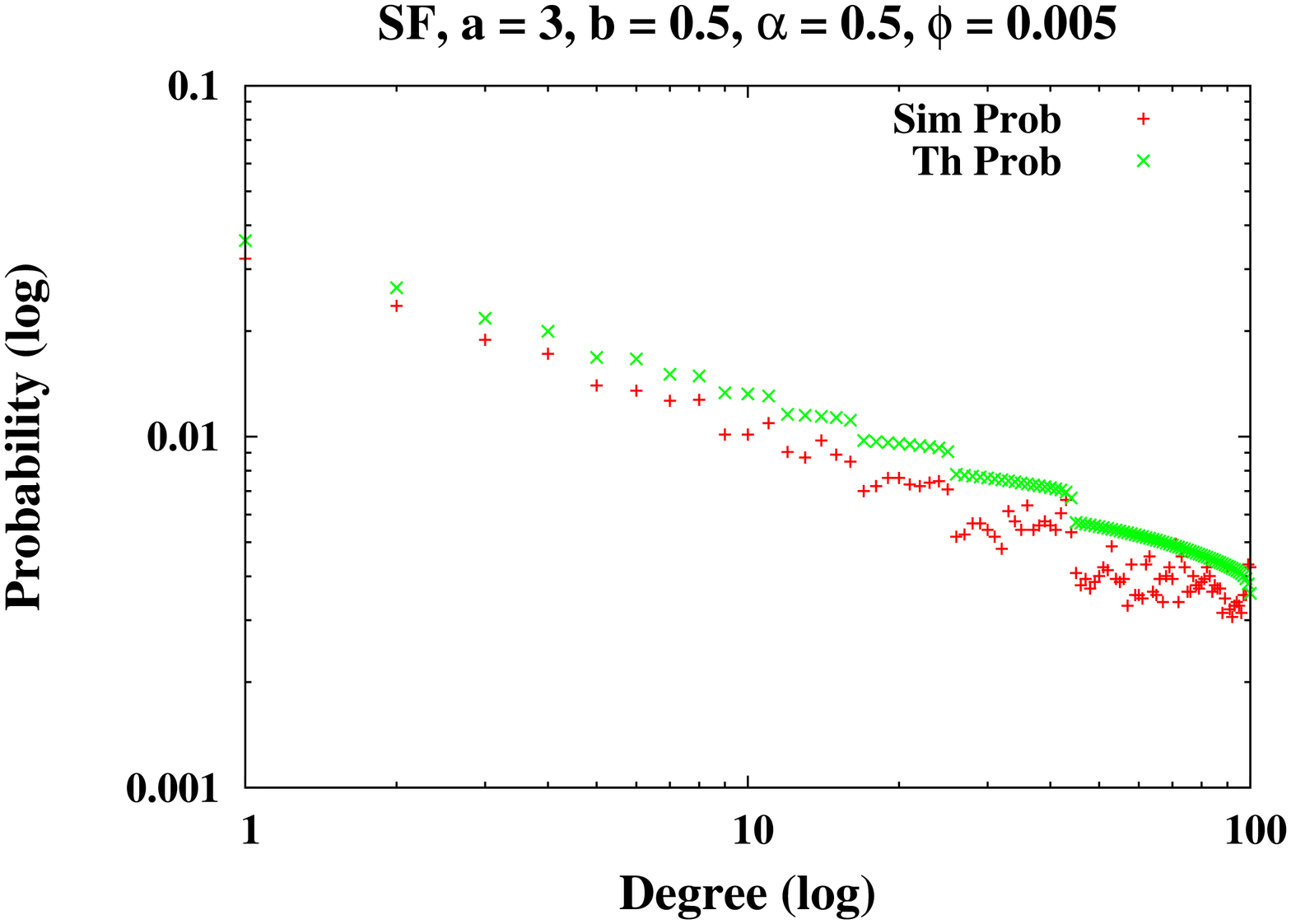}
   \includegraphics[angle=270,width=.45\linewidth]{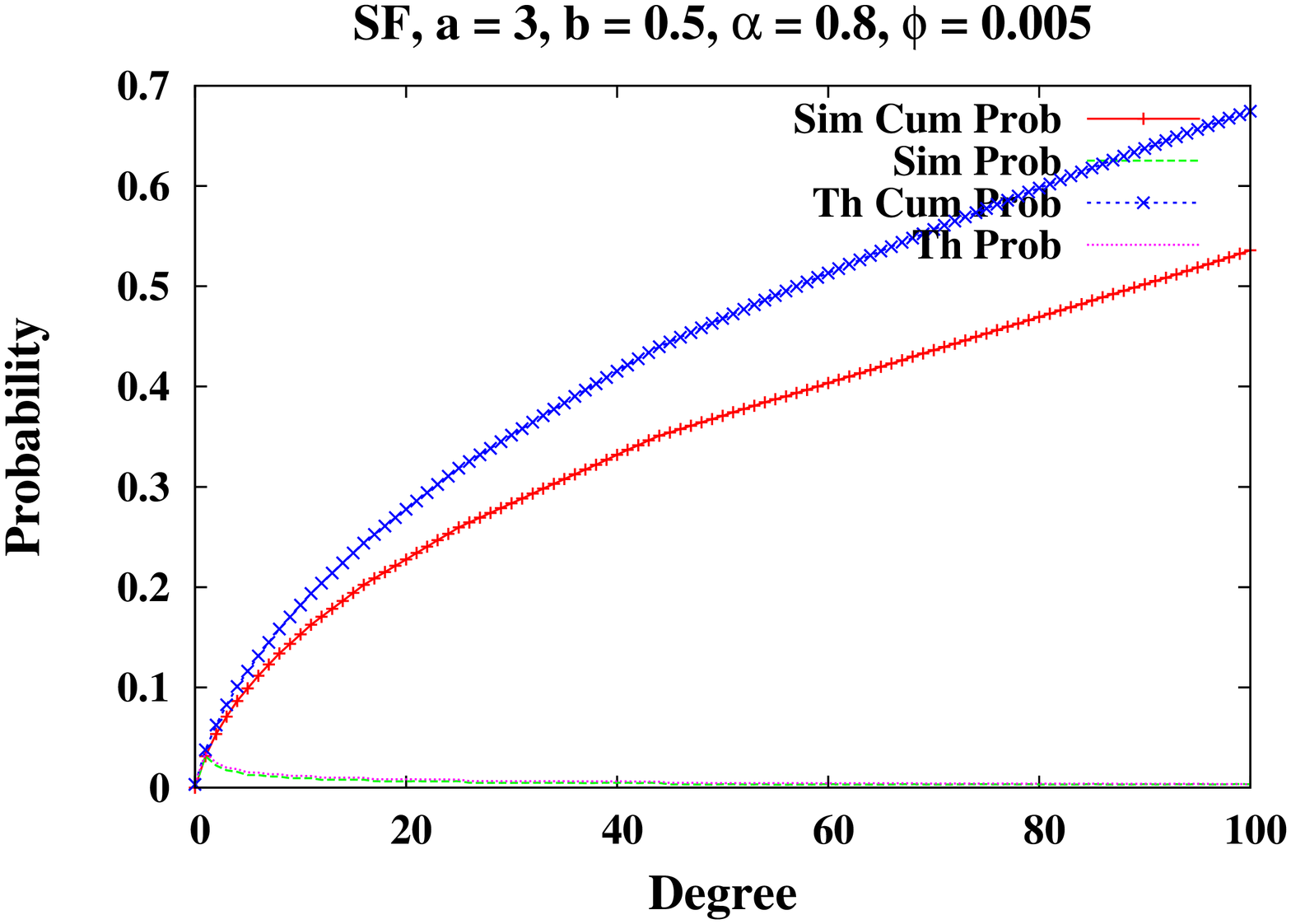}
   \includegraphics[angle=270,width=.45\linewidth]{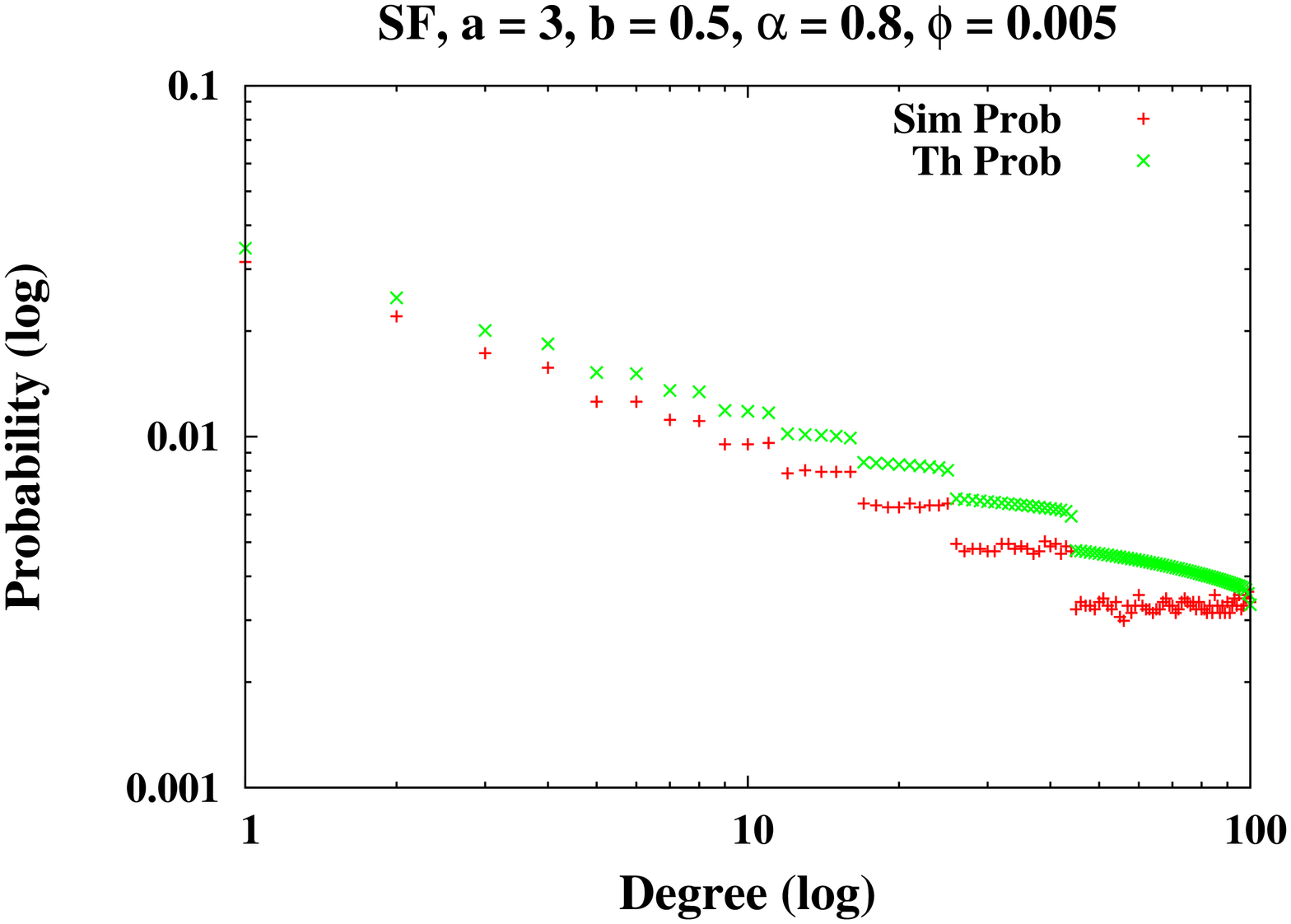}
   \caption{Degree probability and cumulative degree probability varying $\alpha, \phi$ on the left side; degree probability in log scale on the right side; results obtained through simulation (Sim) and the mathematical modeling (Th); Scale Free networks $a=3, b=0.5, |\Pi|=777$}
   \label{fig:fig_sf9}
\end{figure}
%%%%%%%%%%%%%%%%%%%%%%%%%%%%%%%%%%%%%%%%%%%%%%%%%%%%%%%%%%%%%%%%%%%%%%

\begin{figure}
   \centering
   \includegraphics[angle=270,width=.45\linewidth]{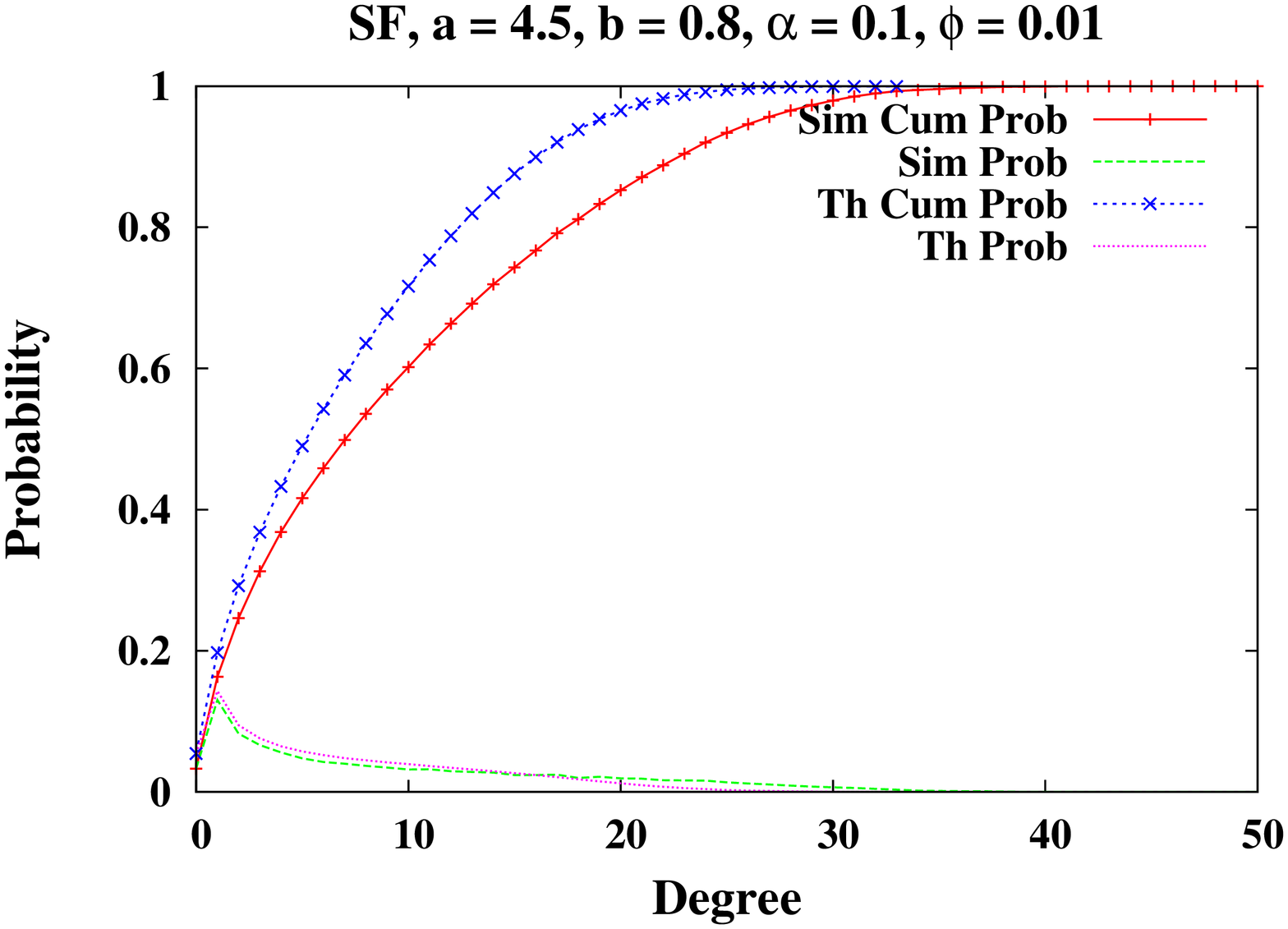}
   \includegraphics[angle=270,width=.45\linewidth]{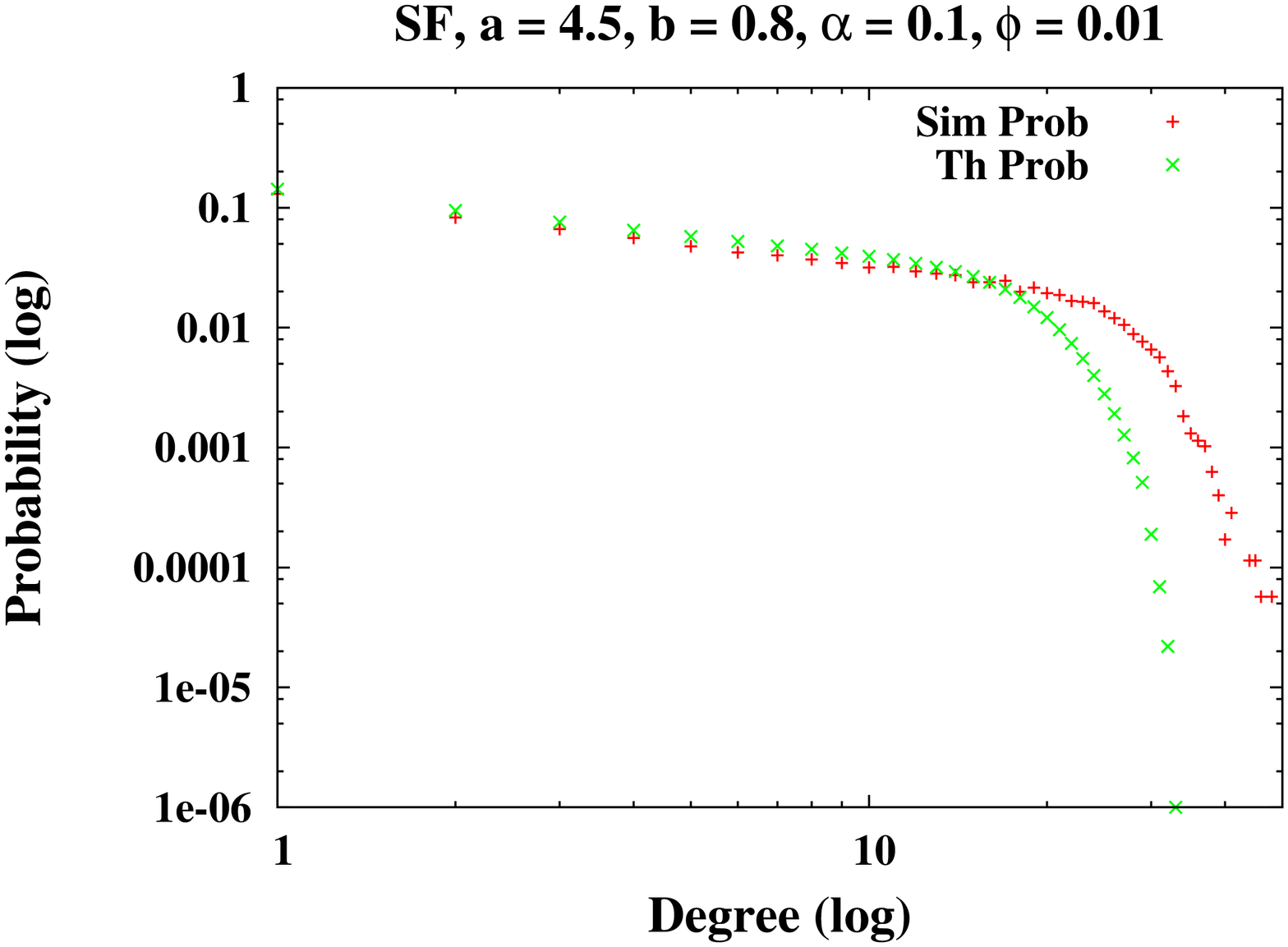}
   \includegraphics[angle=270,width=.45\linewidth]{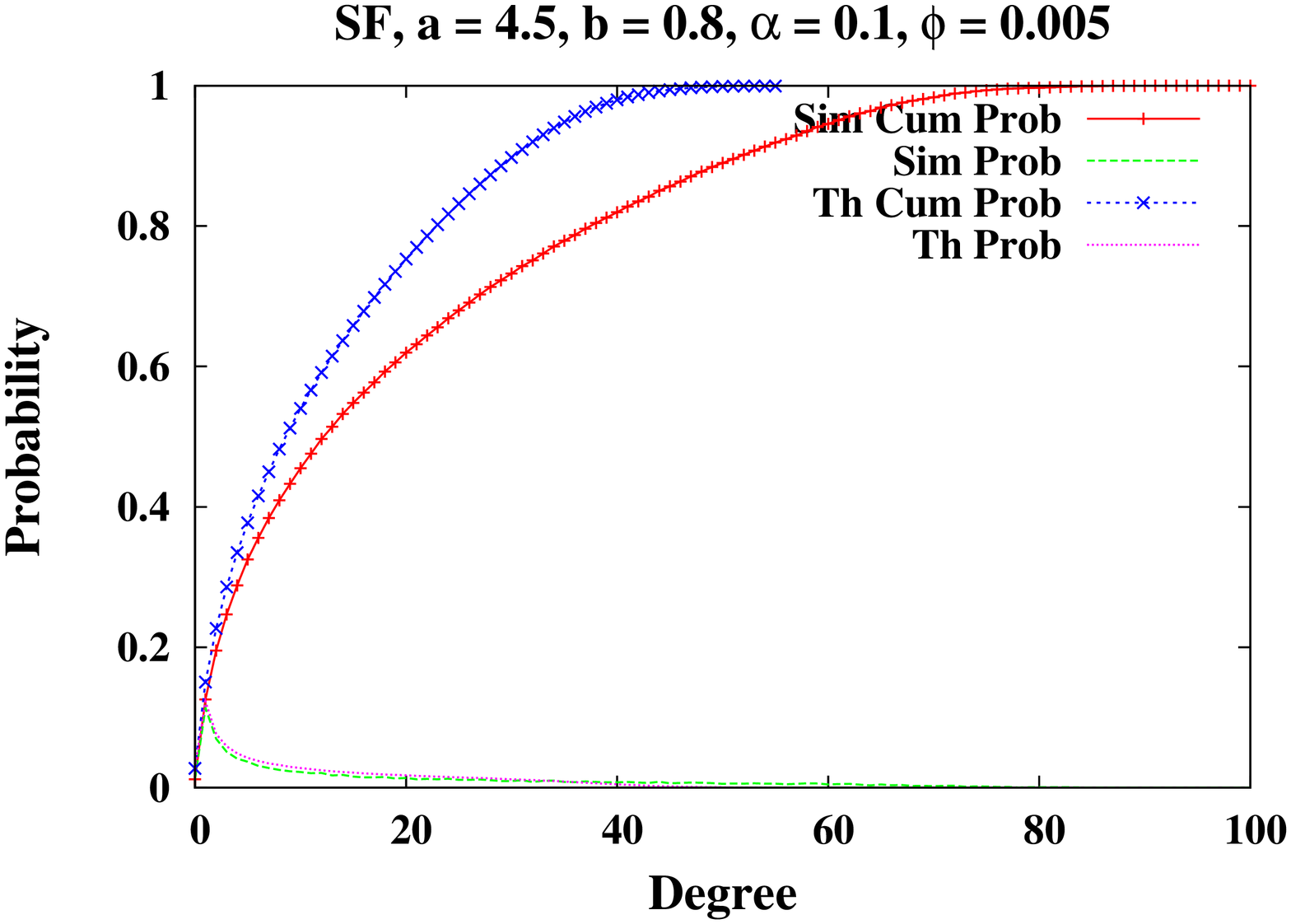}
   \includegraphics[angle=270,width=.45\linewidth]{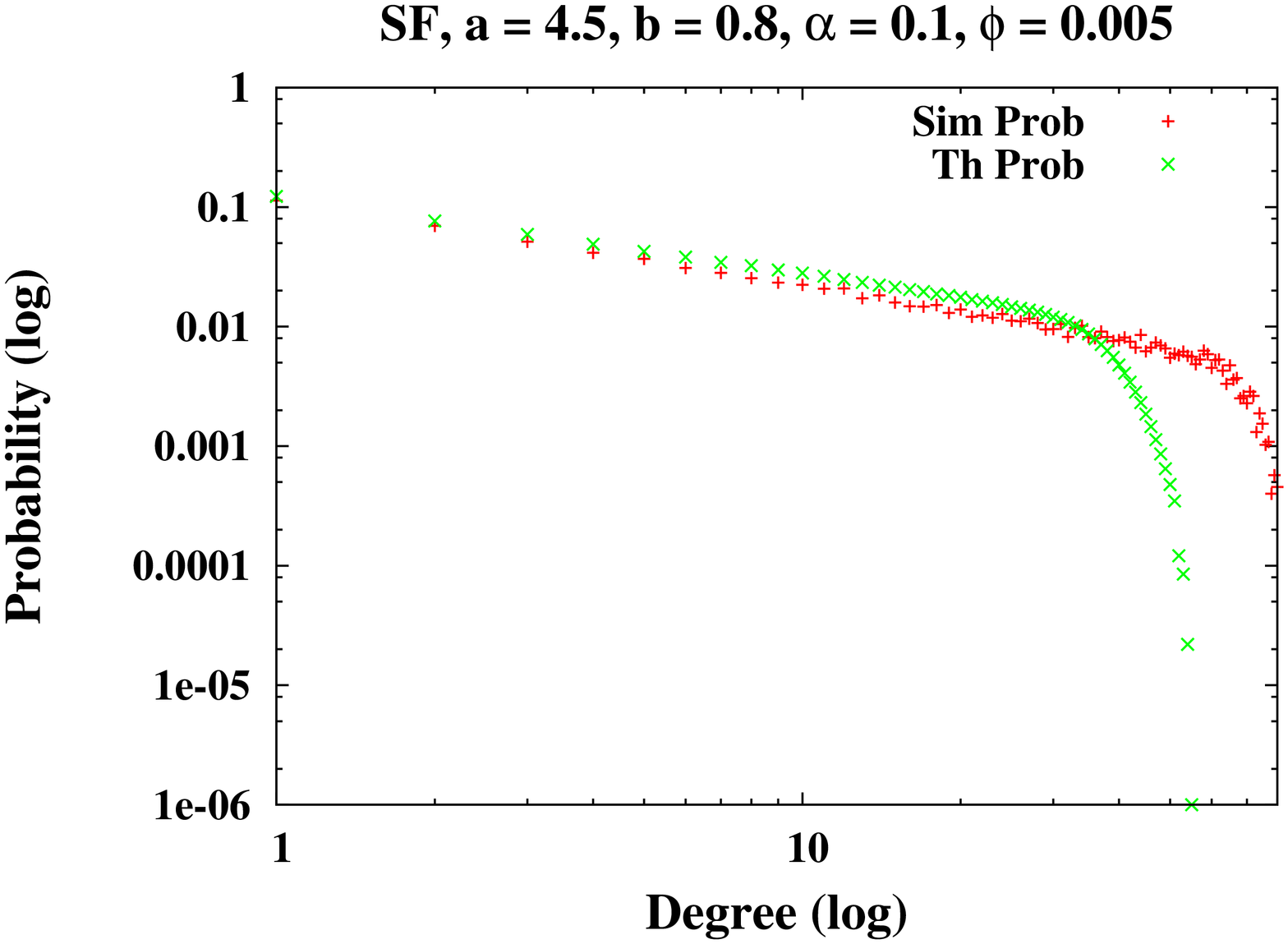}
   \includegraphics[angle=270,width=.45\linewidth]{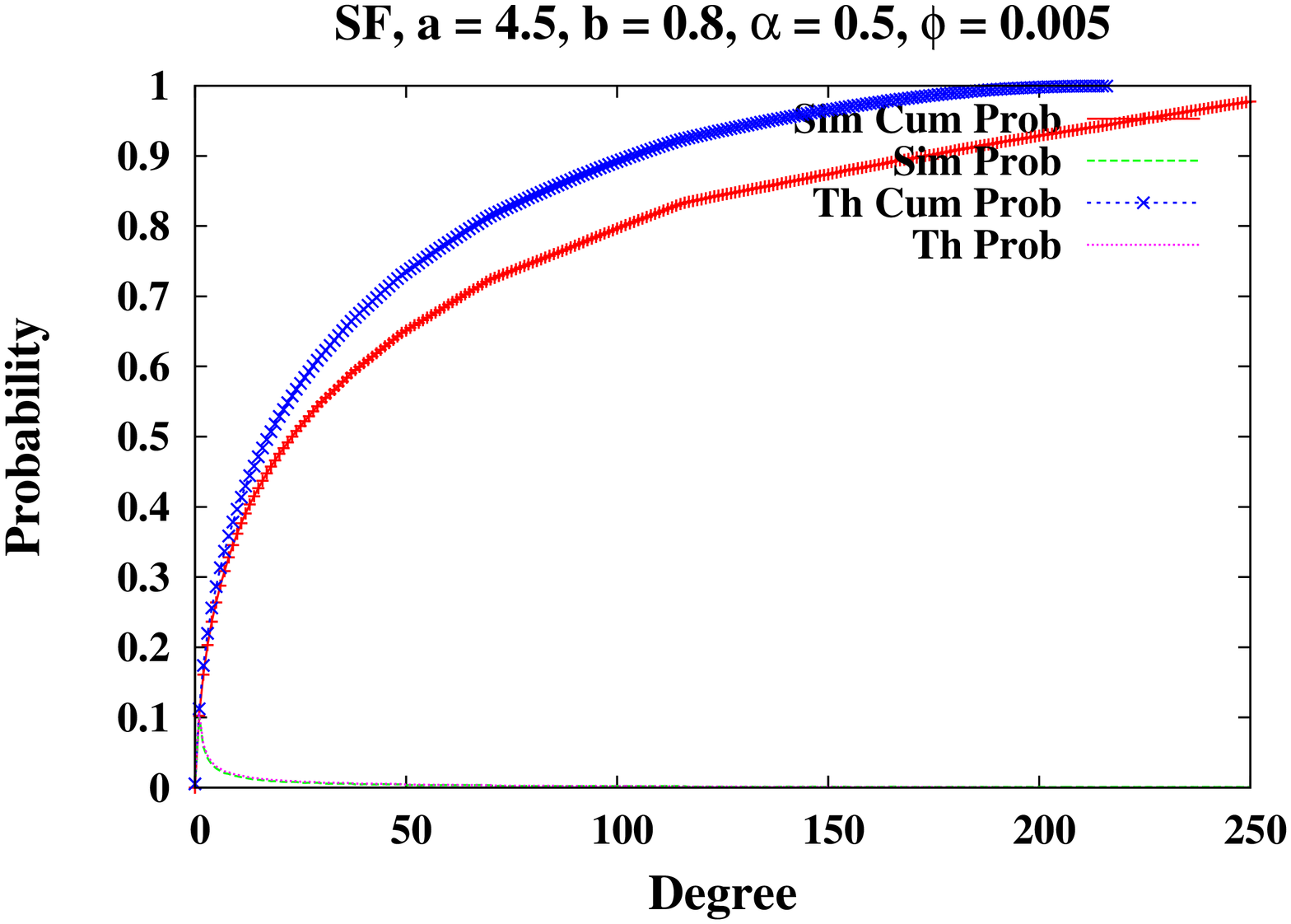}
   \includegraphics[angle=270,width=.45\linewidth]{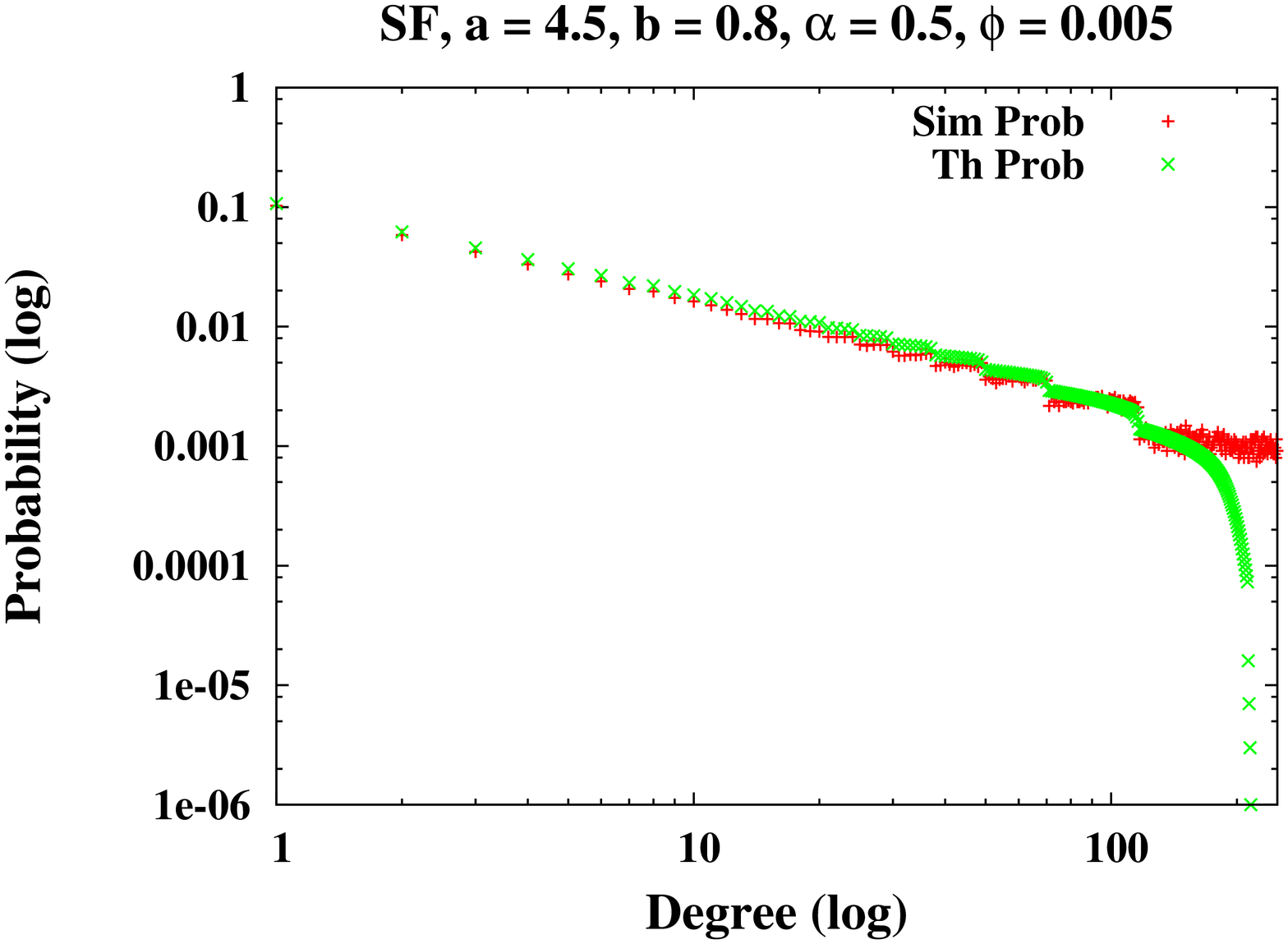}
   \includegraphics[angle=270,width=.45\linewidth]{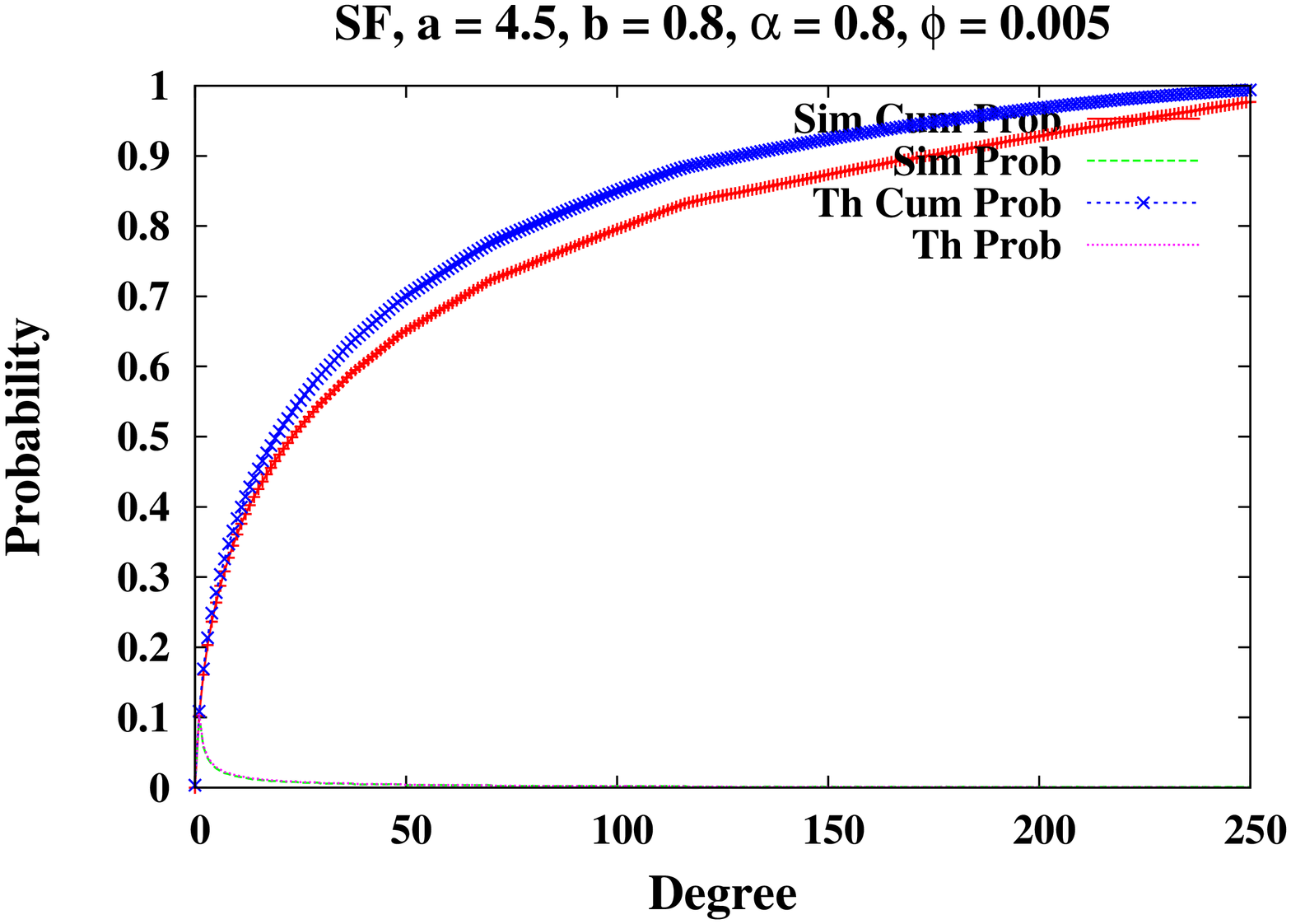}
   \includegraphics[angle=270,width=.45\linewidth]{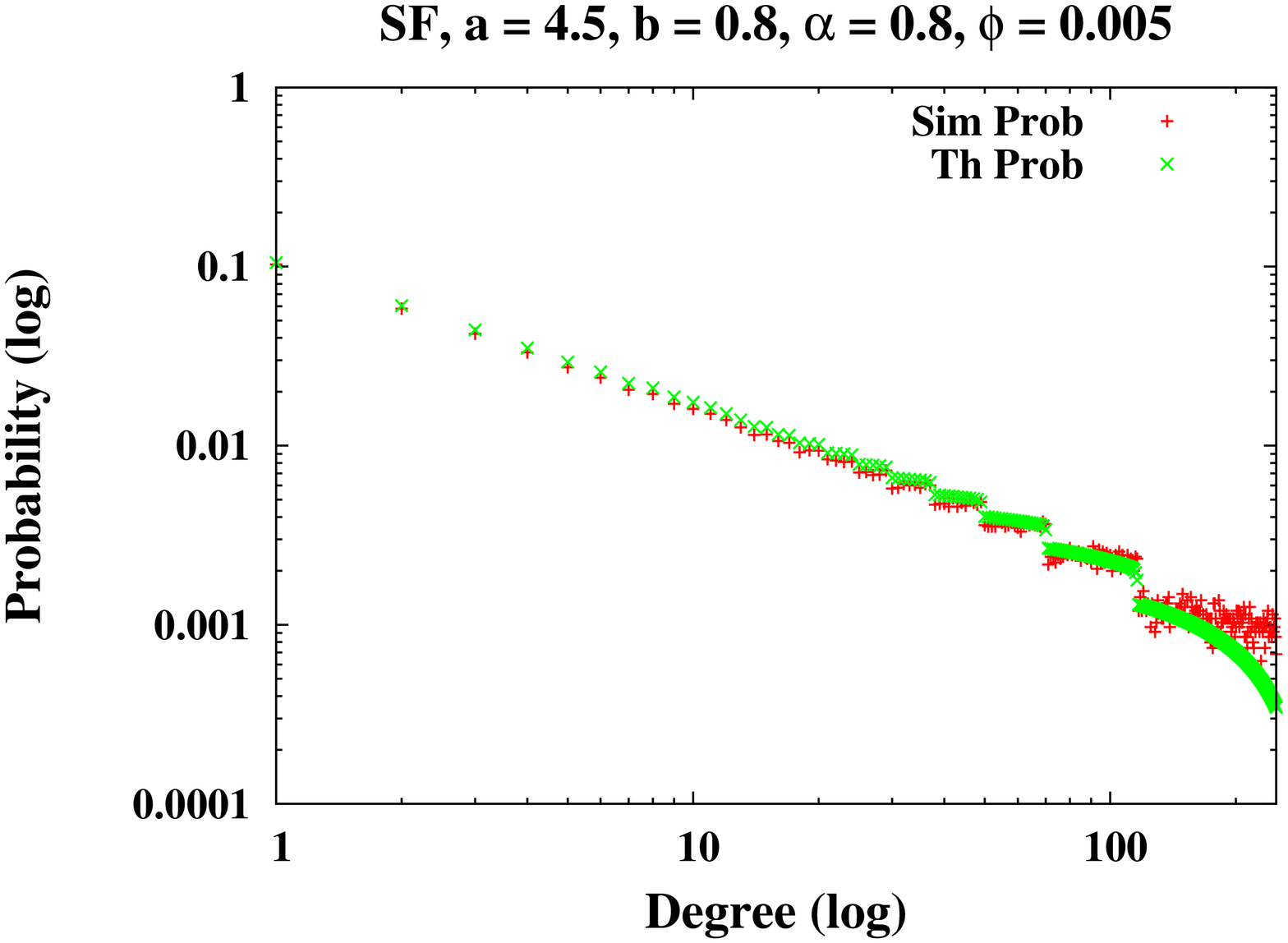}
   \caption{Degree probability and cumulative degree probability varying $\alpha, \phi$ on the left side; degree probability in log scale on the right side; results obtained through simulation (Sim) and the mathematical modeling (Th); Scale Free networks $a=4.5, b=0.8, |\Pi|=876$}
   \label{fig:fig_sf11}
\end{figure}

%%%%%%%%%%%%%%%%%%%%%%%%%%%%%%%%%%%%%%%%%%%%%%%%%%%%%%%%%%%%%%%%%%%%%%%%%
\begin{figure}
   \centering
   \includegraphics[angle=270,width=.45\linewidth]{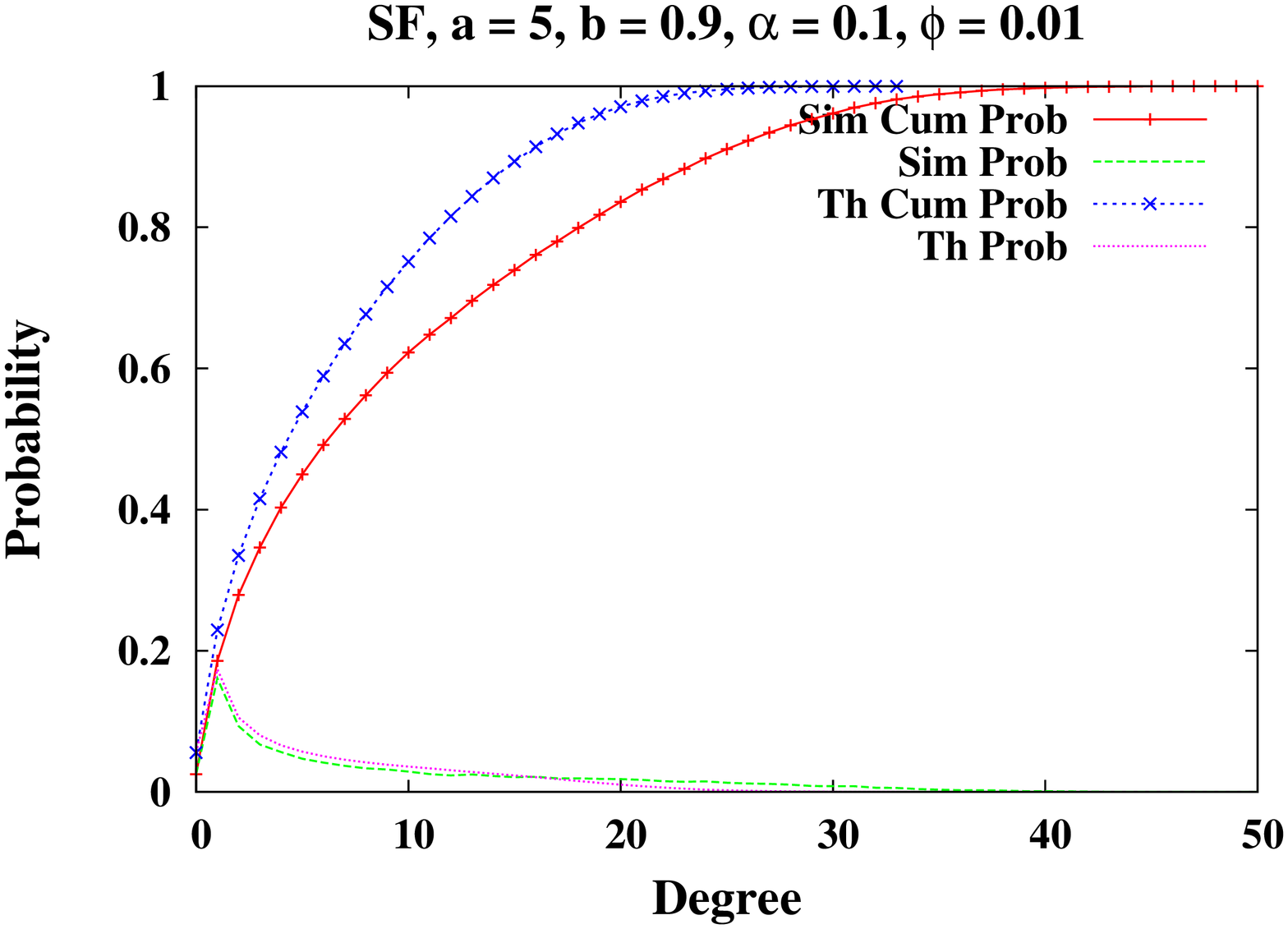}
   \includegraphics[angle=270,width=.45\linewidth]{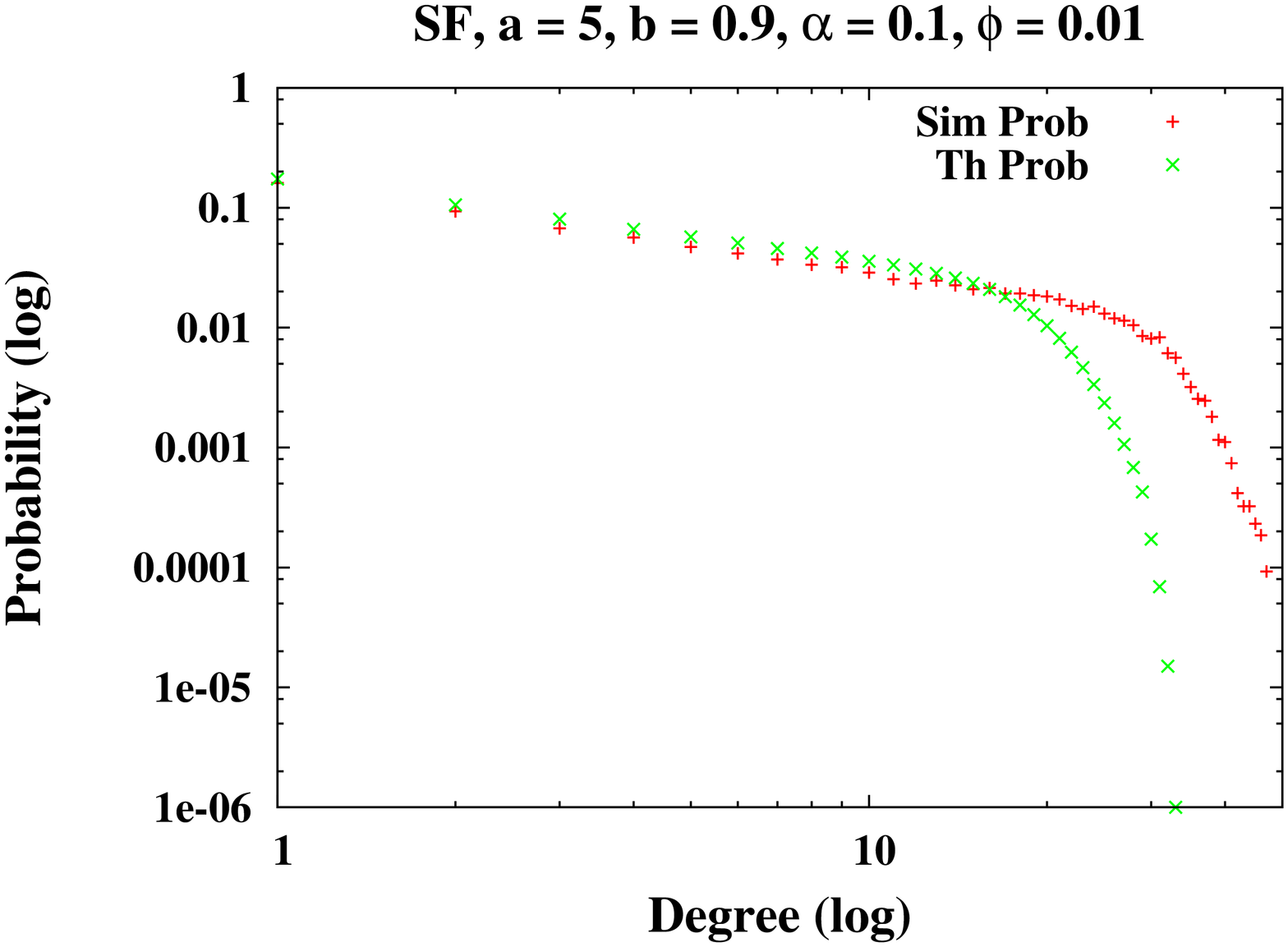}
   \includegraphics[angle=270,width=.45\linewidth]{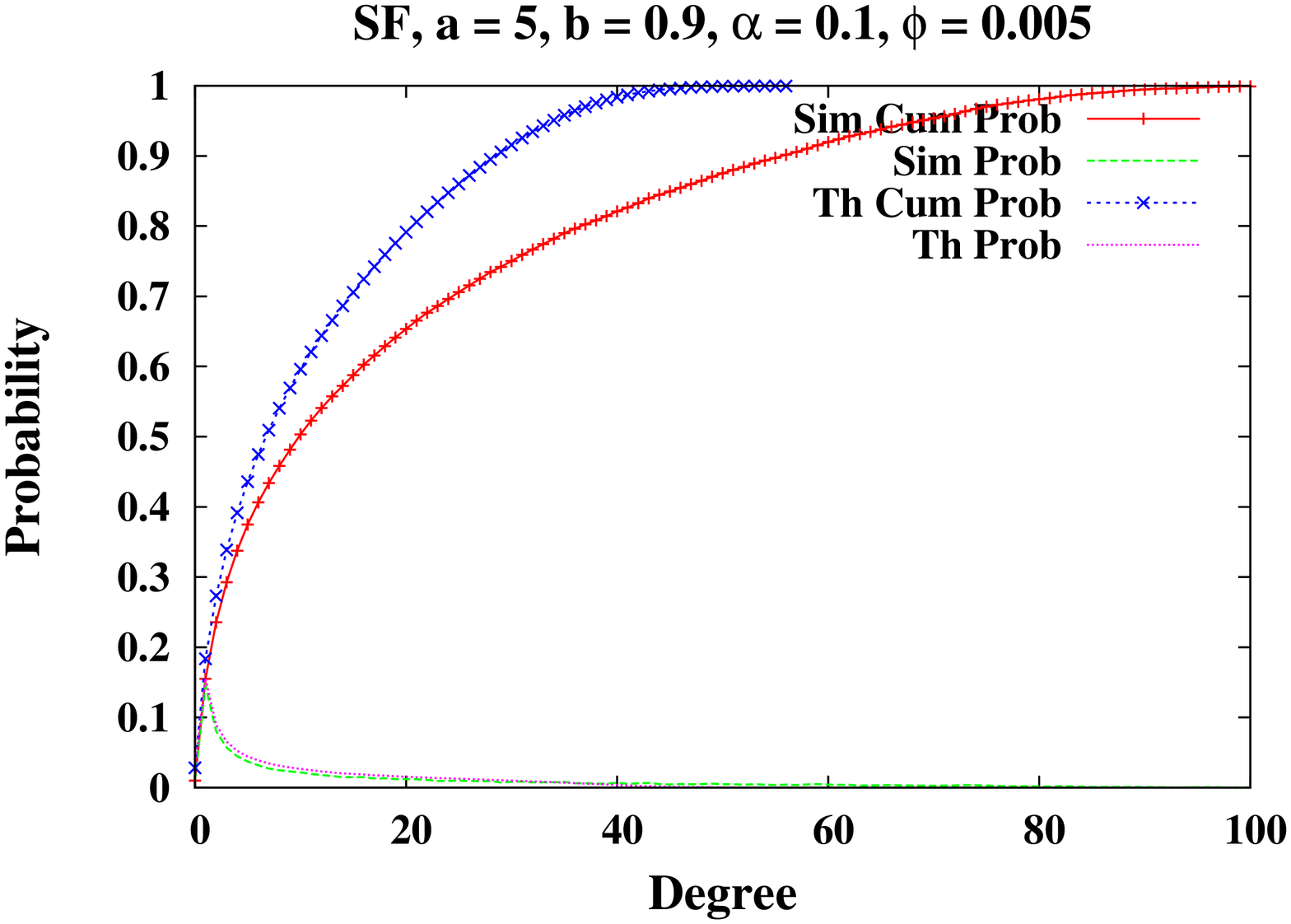}
   \includegraphics[angle=270,width=.45\linewidth]{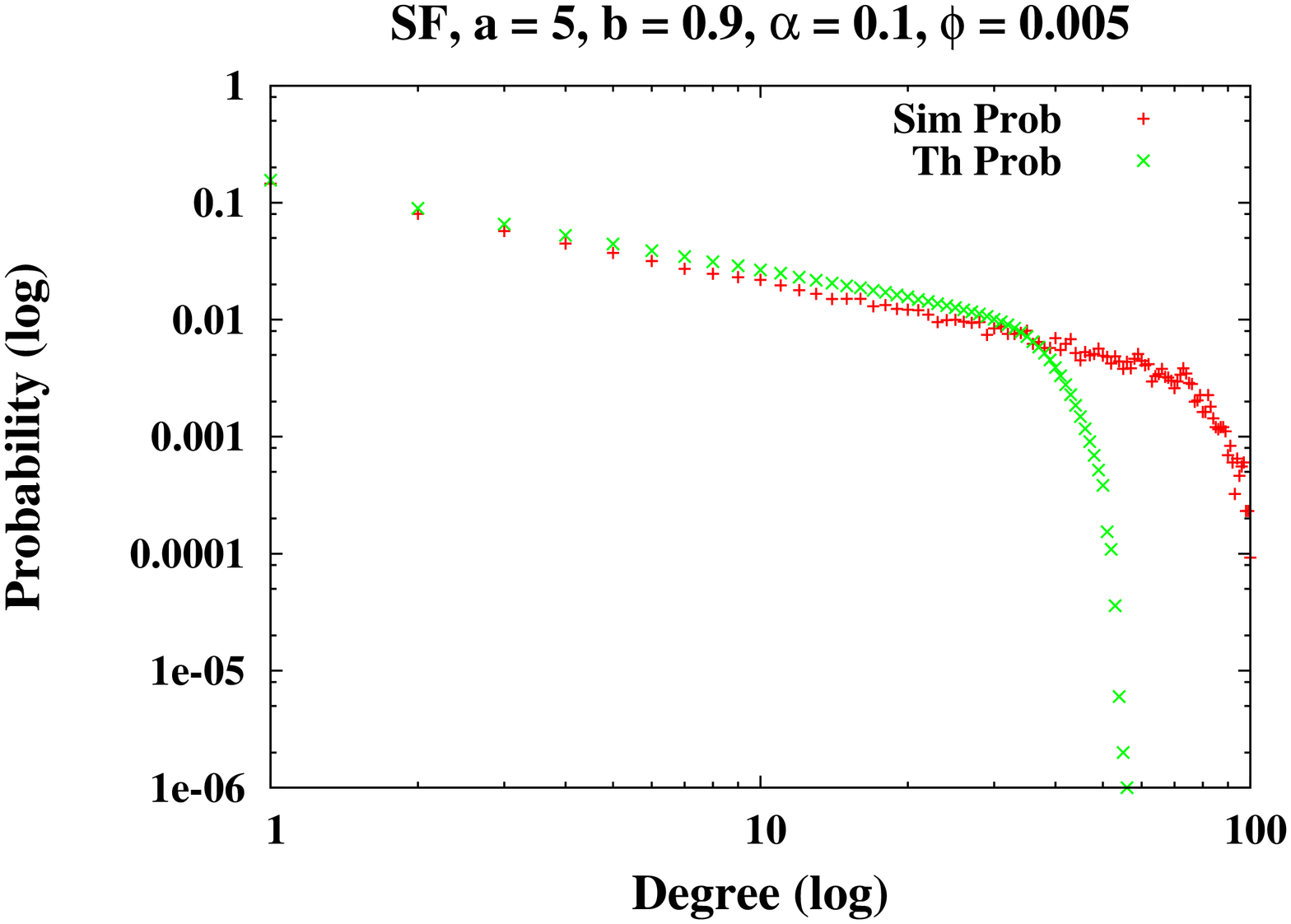}
   \includegraphics[angle=270,width=.45\linewidth]{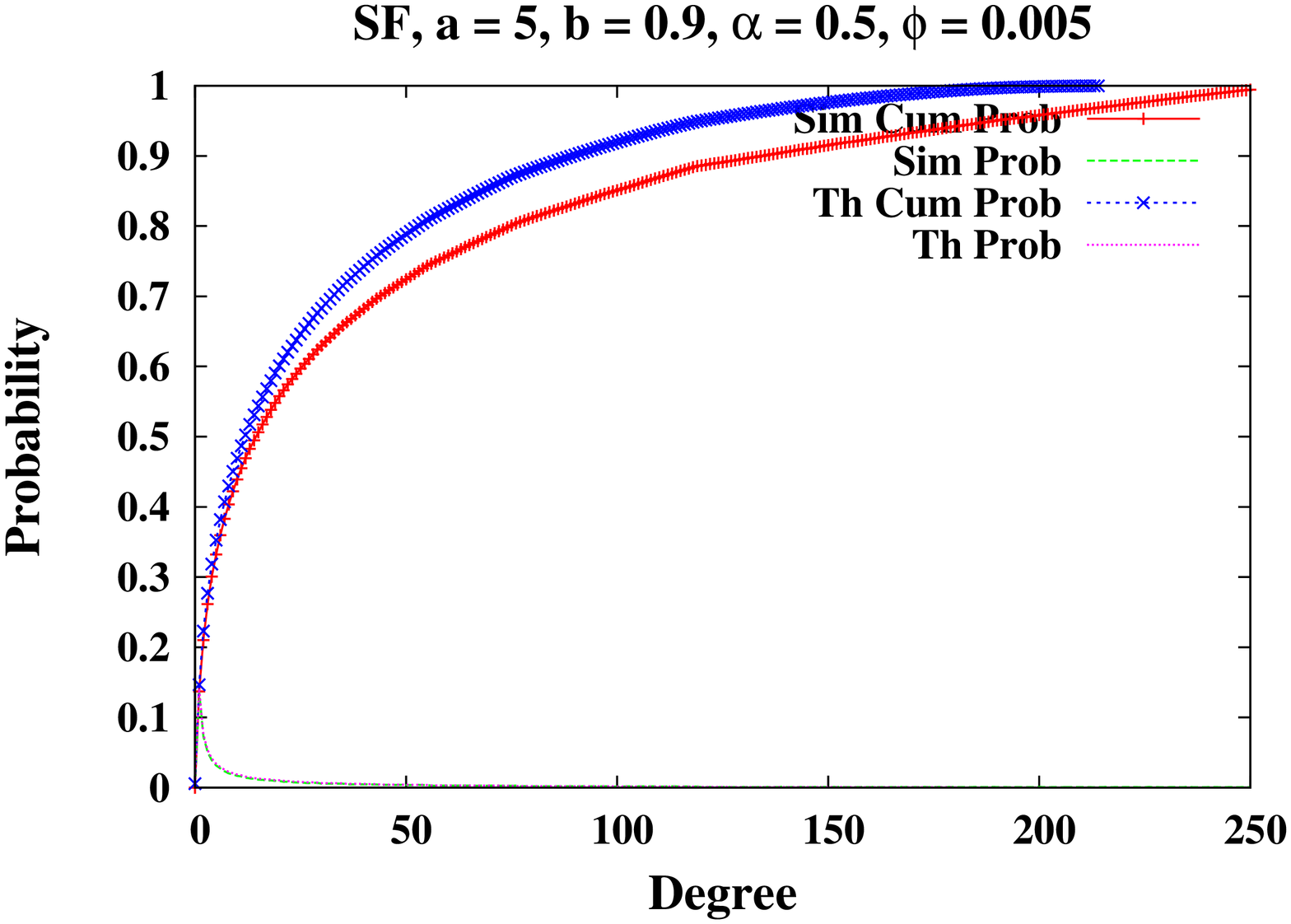}
   \includegraphics[angle=270,width=.45\linewidth]{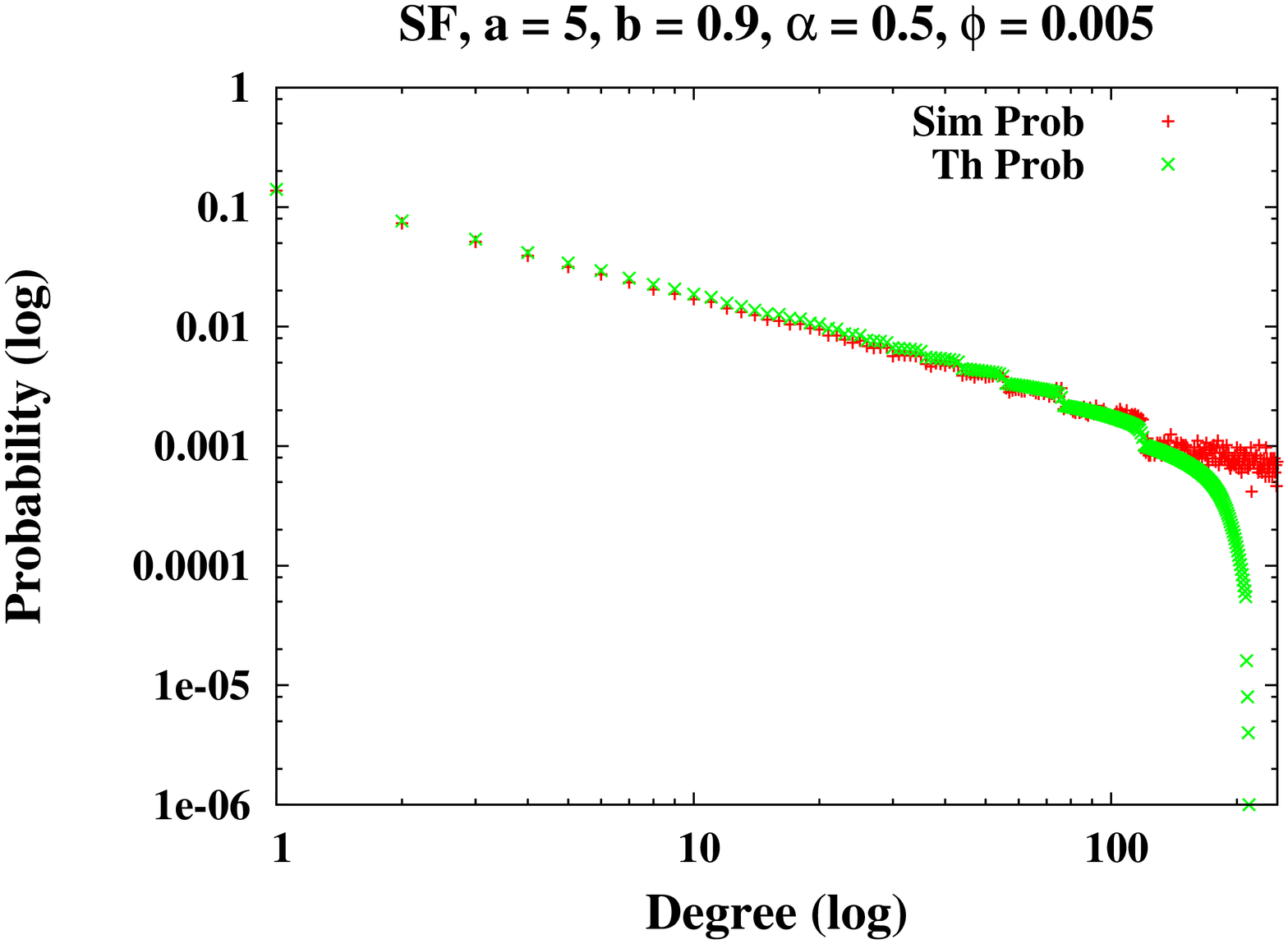}
   \includegraphics[angle=270,width=.45\linewidth]{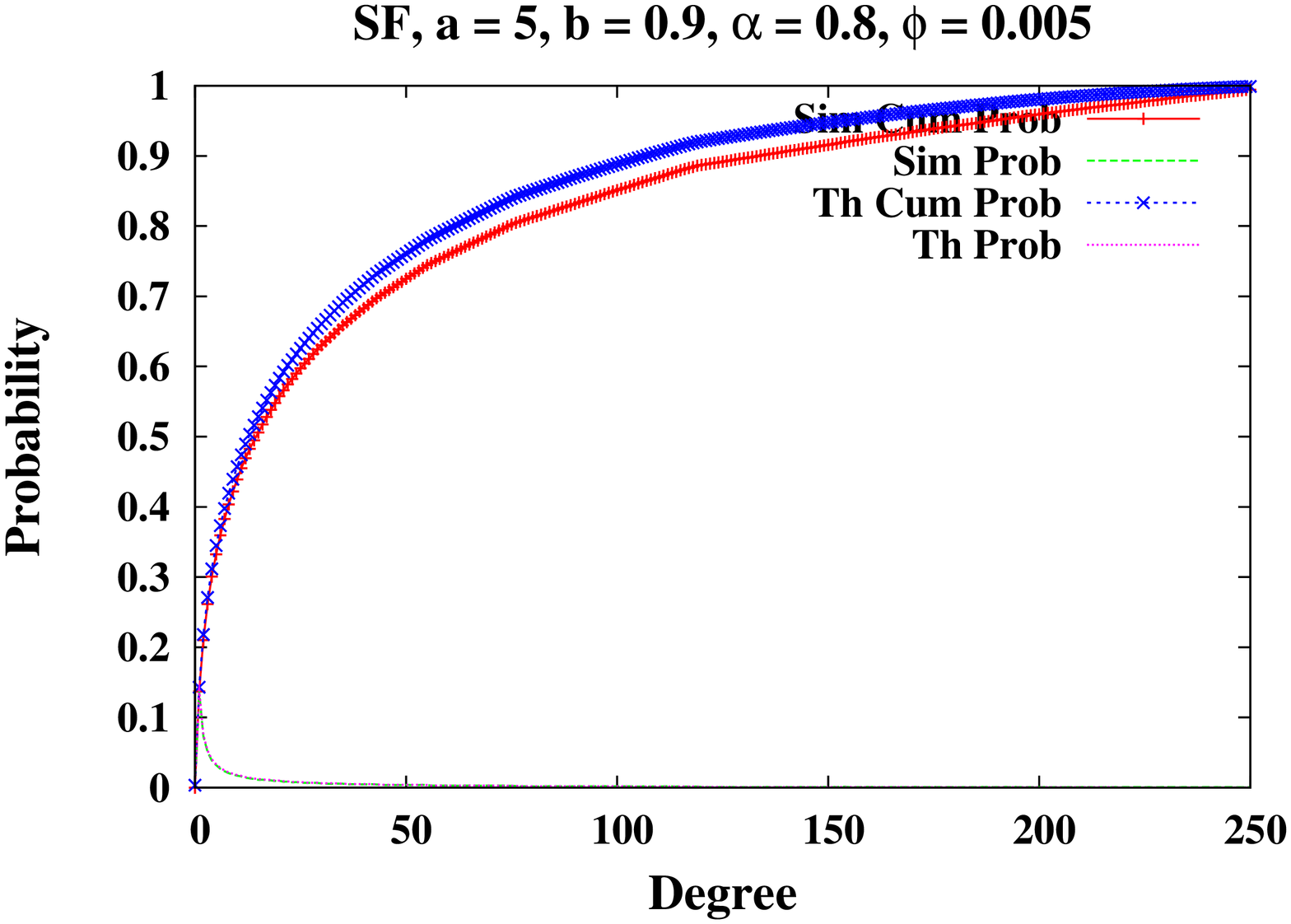}
   \includegraphics[angle=270,width=.45\linewidth]{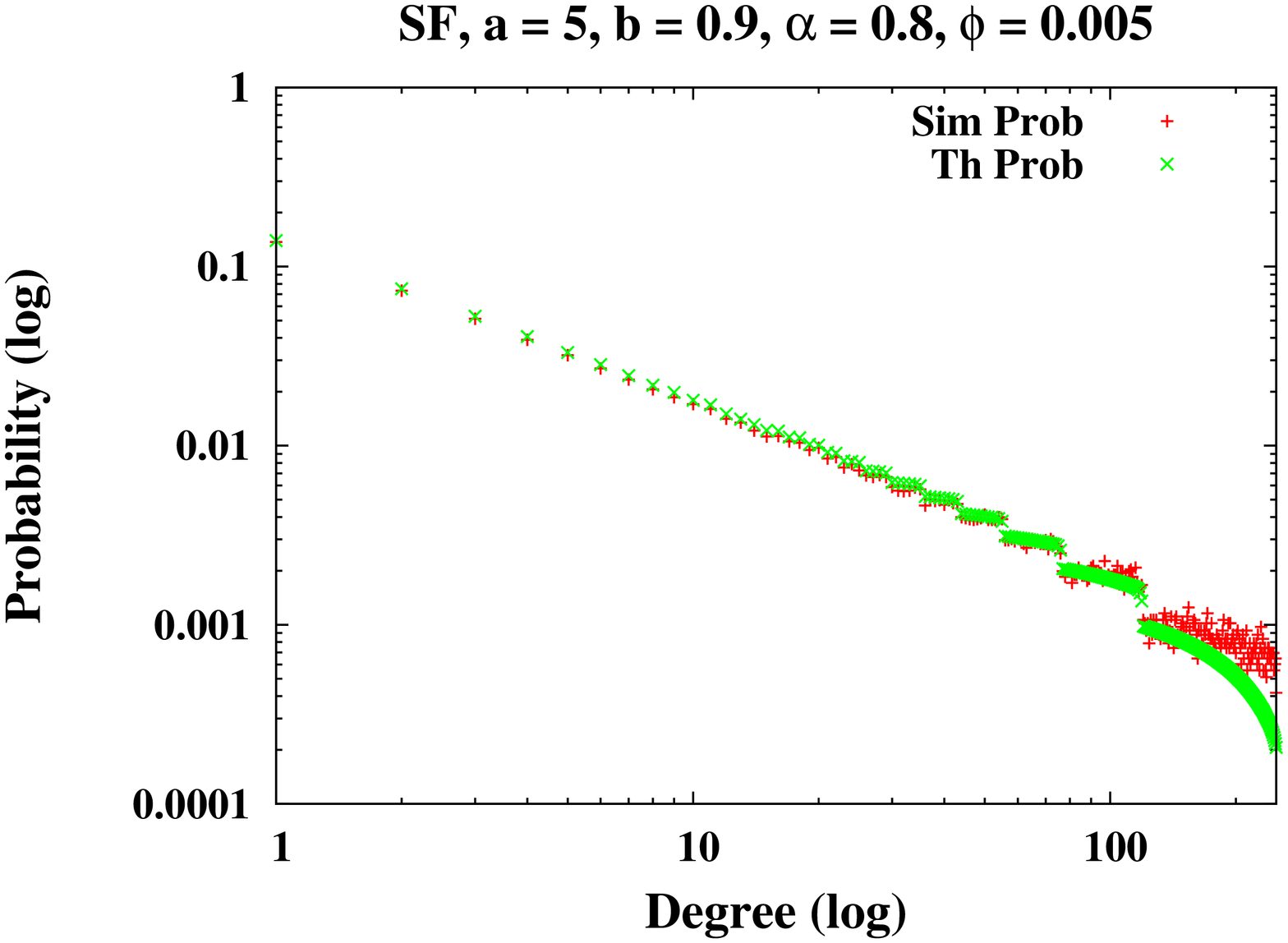}
   \caption{Degree probability and cumulative degree probability varying $\alpha, \phi$ on the left side; degree probability in log scale on the right side; results obtained through simulation (Sim) and the mathematical modeling (Th); Scale Free networks $a=5, b=0.9, |\Pi|=1079$}
   \label{fig:fig_sf12}
\end{figure}

%%%%%%%%%%%%%%%%%%%%%%%%%%%%%%%%%%%%%%%%
\begin{figure}
   \centering
   \includegraphics[angle=270,width=.45\linewidth]{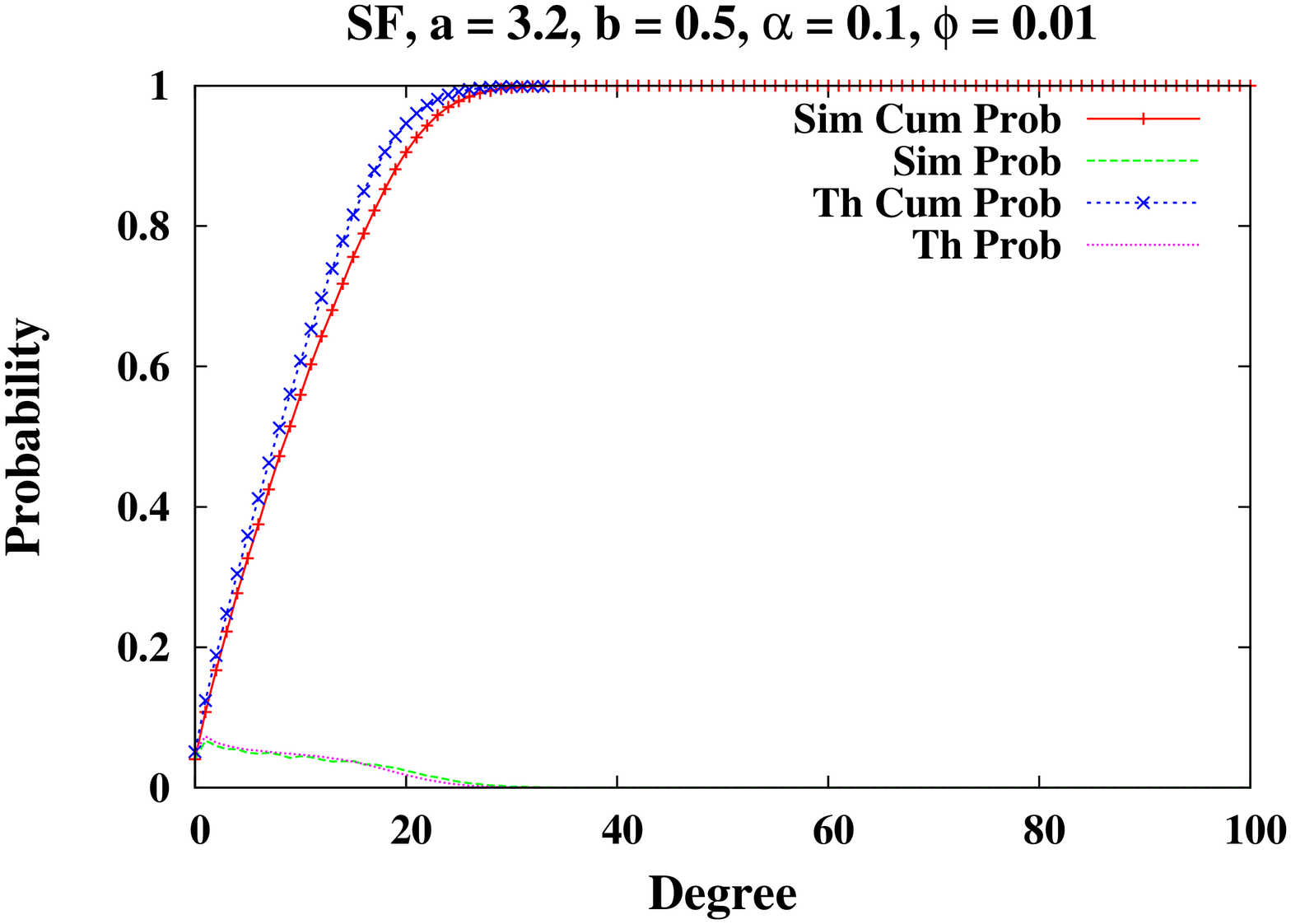}
   \includegraphics[angle=270,width=.45\linewidth]{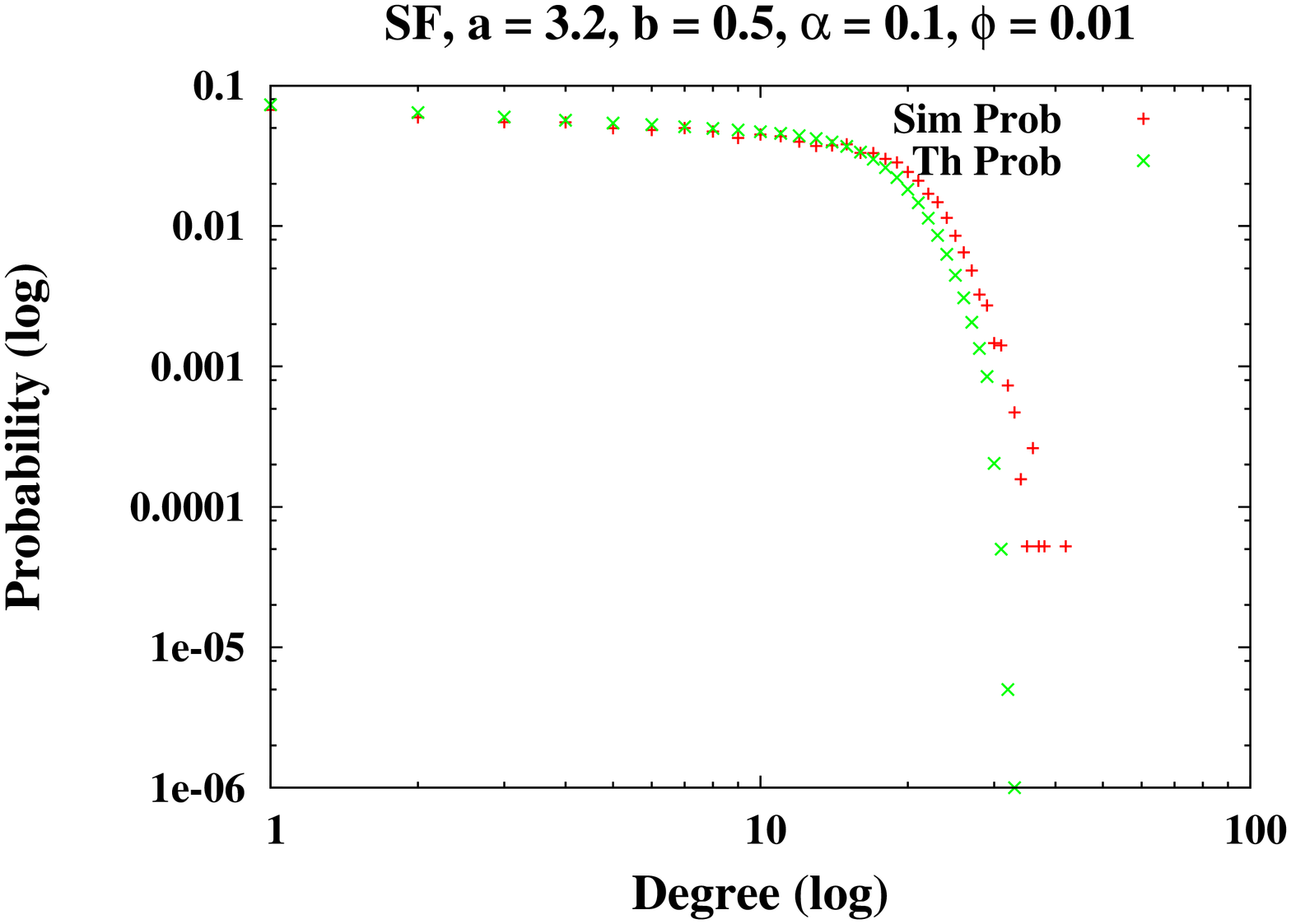}
   \includegraphics[angle=270,width=.45\linewidth]{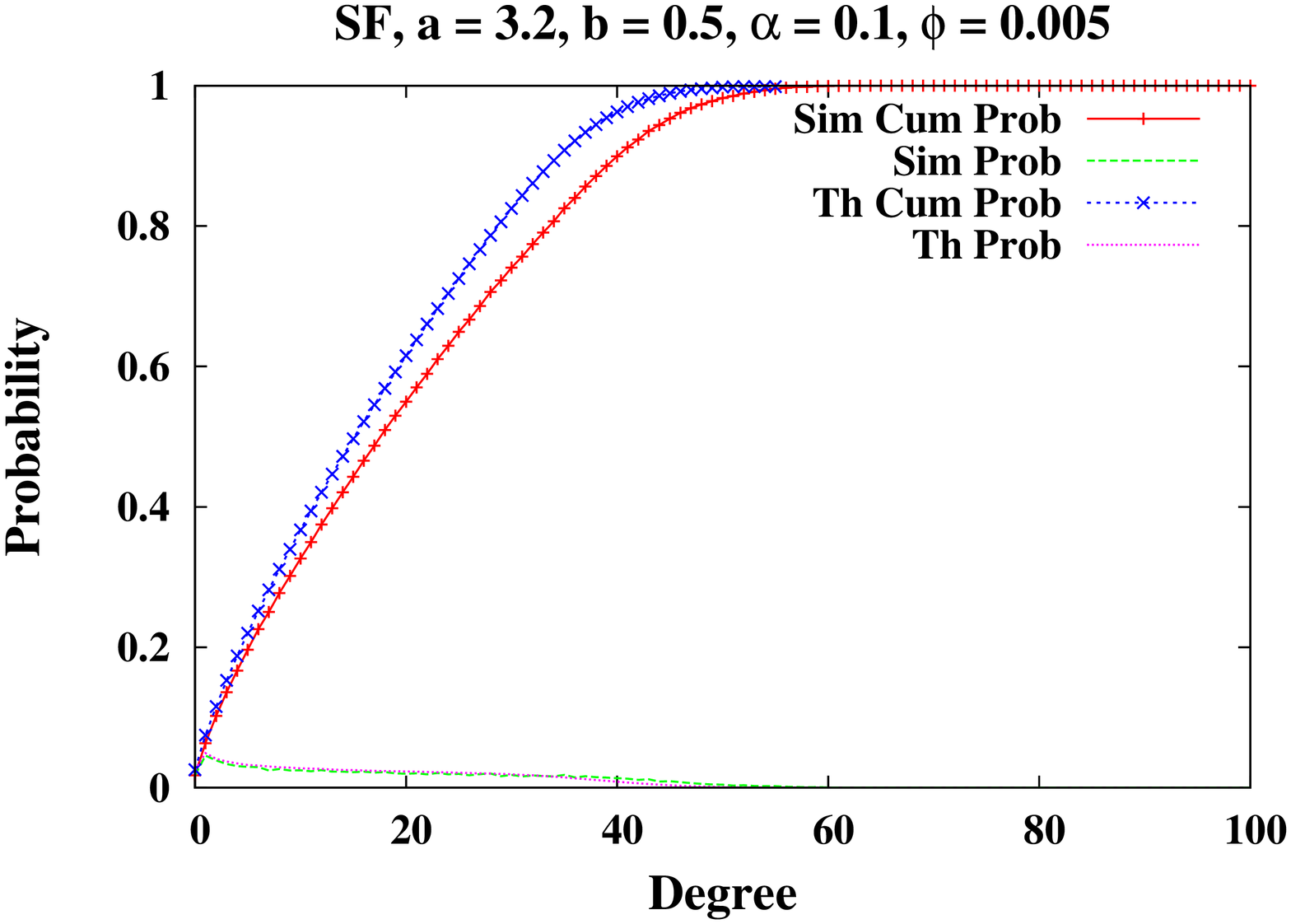}
   \includegraphics[angle=270,width=.45\linewidth]{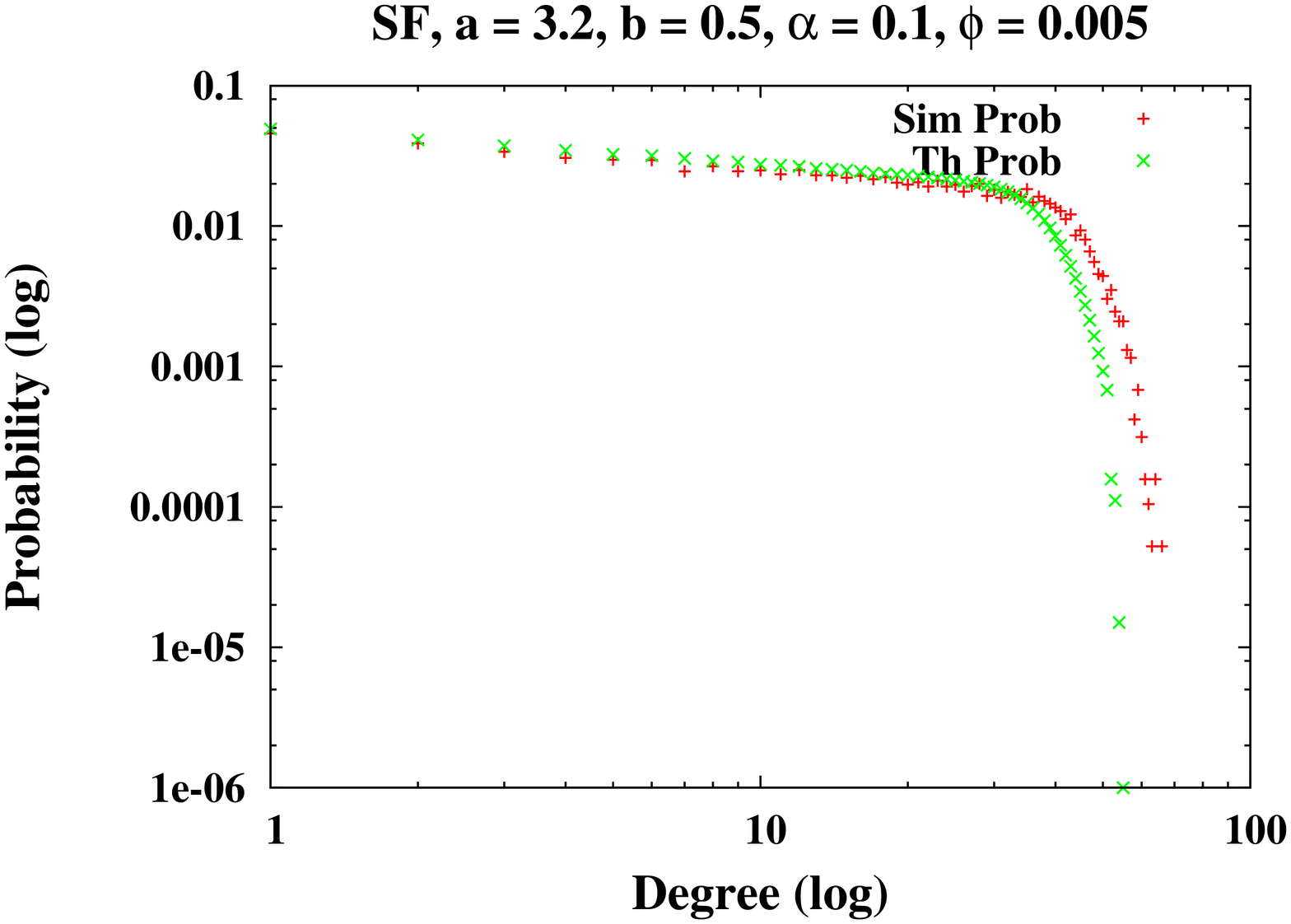}
   \includegraphics[angle=270,width=.45\linewidth]{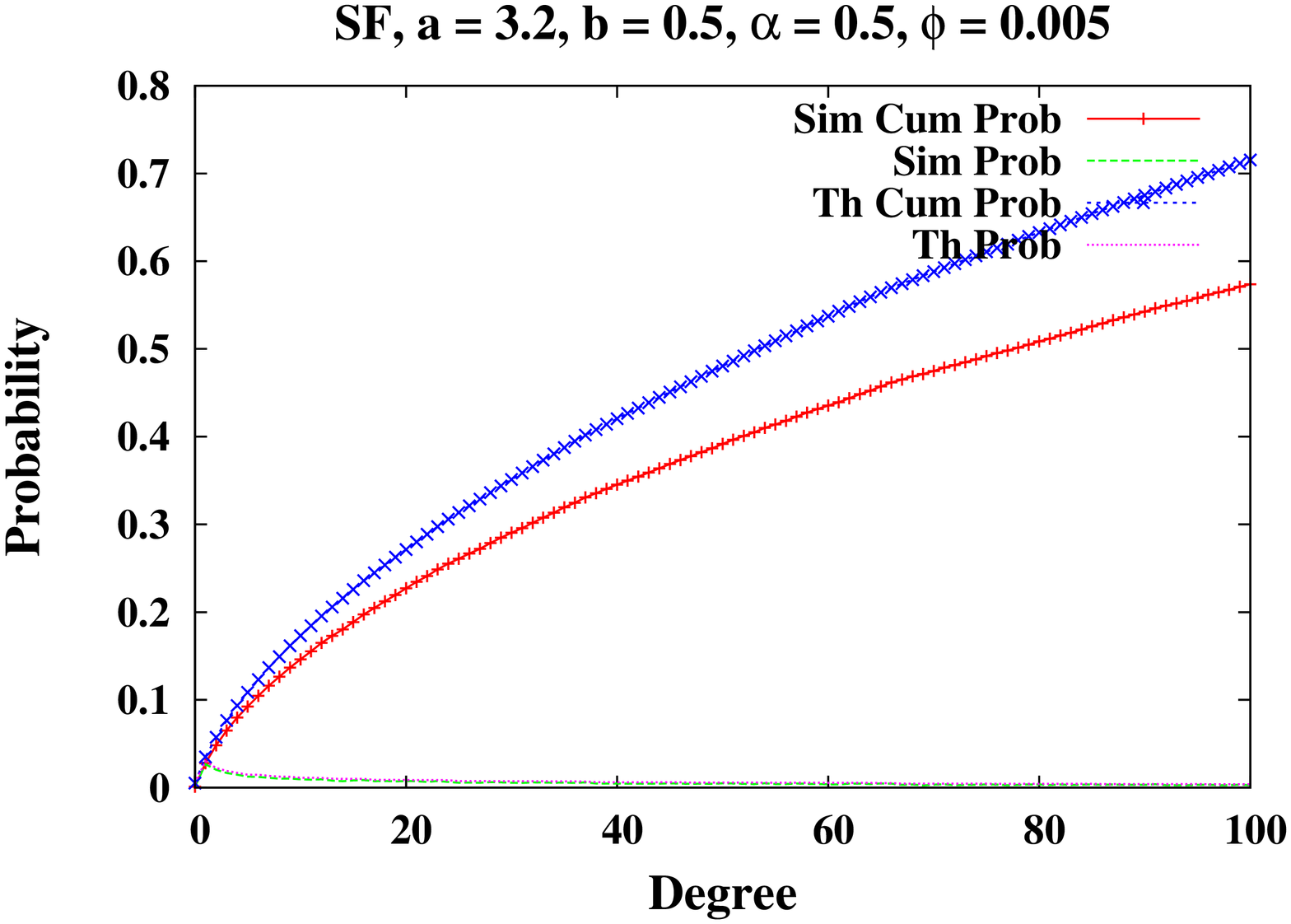}
   \includegraphics[angle=270,width=.45\linewidth]{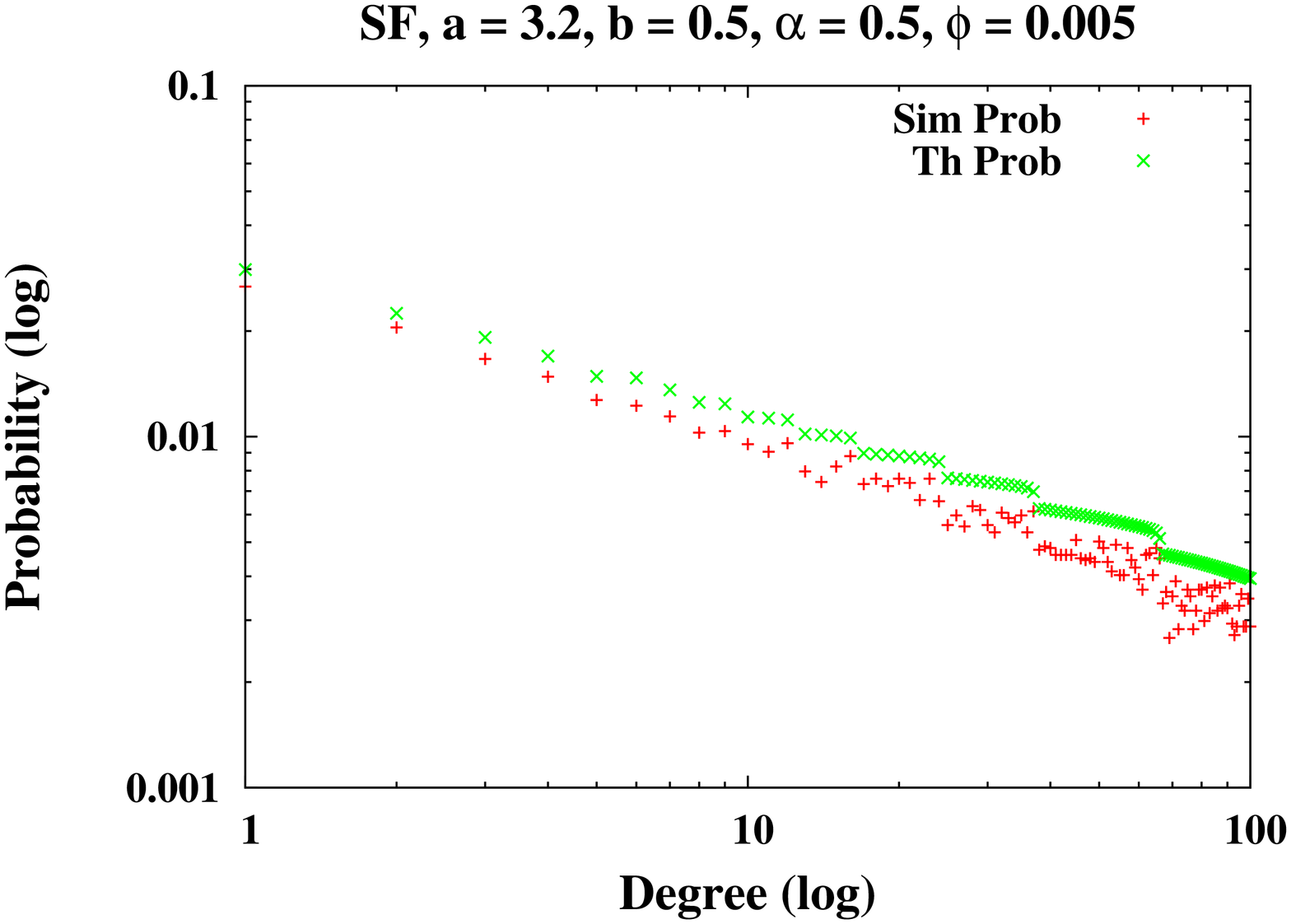}
   \includegraphics[angle=270,width=.45\linewidth]{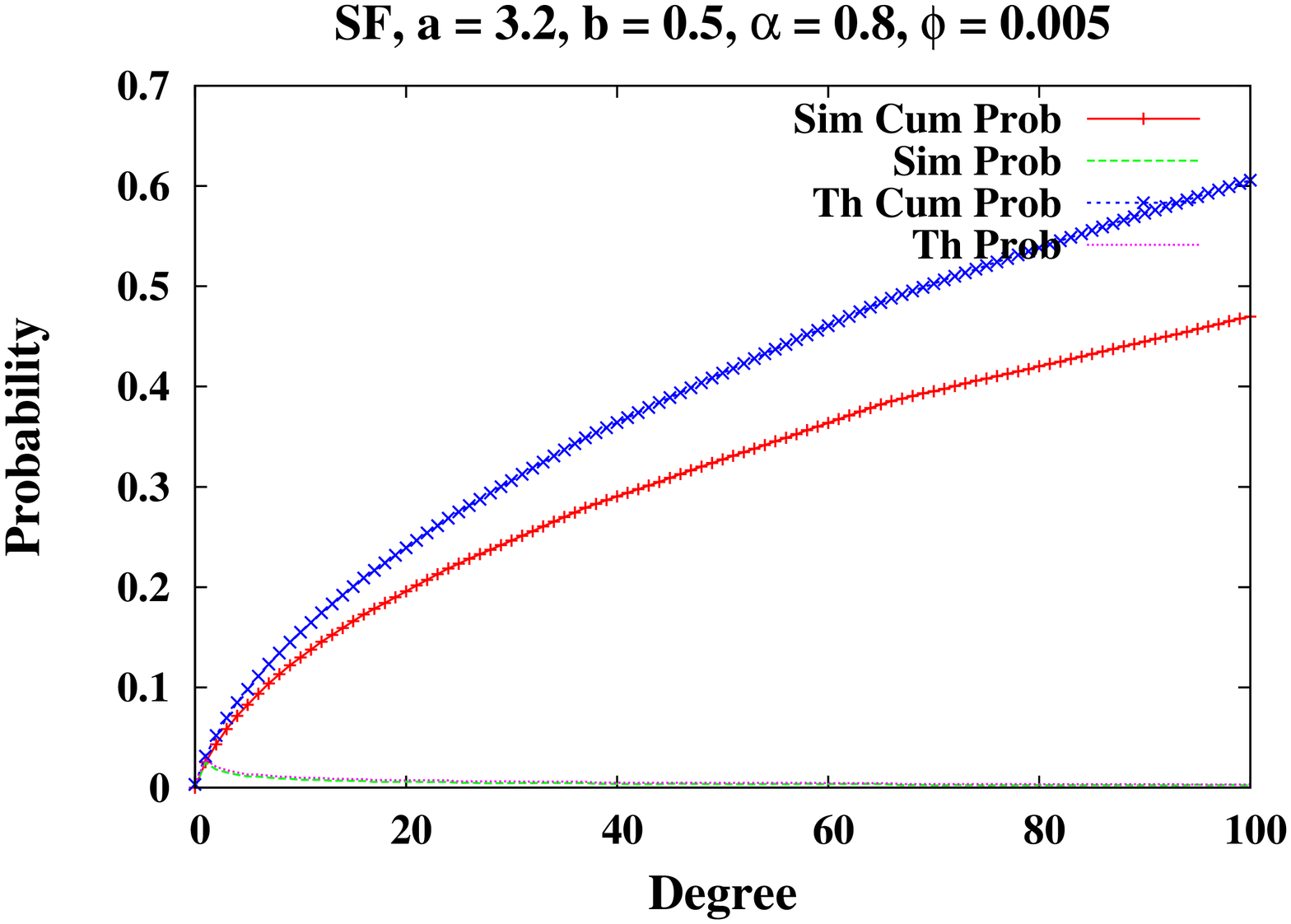}
   \includegraphics[angle=270,width=.45\linewidth]{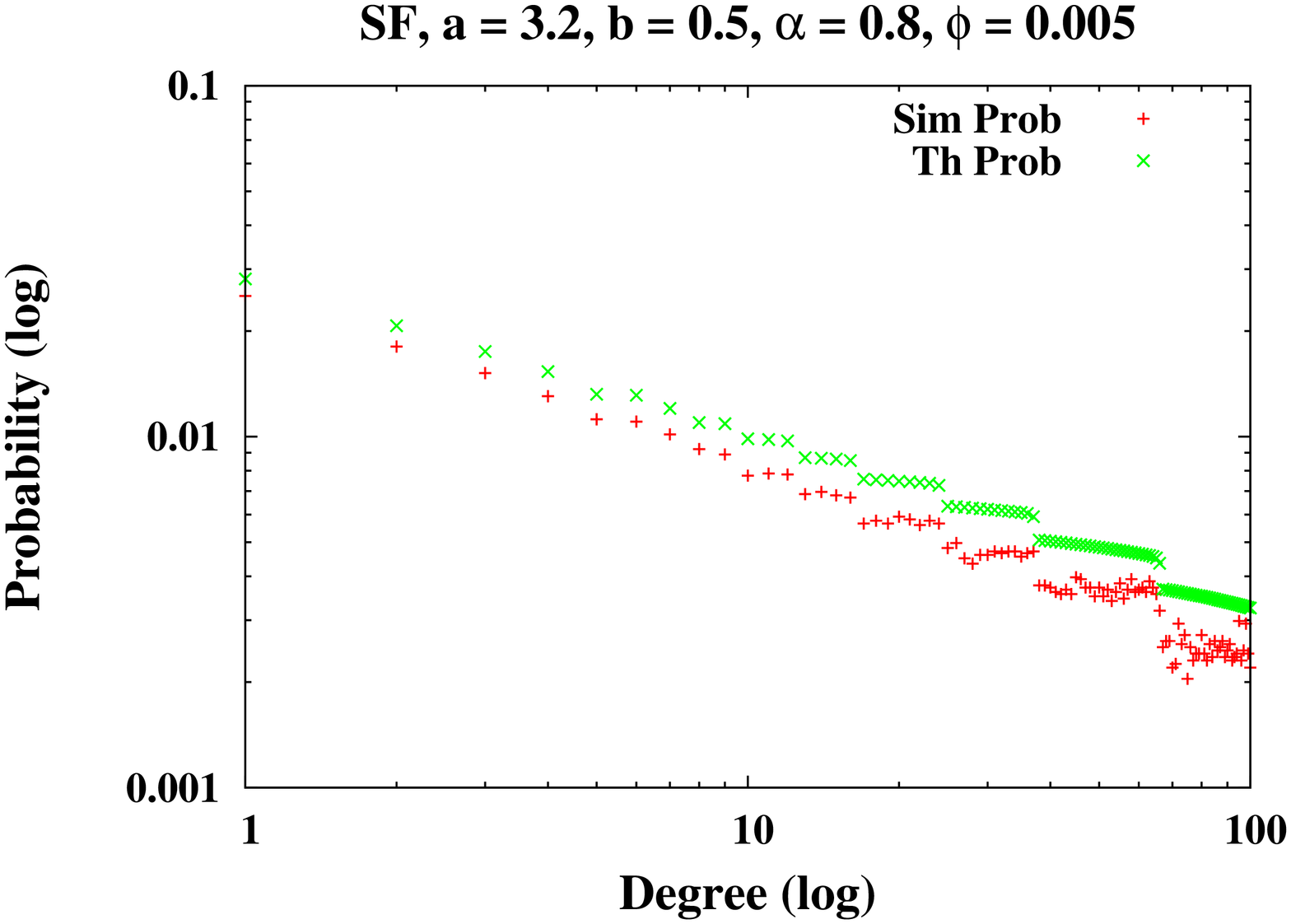}
   \caption{Degree probability and cumulative degree probability varying $\alpha, \phi$ on the left side; degree probability in log scale on the right side; results obtained through simulation (Sim) and the mathematical modeling (Th); Scale Free networks $a=3.2, b=0.5, |\Pi|=1167$}
   \label{fig:fig_sf13}
\end{figure}

%%%%%%%%%%%%%%%%%%%%%%%%%%%%%%%%%%%%%%%%
\begin{figure}
   \centering
   \includegraphics[angle=270,width=.45\linewidth]{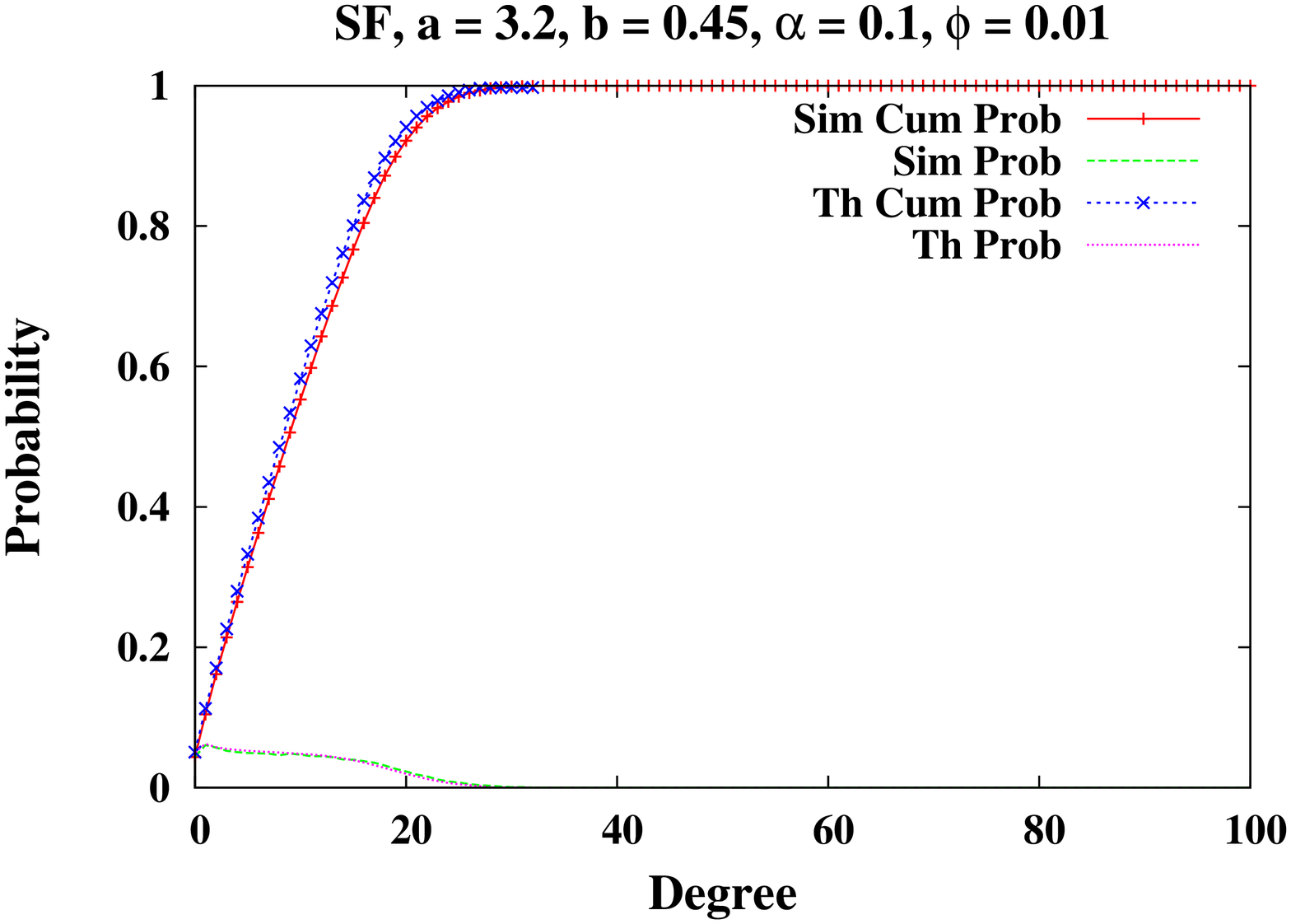}
   \includegraphics[angle=270,width=.45\linewidth]{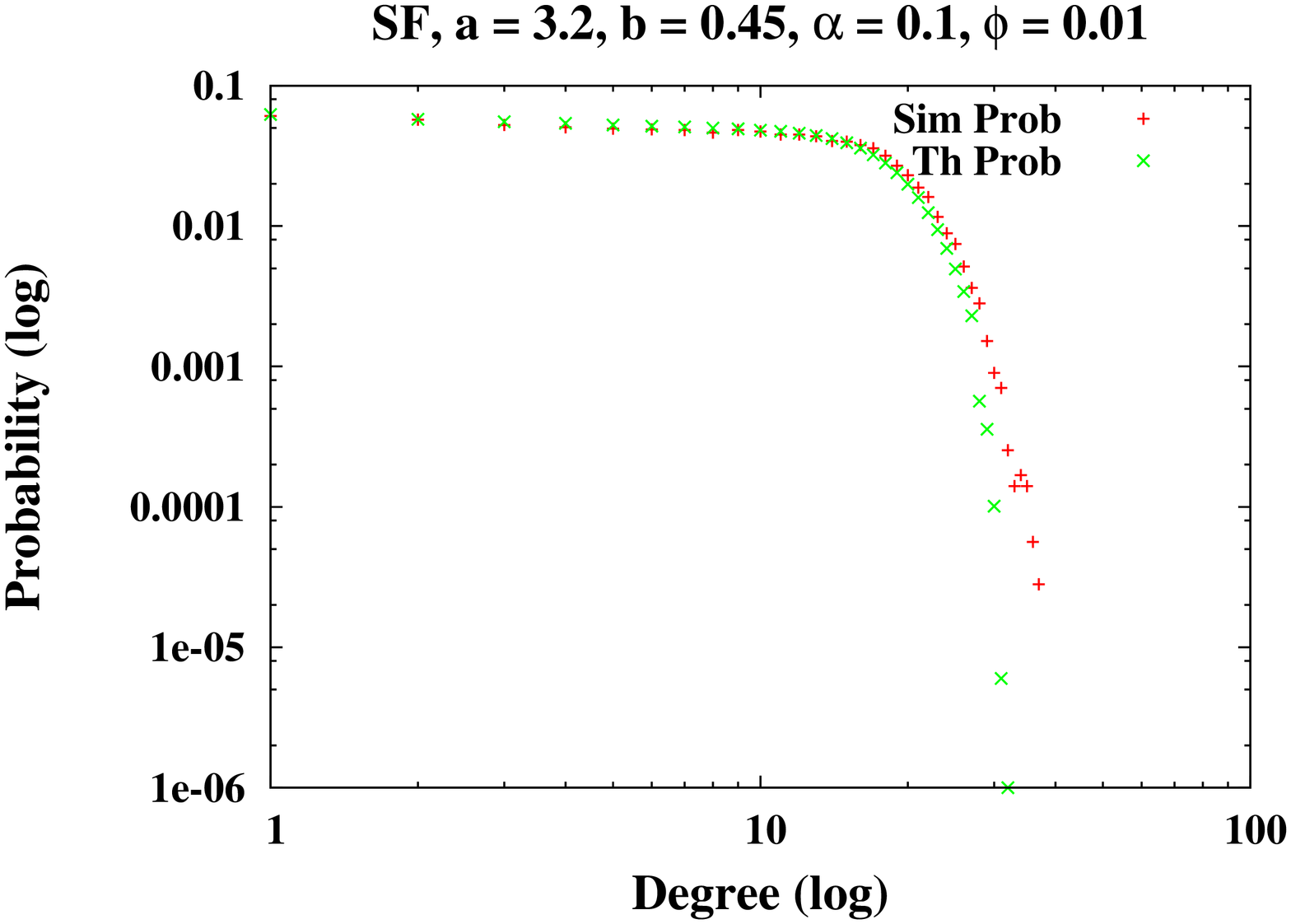}
   \includegraphics[angle=270,width=.45\linewidth]{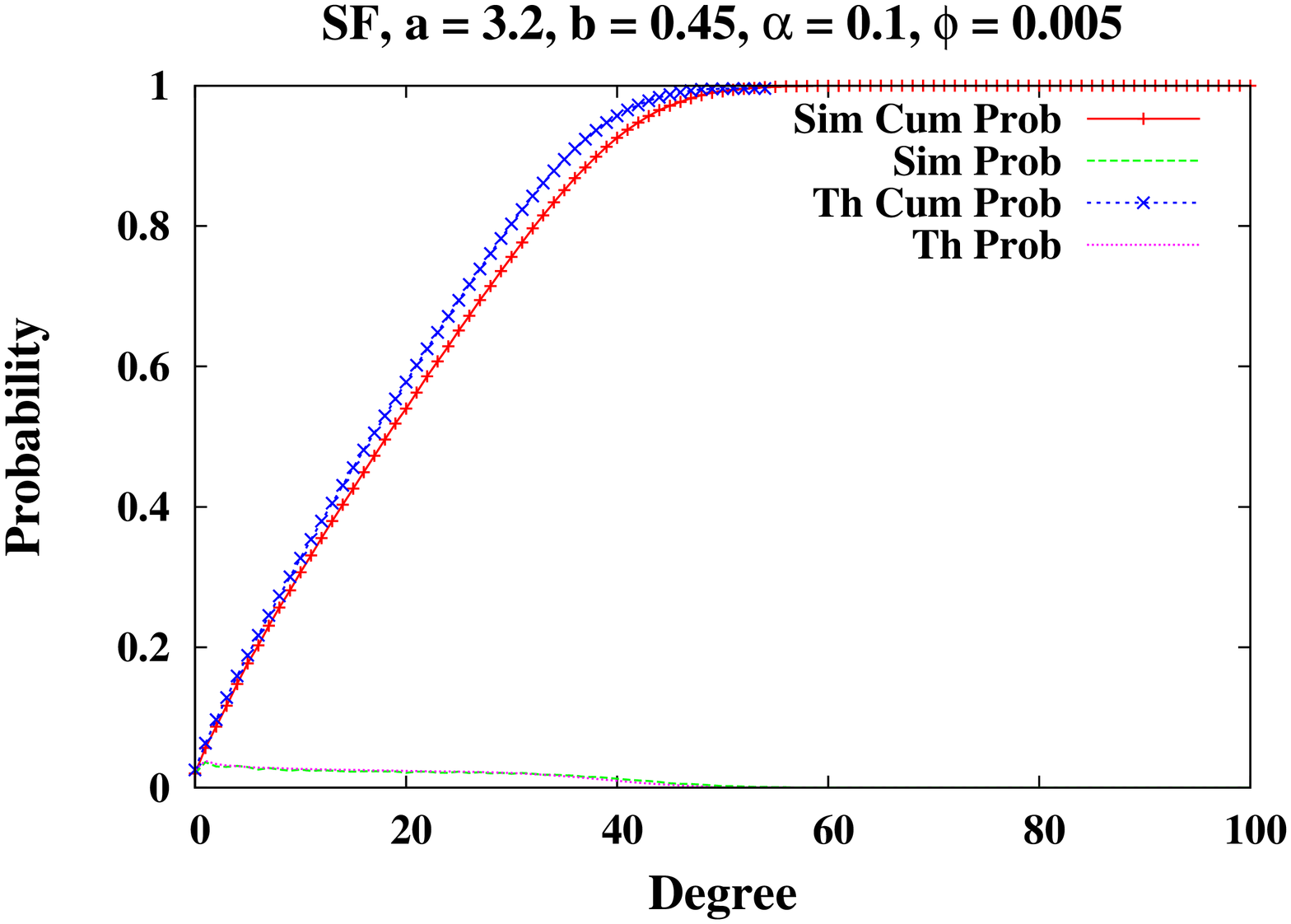}
   \includegraphics[angle=270,width=.45\linewidth]{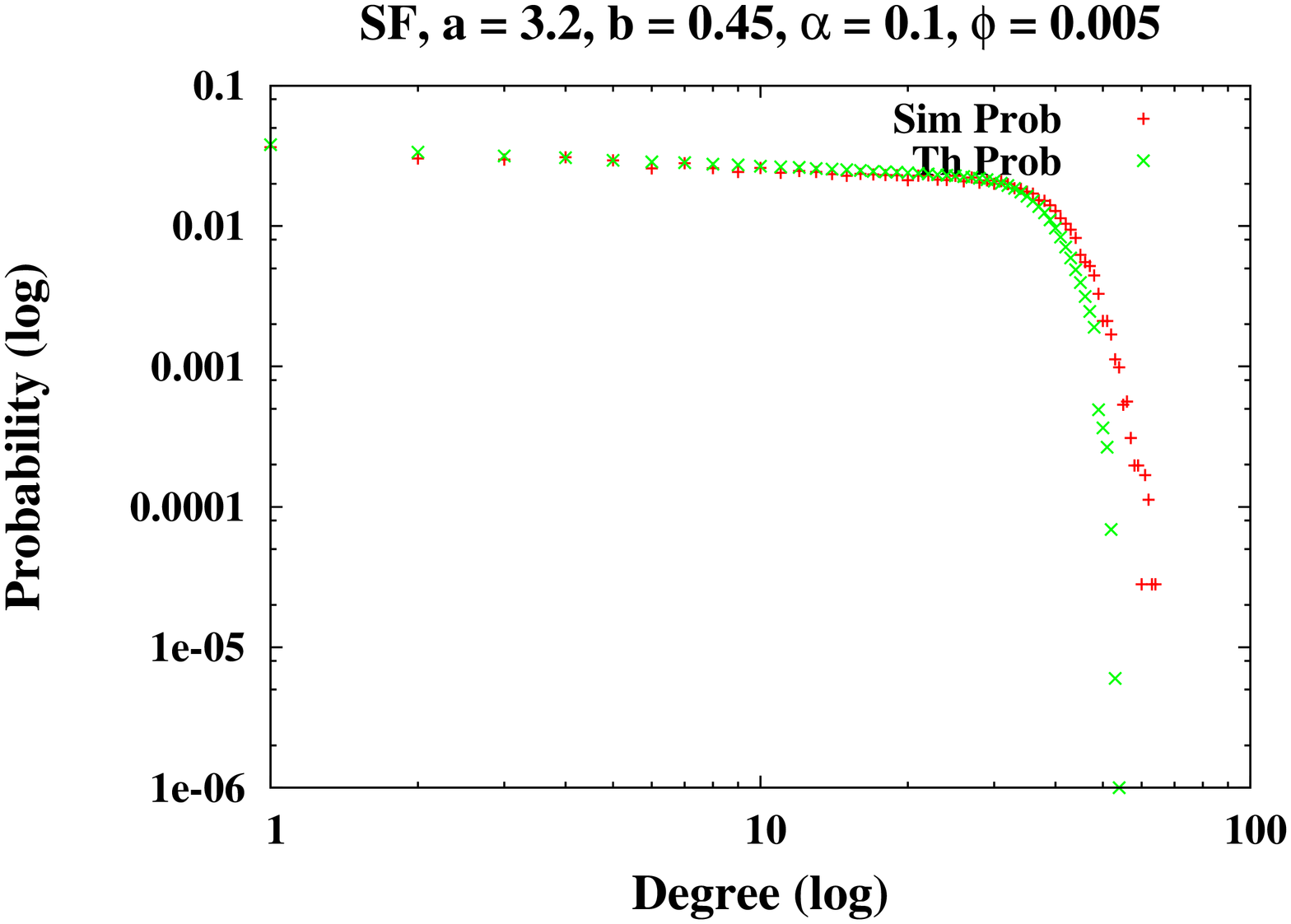}
   \includegraphics[angle=270,width=.45\linewidth]{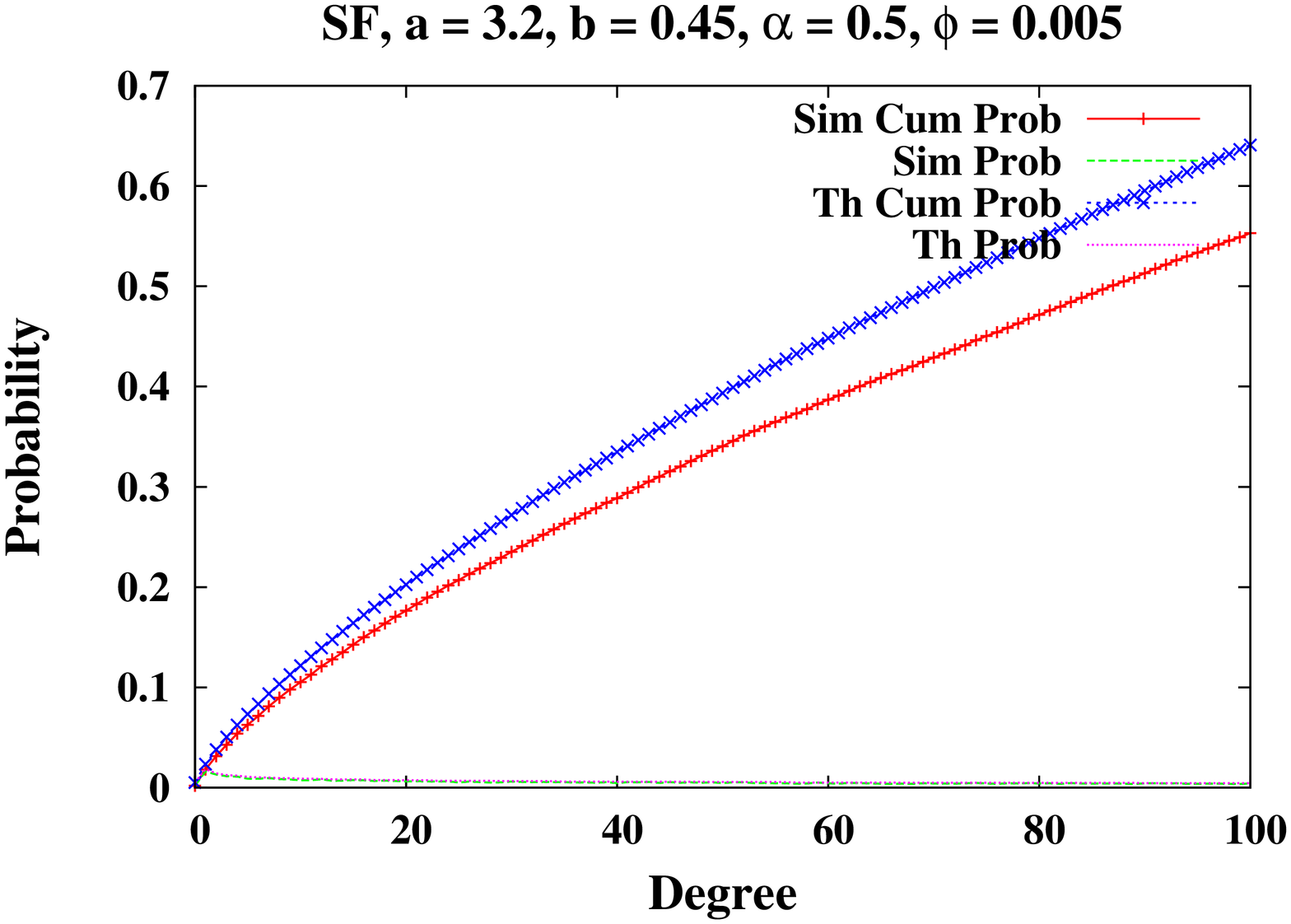}
   \includegraphics[angle=270,width=.45\linewidth]{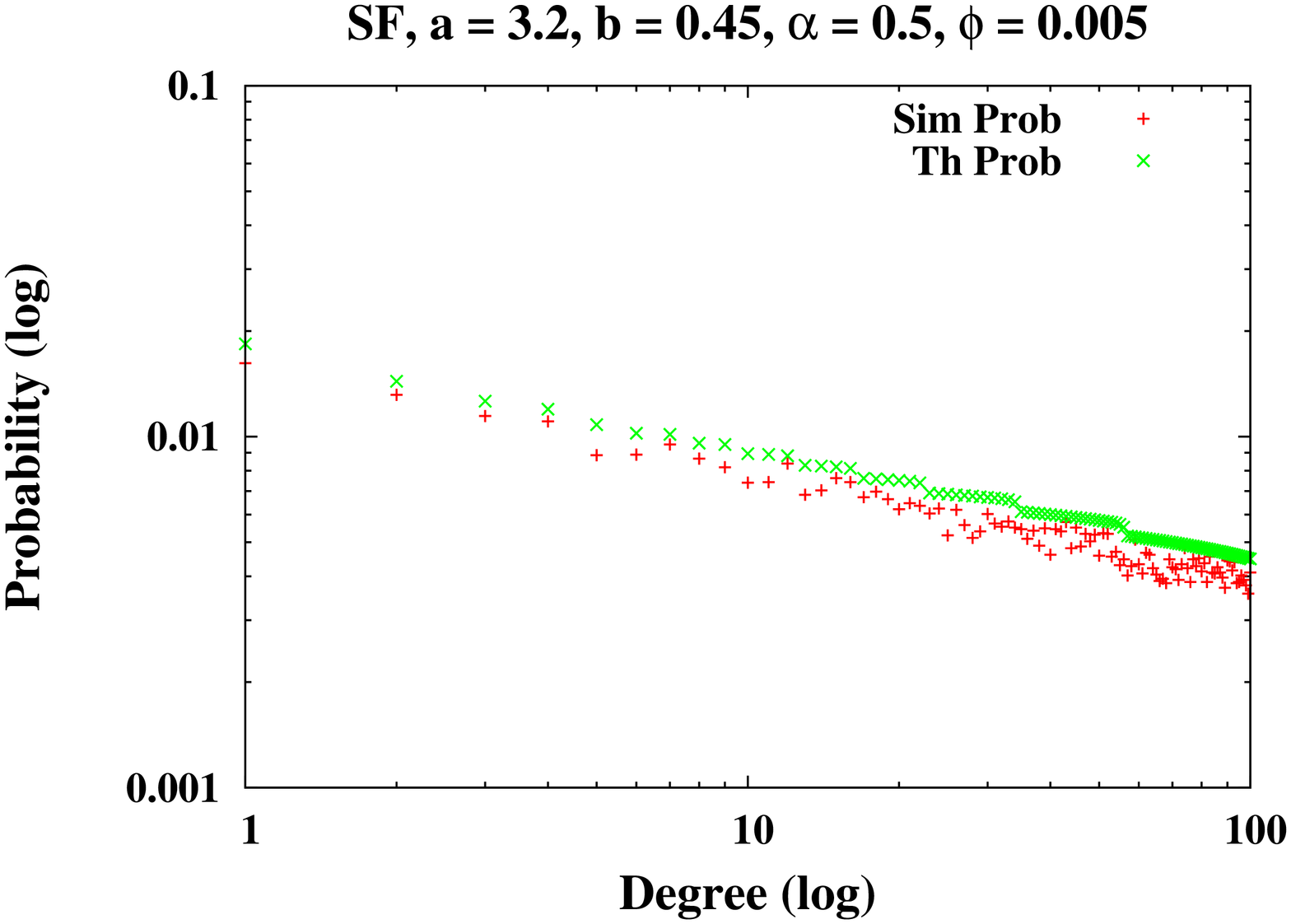}
   \includegraphics[angle=270,width=.45\linewidth]{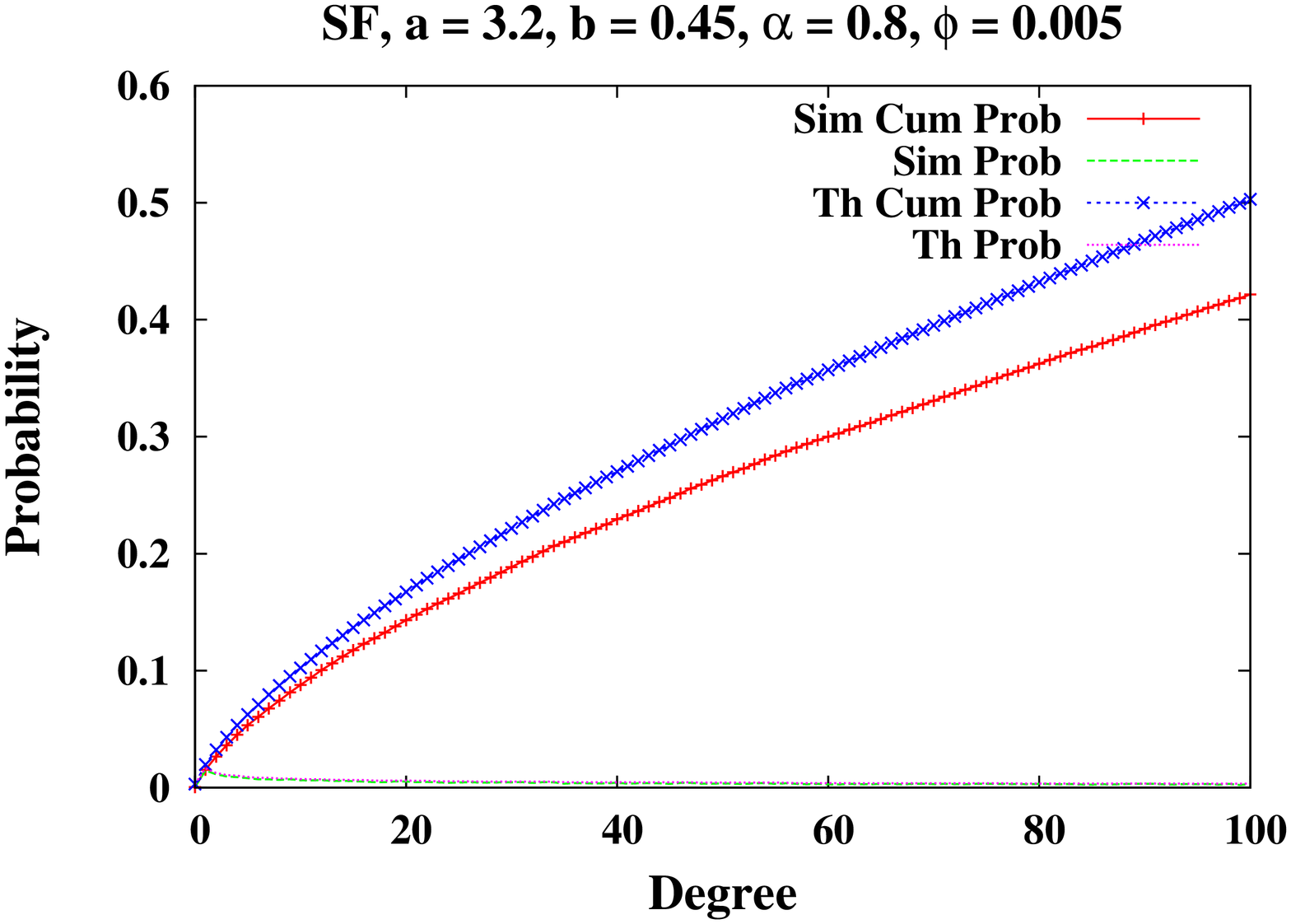}
   \includegraphics[angle=270,width=.45\linewidth]{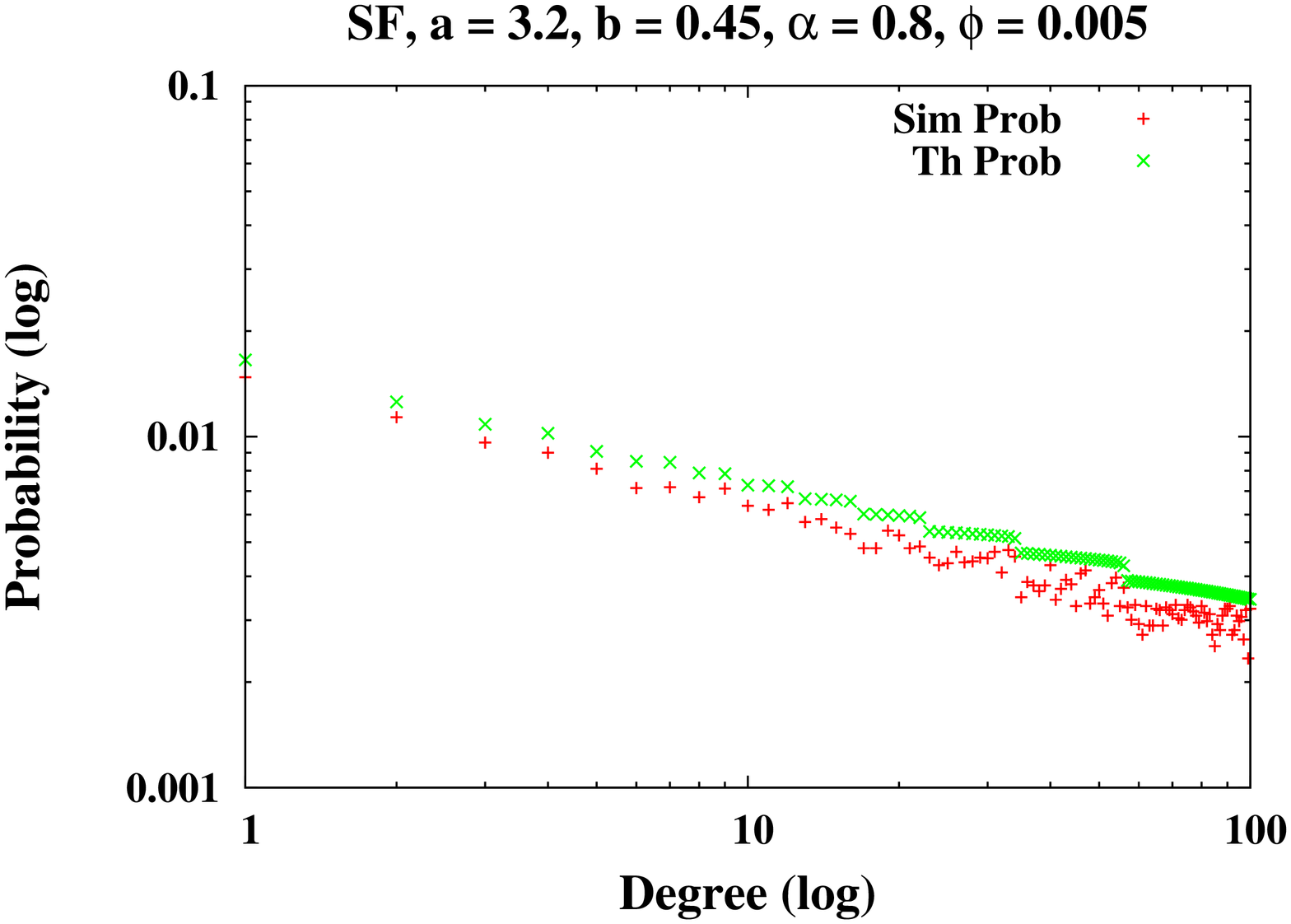}
   \caption{Degree probability and cumulative degree probability varying $\alpha, \phi$ on the left side; degree probability in log scale on the right side; results obtained through simulation (Sim) and the mathematical modeling (Th); Scale Free networks $a=3.2, b=0.45, |\Pi|=2196$}
   \label{fig:fig_sf14}
\end{figure}

\begin{figure}
   \centering
   \includegraphics[angle=270,width=.7\linewidth]{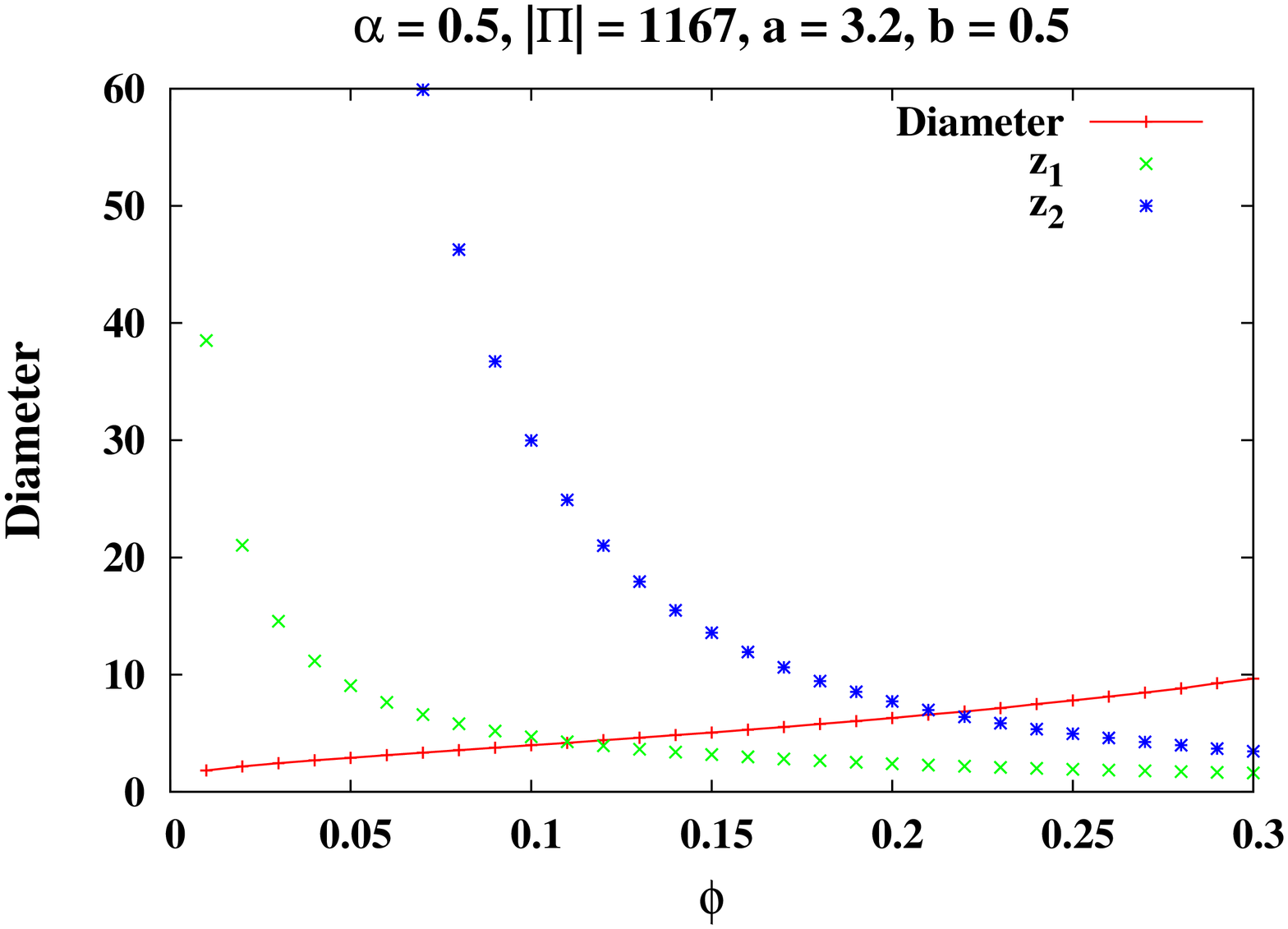}
   \caption{Diameter and average number of first neighbours of scale-free networks, when varying $\phi$, calculated using Equation (\ref{eq:diam})}
   \label{fig:diam_sf}
\end{figure}

Finally, Figure \ref{fig:diam_sf} reports the estimated diameter (together with the average number of first and second neighbours $z_1, z_2$) of scale-free networks, built with $a=3.2, b=0.5, |\Pi| = 1167$, obtained when $\alpha=0.5$, while varying $\phi$.
Also in this case the diameter of the network grows with $\phi$. It is worth noting that, as discussed, in this case the desired topology of these networks is different from that considered for random graphs, being the former a desired topology following a power law distribution, while the latter follows a Poisson distribution. 
Our results show that, with these settings, the average number of first neighbours $z_1$ is (slightly) lower in scale-free networks (even if the number of nodes in the considered network is a bit higher than the $1000$ nodes of random graphs). 
% The trend is however the same, i.e.~the diameter grows with $\phi$, since for higher values of the failure rate $\phi$, in the steady state the networks lose their scale-free properties, as already shown in previous charts. 
% In fact, it is well known that 
It is interesting to observe that theoretical results on scale free nets revealed that, depending on the exponent of the power law characterizing the scale free net, the network diameter ranges from $\log{|\Pi|}/\log \log{|\Pi|}$ down to $\log \log{|\Pi|}$
% , meaning that the diameter remains almost constant with respect to the number of peers $|\Pi|$ in the network 
\cite{Verlag03structuralproperties,simutools,pyun,bollobas_diam}. 
% This is not the case for the networks obtained with high values of the failure rate $\phi$.
In this case, it is worth noticing that the network diameter of the resulting overlay augments with $\phi$, thus confirming that if the attachment rate at a peer is not sufficient, the overlay loses the characteristics of the desired topology.

%%%%%%%%%%%%%%%%%%%%
%% CONCLUSIONS
%%%%%%%%%%%%%%%%%%%%

\section{Conclusions}
\label{sec:conc}

This paper presented a mathematical model of unstructured, self-orga\-ni\-zing overlay networks in faulty peer-to-peer systems. A distributed protocol has been considered, where nodes try to maintain a desired degree, coping with node failures. 
An analysis of the protocol has been provided, and numerical results coming from the obtained mathematical tool have been compared with those obtained through simulation. In essence, the two different approaches provide same outcomes. Different types of network topologies have been considered, i.e.~networks with nodes having the same desired degree, random graphs and scale-free networks.

Results demonstrate that in presence of a non-negligible failure rate, peers need a high attachment rate to cope with node faults. Otherwise, they are not be able to maintain their desired degree. This is important also to control the topology of the evolving network.
Hence, a final remark is that the mathematical tool provided in this paper can be factually exploited to dynamically adapt the peers' attachment rate, based on their desired degree and on the failure rate they are experiencing, so as the preserve the desired topology of the network.

The provided model can be extended in several ways. 
In this model, peers were treated uniformly, all having the same failure and attachment rates. 
A possibility is to replace $\alpha, \phi$ parameters with functions that may depend on several factors like, for instance, the gap between the actual and the desired degree, the actual degree itself, etc. When applied to the attachment rate, these parameters would implement some form of preferential attachment. When applied to the failure rate, forms of targeted attacks may be modeled.
Then, the random selection of novel neighbours could be replaced with mechanisms that employ a local search, e.g.~by limiting the peers' selection over $2^{nd}$, or $3^{rd}$ neighbours.



%% References

%% References with bib TeX database:
% \section*{References}

\bibliographystyle{abbrv}
\bibliography{biblio}






% \end{thebibliography}

\end{document}